\documentclass[a4paper,11pt]{article}
\usepackage[dvipsnames]{xcolor}
\usepackage[utf8]{inputenc}
\usepackage[bbgreekl]{mathbbol}
\usepackage{geometry}

\usepackage{xcolor}
\usepackage{amsmath, array, amssymb, amsfonts,amsthm}
\usepackage{upgreek}
\usepackage{xfrac}
\usepackage[inline]{enumitem}
\setlist{nolistsep}
\usepackage[french, english]{babel}
\usepackage[all]{xy}
\usepackage{txfonts}  
\usepackage{sectsty}
\usepackage{booktabs}
\usepackage{caption}
\usepackage{dsfont}
\usepackage{mathtools}
\usepackage{slashed}
\usepackage[makeroom]{cancel}
\usepackage[hidelinks]{hyperref}
\usepackage{textcomp}
\usepackage{multicol}
\usepackage[title]{appendix}
\usepackage[square,numbers,sort,compress,semicolon,merge]{natbib}
\let\cite\citep 
\usepackage{hyperref}
\hypersetup{colorlinks=true, urlcolor=blue, citecolor=blue, linktoc=page}

\usepackage{tikz}
\usepackage{tikz-cd}
\usetikzlibrary{cd}
\tikzcdset{
arrow style=tikz,
diagrams={>={Straight Barb[scale=0.8]}}
}

\DeclareMathAlphabet{\mathpzc}{OT1}{pzc}{m}{it}

\usepackage{mathrsfs}
\usepackage{footmisc}

\makeatletter
\renewcommand*\env@matrix[1][\arraystretch]{%
  \edef\arraystretch{#1}%
  \hskip -\arraycolsep
  \let\@ifnextchar\new@ifnextchar
  \array{*\c@MaxMatrixCols c}}
\makeatother

\newcommand{\defeq}{\vcentcolon=}
\newcommand{\rdefeq}{=\vcentcolon}
\renewcommand\P{\mathcal{P}}
\newcommand\M{\mathcal{M}}

\newcommand\RR{\mathbb{R}}
\newcommand\CC{\mathbb{C}}
\newcommand\C{\mathcal{C}}
\renewcommand\1{\textbf{1}}
\newcommand\id{\textit{id}}

\newcommand\G{\mathcal{G}}
\renewcommand\H{\mathcal{H}}
\newcommand\A{\mathcal{A}}

\renewcommand\S{\mathcal{S}}

\newcommand\SU{\mathcal{SU}}
\newcommand\U{\mathcal{U}}
\newcommand\SO{\mathcal{SO}}

\newcommand\K{\mathcal{K}}
\newcommand\J{\mathcal{J}}
\renewcommand\O{\mathcal{O}}
\newcommand\GL{\mathcal{GL}}
\newcommand\D{\mathcal{D}}

\newcommand\vphi{\varphi}
\newcommand\vkappa{\varkappa}

\newcommand\sT{\mathsf{T}}

\newcommand\sS{\mathsf{S}}

\renewcommand\epsilon{\varepsilon}

\newcommand\rarrow{\rightarrow}

\newcommand\so{\mathfrak{so}}

\renewcommand\t{\tilde}

\renewcommand\b{\bar }
\newcommand\w{\wedge}
\renewcommand\d{\partial}
\newcommand\s{\sigma}
\newcommand\bs{\boldsymbol}

\renewcommand\-{^{-1}}
\newcommand\Ad{\text{Ad}}
\newcommand\ad{\text{ad}}
\renewcommand\id{\text{id}}
\renewcommand\1{\mathds{1}}

\def\munderline#1#2{\color{#1}\underline{{\color{black}#2}}\color{black}}

\makeatletter

\newcommand{\Rmnum}[1]{\expandafter\@slowromancap\romannumeral #1@}
\makeatother

\makeatletter
\newcommand{\leqnomode}{\tagsleft@true\let\veqno\@@leqno}
\newcommand{\reqnomode}{\tagsleft@false\let\veqno\@@eqno}
\makeatother

\DeclareMathOperator{\Diff}{Diff}
\DeclareMathOperator{\Aut}{Aut}
\DeclareMathOperator{\Tr}{Tr}

\DeclareMathOperator{\im}{Im}

\theoremstyle{definition}

\makeatother

\begin{document}


\title{ 
Note on the bundle geometry of field space, \\ variational connections, the dressing field method, \\ \&  presymplectic structures of  gauge theories over bounded regions 
}
\author{J. François${\,}^{a}$, N. Parrini${\,}^{a}$, N. Boulanger${\,}^{a}$}
\date{}

\maketitle
\begin{center}
\vskip -0.8cm
\noindent
${}^a$ Service de Physique de l'Univers, Champs et Gravitation, Universit\'e de Mons -- UMONS\\
20 Place du Parc, B-7000 Mons, Belgique.
\end{center}
%
\vspace{-3mm}

\begin{abstract}
In this note, we consider how the bundle geometry of field space interplays with the covariant phase space methods so as to allow to write results of some generality on the presymplectic structure of invariant gauge theories coupled to matter. We obtain in particular the generic form of Noether charges associated with field-independent and field-dependent gauge parameters, as well as their Poisson bracket. We also provide the general field-dependent gauge transformations of the presymplectic potential and 2-form, which clearly highlight the problem posed by boundaries in generic situations. 
We then conduct a comparative analysis of  two  strategies recently considered to evade the boundary problem and associate a modified symplectic structure to a gauge theory over a bounded regions: namely the use of edge modes on the one hand, and of variational connections on the other. To do so, we first try to give the clearest geometric account of both, showing in particular that edge modes are a special case of differential geometric tool of gauge symmetry reduction known as the ``dressing field method". Applications to Yang-Mills theory and General Relativity reproduce or generalise several results of the recent literature.
\end{abstract}


\vspace{-3mm}

\tableofcontents

\bigskip

 
\section{Introduction}  
\label{Introduction}  

One motivation for the introduction of covariant phase space methods, at least in some of the foundational literature  \cite{Zuckerman1986, CrnkovicWitten1986, Crnkovic1987}, is to associate a symplectic structure to a gauge field theory -- given by a Lagrangian $L$ -- over some region $\Sigma$ of spacetime $M$, and so doing while keeping spacetime symmetries manifest (with in mind a possible covariant canonical or geometric quantization). Usually the field space $\Phi$ is considered the configuration space of the theory,  the space of solutions $\S$ is the phase space while its quotient by the gauge group $\H$ of the theory, $\M_\S\defeq \S/\H$ of the field equations, is the reduced phase space. The symplectic structure is complete once a symplectic 2-form on $\M_\S$ is given. From the functional variation of $L$, one derives the presympletic potential $\bs\theta_\Sigma$ which is a 1-form on field space, and its  functional variation defines the associated presymplectic  2-form  $\bs\Theta_\Sigma$ on $\Phi$. When, on-shell, the gauge directions are in its kernel, the latter descends  to a symplectic 2-form on $\M_\S$. This is the case when the region $\Sigma$ has no boundary or when adequate fall-off conditions for the fields $\phi$ are specified. 
But  when $\d\Sigma \neq \emptyset$ and there are no good reasons to assume $\phi=0_{|\d\Sigma}$, neither $\bs\theta_\Sigma$ nor $\bs\Theta_\Sigma$  descend on the reduced phase space.  We may refer to this as the \emph{boundary problem}. It naturally arises e.g. when one considers the decomposition of a well-defined symplectic structure for a boundaryless region $\Sigma$ into symplectic sub-structures associated with an arbitrary partition of $\Sigma$ into  subregions $\cup_i \Sigma_i$  sharing  fictitious boundaries $\d\Sigma_i$ -- a classical analogue of the problem of factorising  the Hilbert space of a quantum system into Hilbert subspaces, which in turn is closely related to the topic of entanglement entropy. 

In the wake of renewed interest in this issue, over the past few years  two strategies have been proposed to deal with this boundary problem in Yang-Mills theory and General Relativity: the ``\emph{edge modes}" strategy as introduced by Donnelly \& Freidel  in \cite{DonnellyFreidel2016}, and the use of variational connections on field space as advocated by Gomez \& Riello first in \cite{Gomes-Riello2017} and further developed  in \cite{Gomes-Riello2018, Gomes-et-al2018, Gomes-Riello2021} (see also \cite{Gomes2019, Gomes2019-bis, Riello2020, Riello2021}). Both essentially aim at providing a modified presymplectic structure that descends onto $\M_\S$ and that we will call \emph{basic} for reasons to be made clear in due time. 

 The former strategy, explored in various contexts \cite{Geiller2017, Geiller2018, Speranza2018, Geiller2019}, considers an \emph{extended phase spaces} comprising new degrees of freedom at $\d\Sigma$ -- the edge modes -- entering boundary counterterms added to $\bs\theta_\Sigma$ and $\bs\Theta_\Sigma$ enforcing their vanishing along gauge directions. According to its proponents, the main virtue of edge modes beyond addressing the boundary issue is that they reveal new \emph{physical} symmetries, sometimes called ``surface symmetries", to which charges can be associated with and whose Poisson algebra encodes important kinematical information.\footnote{For other works on extended phase spaces by edge modes, or edge states, in another sense of the term -- i.e. true boundary d.o.f. -- see e.g. \cite{Balachandran-et-al1995, Balachandran-et-al1996, Asorey-et-al2015}, and in the context of asymptotic boundaries and gravitational waves memory effects see \cite{Adami-et-al2020, Seraj2021}} For this reason,  it has been suggested that edge modes provide a new angle for a quantum gravity program \cite{Freidel-et-al2020-1, Freidel-et-al2020-2, Freidel-et-al2020-3}. 
 
 The use of connections received less systematic attention, but it has the  advantage of possibly relying only on the resources of the original  field space, without the need for introducing extra d.o.f. Its principal advocates also remarked that  modified presymplectic structure obtained via a connection seems to generalise the one obtained via  edge modes. No equivalent of the surface symmetries of edge mode seems to exist in this approach though. However, it was noted that its own special merit was to connect with the literature on dressings in gauge theory -- à la Dirac, see e.g. \cite{McMullan-Lavelle97, Lavelle-McMullan-Bagan2000} --  and in particular  to reproduce the so-called DePaoli-Speziale (DPS) ``dressing 2-form" \cite{De-Paoli-Speziale2018, Oliveri-Speziale2020, Oliveri-Speziale2020-II} relating the presymplectic potentials of the tetrad and metric formulations of GR. 
 
 It would seem that a careful comparative analysis of both  approaches weighting their respective strengths and their overlaps would be useful, clarifying the conceptual landscape and establishing clear bridges within the existing literature. Doing so is one of the objectives of this paper. 
 \medskip
 
  The difference in degree of formalisation would be a first hurdle to clear, the second approach being a priori much more geometric than the first. 
Fortunately, philosophers of physics interested in foundational issues in gauge field theories first remarked \cite{Teh2020} -- see also \cite{Teh-et-al2020} -- that the edge mode strategy could be seen as a special application of a geometric tool designed to reduce gauge symmetries called the ``\emph{dressing field method}" (DFM). In \cite{Francois2021}, the case was made thoroughly  that it is indeed so, and that many conceptual clarifications ensue.

A second hurdle is that both approaches have been mainly detailed via applications to specific examples. Even though one could compare the outputs of both methods in common examples, e.g. in the YM case as was done in \cite{Gomes-Riello2021} or \cite{Riello2021}, we aim for a more ambitious goal: We want to confront these two approaches in their most general  versions. To do so, we must first conduct the most general analysis possible of the relevant presymplectic structure of  gauge theories coupled to matter, restricting ourselves -- for reasons to be clarified later -- to theories that are strictly gauge invariant. This is our second goal in this note. 
To be specific, we will provide the generic form of the Noether charges associated with both field-independent and field dependent gauge parameters, as well as their off-shell Poisson bracket. We also derive the general field-dependent gauge transformations of the presymplectic potential and 2-form, from which the boundary problem will appear in its most general form. Doing all this, we will have generalised some of the results of  \cite{Francois2021} valid only for pure gauge theories.\footnote{Where was also conducted the analysis of pure gauge theories that are non-invariant, but whose classical gauge anomaly is $d$-exact. See the conclusion for further comments.} 
\medskip

To achieve this, we must start  one step higher in generality. Indeed, we find indispensable for conceptual clarity and technical efficiency to clearly articulate how the bundle geometry of field space interplays with covariant phase space methods.
We are thus lead to spell out the  $\H$-principal bundle structure of field space $\Phi$, whose base is the moduli space $\M\defeq\Phi/\H$ where gauge-invariant quantities live. Variational forms on $\Phi$ projecting to well-defined objects on the base are the so-called \emph{basic} forms. Solving the boundary problem thus means identifying strategies to construct basic versions of the initial presymplectic structure of a gauge theory. The broad issue is then to build the basic counterpart of a given variational form on $\Phi$. Variational principal connections and the DFM are  two methods to achieve this goal, as we propose to show. 
\bigskip

So, let us  recap what we intend to do in this note: 
In Section \ref{Geometry of field space}, we  describe in some details the bundle geometry of field space to give the most conceptual clarity on the kinematics at play. We  remind the definition of variational Ehresmann connections, which are fit for the analysis of invariant theories -- our main focus. As an hopefully informative aside, we  also define a generalisation know as \emph{twisted} connections that are well adapted for non-invariant theories. Then we  review the DFM, introducing the notion of field-dependent dressing fields. We show how basic forms are built via connections and dressings, and we stress the relations between the two approaches.

In Section \ref{Presymplectic structures of matter coupled gauge theories over bounded regions} we show how the bundle geometry of $\Phi$ interlaces with covariant phase space methods, thereby allowing to write down the above mentioned general results on the presymplectic structure of invariant  matter coupled gauge theories. We remark that, once such general formulae are given, the only computation needed for applications to specific examples is to derive the field equations and presymplectic potential from the Lagrangian at hand. We~illustrate the procedure in YM theory and in a Cartan geometric formulation of GR,  and recover standard results. 
We  also say a word about the generic physical interpretation of a Noether charge, relying on the affine structure of the space of connections to split it as a background contribution plus a measurable contribution, thereby connecting with the definition of charges of Abbott \& Deser \cite{Abbott-Deser1982, Abbott-Deser1982bis}.\footnote{Even though this discussion is somewhat in tension w.r.t. the goal --  discussed e.g. in \cite{Witten1986, Zuckerman1986, CrnkovicWitten1986, Crnkovic1987, Lee-Wald1990}  -- of producing basic presymplectic structures, which precisely requires to kill  all Noether charges.}
We therefore emphasize a geometric aspect of (generalised) Noether charges, as is typically done with covariant phase space methods, which is complementary to the  intrinsically cohomological nature of charges \cite{Barnich-Brandt2002}.

Finally, in Section \ref{Basic presymplectic structures} we give the most general versions of the basic presymplectic structure obtained respectively via connections and via the DFM. We  highlight the geometric origin of their structural similarities and stress the crucial differences, in particular regarding how ambiguities arise in both schemes. This last point is relevant to better understand the meaning of  surface symmetries  in the edge mode literature, as we show that these and coordinate  transformations in GR are on the same conceptual footing from the DFM viewpoint. Applications to YM theory and  GR allow to recover many results of the literature cited above. In particular we show how the DFM gives from first principles the unambiguous link between the presymplectic structures of GR in the tetrad formulation and in the metric formulation, thereby generalising the DPS dressing-2 form. 

In our Conclusion \ref{Conclusion}, beside giving a quick review of our results, we hint at  generalisations that we intend to pursue. 
Technical details are completed in Appendix \ref{Appendix}. For comparison with the body of the text,  we also give  a reformulation in terms of differential forms of the Abbott-Deser algorithm for defining charges in YM theory.


\section{Geometry of field space}
\label{Geometry of field space}

In this section we give a sense of the bundle geometry of field space. As it involves
 infinite dimensional vector spaces and manifolds, we defer to the relevant literature \cite{Frolicher-Kriegl1988, Kriegl-Michor1997} to back the soundness of extending any standard notion defined in the finite dimensional context to its infinite dimensional counterpart.  We aim for a correct conceptual picture rather than  perfectly mathematically rigorous one. 

\subsection{Field space as a principal bundle}
\label{Field space as a principal bundle}

The configuration space of a gauge theory is the space of gauge fields $\Phi =\A \times \Gamma(E)$, where $\A$ is the space of (Ehresmann of Cartan) connections on a $H$-bundle $\P$ over spacetime $M$, i.e. gauge potentials, and $\Gamma(E)\simeq \Omega^0_\text{eq}(\P, V)$ is the space  of sections of bundles $E$ associated with $\P$ via representations $(V, \rho)$  of  $H$, i.e. matter fields. 

The space $\Phi$ is an infinite dimensional Banach manifold, so is the gauge group $\H$ as an infinite dimensional  Lie group. Under proper restrictions (on either $\Phi$ or $\H$ \cite{Singer1978, Singer1981, Mitter-Viallet1981, Cotta-Ramusino-Reina1984, Abbati-et-al1989, Fuchs1995}), the moduli space $\M=\Phi/\H$ is  well-behaved as a manifold. Then,  $\Phi$ is a principal bundle over $\M$  with structure group $\H$, whose right action we denote $(\phi, \gamma) \mapsto R_\gamma \phi \defeq \phi^\gamma$.\footnote{Since $(\phi^\gamma)^{\gamma'}=(\phi^{\gamma'})^{\gamma^{\gamma'}}=(\phi^{\gamma'})^{\gamma'^{-1} \gamma \, \gamma'}=\phi^{\gamma\gamma'}$, this is indeed a right action: $R_{\gamma'} \circ R_\gamma=R_{\gamma\gamma'}$.} 
Explicitly of course, $\phi^\gamma=(A^\gamma, \vphi^\gamma)=(\gamma\-A\gamma + \gamma\-d\gamma,  \rho(\gamma)\-\vphi)$, as $\A$ and $\Gamma(E)$ are separately $\H$-principal bundles.  
The gauge orbit $\O_\H[\phi]$ of $\phi\in \Phi$ is a fiber over the gauge class $[\phi]\in \M$. The projection $\pi:\Phi \rarrow \M$, $\phi \mapsto \pi(\phi)=[\phi]$, is s.t. $\pi \circ R_\gamma = \pi$.

As a bundle, $\Phi$ is locally trivial:  given $\U \subset \M$, $\Phi_{|\U}\simeq \U \times \H$.  A trivialising (or local) section $\bs\s:\U \rarrow \Phi$, $[\phi] \mapsto \bs\s([\phi])$, is a  section of $\pi$, so that $\pi \circ \bs\s=\id_\U$.  If $\exists$ a global section $\bs\s:\M \rarrow \Phi$, then the bundle is trivial, $\Phi \simeq \M \times \H$.  
Given $\bs\s_i$ and $\bs\s_j$ sections over  $\U_i,\U_j\subset \M$ s.t. $\U_i \cap \U_j \neq \emptyset$, on the overlap $\bs\s_j=R_{ \bs h_{ij}}\bs\s_i=  \bs\s_i^{\bs h_{ij}}$ where $\bs h_{ij}:  \U_i \cap \U_j \rarrow \H$ is a transition function. The set $\{\bs h_{ij}\}$ of transition functions  subordinated to a covering $\{\U_i\}_{i\in I \subset\mathbb{N}}$ of $\M$ are local data from which it is possible to reconstruct the bundle $\Phi$. 
A trivialising section $\bs\s$ selects a single representative by gauge orbit $\bs\s([\phi])=\phi \in \O_\H[\phi]$, $\forall [\phi] \in \U$, it is thus a gauge choice (a gauge fixing).\footnote{If the factor $\A$ in $\Phi$ is the space of connections of a $S\!U(N)$-bundle $P$ over a compact spacetime manifold $M$, then no such global section exists (this is the Gribov ambiguity \cite{Gribov}). As Singer says in the abstract of his paper  \cite{Singer1978}, in this case ``no gauge fixing is possible".} A transition function $\bs h([\phi]) \in \H$, that allows to switch from  one gauge choice to another, is a gauge transformation that depends only on the gauge class $[\phi]$ of the gauge fields $\phi$. This is close to the notion of \emph{field-dependent gauge transformations} often encountered in the literature, but not quite the same thing yet. Another natural geometric concept, described next, better captures this notion.

 The natural transformation group of $\Phi$ is its automorphism group 
 $\bs{\Aut}(\Phi)\defeq\left\{\,  \bs{\Psi}: \Phi \rarrow \Phi\, |\,  \bs{\Psi} \circ R_\gamma =R_\gamma \circ \bs{\Psi} \,\right\}$. Only $\bs{\Psi} \in \bs{\Aut}(\Phi)$ project to  well-defined $\bs{\psi} \in \bs{\Diff}(\M)$,   which is the \emph{physical} transformation group permuting physical states. As usual, the subgroup of vertical automorphisms 
 $\bs{\Aut}_v(\Phi) \defeq \left\{\,  \bs{\Psi} \in \bs{\Aut}(\Phi)  \, |\,  \pi \circ \bs{\Psi} = \pi\,\right\}$ is isomorphic to the gauge group $\bs{\H}\defeq \left\{ \bs{\gamma}:\Phi \rarrow \H \, |\, R^\star_\gamma \bs{\gamma}(\phi)=\gamma\- \bs{\gamma}(\phi) \gamma \right\}$ by the correspondance $\bs{\Psi}(\phi)=R_{\bs{\gamma}(\phi)}\phi=\phi^{\bs{\gamma}(\phi)}$. 
 Now, the gauge group $\bs{\H}$ is indeed the geometric underpinning of the notion of \emph{field-dependent gauge transformations}.
 We have the characteristic short exact sequence (SES) of groups associated with the bundle $\Phi$,  
 \vspace{-1mm}
 \begin{align}\label{SESgroups-inf}
\makebox[\displaywidth]{
\hspace{-18mm}\begin{tikzcd}[column sep=large, ampersand replacement=\&]
\&0     \arrow [r]         \& \bs{\Aut}_v(\Phi) \simeq \bs{\H}     \arrow[r, "\iota"  ]          \& \bs{\Aut}(\Phi)       \arrow[r, "\t\pi"]      \&  \bs{\Diff}(\M)        \arrow[r]      \& 0.  
\end{tikzcd}}  \raisetag{3.4ex}
\end{align}
Without a splitting of this SES, one cannot decompose uniquely an element of $\bs\Aut(\Phi)$ into a vertical automorphism and a diffeomorphism of the physical configuration space $\M$.

\paragraph{Tangent and vertical bundles} 

The connection space $\A$ is an affine space modelled on $\Omega^1_\text{tens}(\P, \text{Lie}H)$, while $\Gamma(E)\simeq \Omega^0_\text{tens}(\P, V)$ is a vector space. Therefore, the tangent space at $\phi \in \Phi$ is $T_\phi\Phi =T_A\A \oplus T_\vphi \Gamma(E)\simeq \Omega^1_\text{tens}(\P, \text{Lie}H) \oplus \Omega^0_\text{tens}(\P, V)$. 
A generic vector $\bs{X}_\phi \in T_\phi\Phi$ with flow $f_\tau : \Phi \rarrow \Phi$, $\phi \mapsto f_\tau(\phi)=\left( f^A_\tau(\phi), f^\vphi_\tau(\phi) \right)$ and $f_{\tau=0}(\phi)=\phi$, is s.t.   $\bs{X}_\phi =\tfrac{d}{d\tau}f_\tau(\phi)\big|_{\tau=0}$. 
Formally, we can  write a vector field $\bs{X} \in \Gamma(T\Phi)$  as a variational differential operator on $C^\infty(\Phi)$: $\bs{X}_\phi= \bs{X}(\phi) \tfrac{\delta}{\delta \phi}$, with $\bs{X}(\phi)=\tfrac{d}{d\tau}f_\tau(\phi)\big|_{\tau=0}= \left(\tfrac{d}{d\tau}f^A_\tau(\phi)\big|_{\tau=0}, \tfrac{d}{d\tau}f^\vphi_\tau(\phi)\big|_{\tau=0}\right) =\bs{X}^A(\phi) + \bs{X}^\vphi(\phi)  \in \Omega^1_\text{tens}(\P, \text{Lie}H) \oplus \Omega^0_\text{tens}(\P, V)$ the `components' of $\bs{X}$.\footnote{So that $\bs{X}_\phi= \bs{X}(\phi) \tfrac{\delta}{\delta \phi}= \bs{X}^A(\phi) \tfrac{\delta}{\delta A}+ \bs{X}^\vphi(\phi) \tfrac{\delta}{\delta \vphi}$. Integration over domains is tacit, and we also avoid  generalised indices à la DeWitt \cite{DeWitt2003}.}
%
Only right-invariant vector fields, $\Gamma_\H(T\Phi)\defeq\left\{ \bs X \in \Gamma(T\Phi)\, | \, R_{\gamma\star}\bs{X}_\phi=\bs{X}_{\phi^\gamma} \right\}$, project to well-defined vector fields on the base, and $\pi_\star : \Gamma_\H(T\Phi) \rarrow \Gamma(T\M)$ is a morphism of Lie algebras. The flow of a right-invariant vector field belongs to $\bs{\Aut}(\Phi)$, so that $\Gamma_\H(T\Phi)\simeq \text{Lie}\bs{\Aut}(\Phi)$.

Any element $\chi \in \text{Lie}\H$ induces a vector $\chi^v_\phi \defeq \tfrac{d}{d\tau}  \phi^{\exp(\tau \chi)} \big|_{\tau=0} = \left( \tfrac{d}{d\tau}  A^{\exp(\tau \chi)} \big|_{\tau=0} , \tfrac{d}{d\tau}  \vphi^{\exp(\tau \chi)} \big|_{\tau=0}  \right) =\left( D^A\chi, -\rho_*(\chi)\vphi \right) $ tangent to the fiber $\O_\H[\phi]$ at $\phi=(A, \vphi)$. 
All such vectors span the vertical subbundle $V\Phi \subset T\Phi$, and a  \mbox{vertical} vector field $\chi^v \in \Gamma(V\Phi)$ is s.t. $\pi_\star \chi^v=0$ and $R_{\gamma\star}\chi_\phi^v=(\gamma\- \chi \gamma)^v_{\phi^\gamma}$. We have the injective morphism of Lie algebras  Lie$\H \rarrow \Gamma(V\Phi)$. Similarly, elements of the Lie algebra of the gauge group Lie$\bs{\H} \defeq \left\{ \, \bs{\chi}:\Phi \rarrow \text{Lie}\H\, |\,  R^\star_\gamma \bs{\chi}= \gamma\- \bs{\chi} \gamma    \, \right\}$ induce  $\H$-right invariant vertical vector fields $\bs{\chi}^v_\phi\defeq \tfrac{d}{d\tau}  \phi^{\exp(\tau \bs{\chi}(\phi))} \big|_{\tau=0}=\left( D^A \bs \chi(\phi), \, -\rho_*\left(\bs\chi(\phi)\right)\vphi \right)$, s.t.  $R_{\gamma\star} \bs{\chi}^v_\phi=\bs{\chi}^v_{\phi^\gamma}$, so that the map Lie$\bs{\H} \rarrow \Gamma_\H(V\Phi)$ is a Lie algebra \emph{anti}-isomorphism. Corresponding to \eqref{SESgroups-inf} we have, 
\begin{align}
\label{SESLieAlg-inf}
\makebox[\displaywidth]{
\hspace{-18mm}\begin{tikzcd}[column sep=large, ampersand replacement=\&]
\&0     \arrow [r]         \& \Gamma_\H(V\Phi) \simeq \text{Lie}\bs{\H}     \arrow[r, "\iota"  ]          \&  \Gamma_\H(T\Phi)     \arrow[r, "\pi_\star"]      \&  \Gamma(T \M)         \arrow[r]      \& 0.
\end{tikzcd}}  \raisetag{3.4ex}
\end{align}
A splitting of this SES would allow to define (non-canonically) a notion of horizontality on $\Phi$ complementary to the verticality canonically given by the action of $\H$. This is what a
 choice of variational Ehresmann connection $\bs{\omega} \in \bs{\A}_\Phi$ on $\Phi$ achieves. But depending on the problem at hand, one may instead need to endow $\Phi$ with  a twisted variational connection  ${\bs{\t \omega}} \in \bs{\t \A}_\Phi$. See  section \ref{Variational  connections on field space} below.
 

A local   section  doesn't provide such a (global) splitting, yet it is useful to record its action: we have $\bs\s_\star : T_{[\phi]}\M \rarrow T_{\bs\s([\phi])}\Phi$, $\bs X_{[\phi]} \mapsto \bs\s_\star \bs X_{[\phi]}$, and  the pushforwards of $\bs X_{[\phi]} \in T_{[\phi]}\M$ by two sections $\bs \s$ and $\bs\s'$ s.t. $\bs\s'=R_{\bs h} \bs \s$, are related by
\begin{align}
\label{Push-X-local}
\bs\s'_\star \bs X_{[\phi]} = R_{\bs h([\phi])\star} \left( \bs\s_\star \bs X_{[\phi]} \right) +  \left\{  \bs h\-\bs d \bs h_{|[\phi]}(\bs X_{{\phi}})  \right\}^v_{\bs \s'([\phi])}= R_{\bs h([\phi])\star} \left( \bs\s_\star \bs X_{[\phi]} +  \left\{ \bs{dhh}\-_{|[\phi]}(\bs X_{[\phi]})  \right\}^v_{\bs \s([\phi])} \right),
\end{align}
where in the second equality we used the equivariance property of fundamental vertical vector fields. This result allows to define the gluing relations between local representatives on $\M$ of variational forms on $\Phi$, as we are about to see. To define the \emph{field-dependent} gauge transformations of such forms, below, we need to record the action of $\bs\Aut_v(\Phi)\simeq \bs{\H}$ on $\Gamma(T\Phi)$: the action of $\bs{\Psi}\in \bs{\Aut}_v(\Phi)$ on a generic $\bs{X} \in \Gamma(T\Phi)$  is
\begin{align}
\label{Pushforward-X-inf}
\bs{\Psi}_\star \bs{X}_\phi&= R_{\bs{\gamma}(\phi)\star} \bs{X}_\phi + \left\{ \bs{\gamma}\- \bs{d}\bs{\gamma}_{|\phi}(\bs{X}_\phi)\right\}^v_{R_{\bs{\gamma}(\phi)} \phi}
               = R_{\bs{\gamma}(\phi)\star} \left( \bs{X}_\phi + \left\{ \bs{d}\bs{\gamma} {\bs{\gamma}\- }_{|\phi}(\bs{X}_\phi)\right\}^v_\phi \right).
\end{align}
Equations \eqref{Push-X-local} and \eqref{Pushforward-X-inf} are the exact analogues of their finite dimensional counterpart, and are proved in much the same way (the proof valid on $\A$ given in appendix B of \cite{Francois2021} is easily adapted to $\Phi$).

\paragraph{Variational differential forms} 

The de Rham complex on $\Phi$ is $\left( \Omega^\bullet(\Phi), \bs{d}\right)$ with $\bs d$ the variational exterior derivative of degree 1, defined via a Kozsul formula. The interior product  $\iota: \Gamma(T\Phi) \times \Omega^\bullet(\Phi) \rarrow \Omega^{\bullet-1}(\Phi)$,  $(\bs{X}, \bs{\alpha}) \rarrow \iota_{\bs X} \bs{\alpha}=\bs\alpha(\bs X, \ldots)$, is a degree $-1$ derivation $\forall \bs X$. The Lie derivative  $\bs{L_X}\defeq [\iota_{\bs X}, \bs{d}]$, is thus a degree 0 derivation.\footnote{The vector space of derivation Der$(\sf A)$ of an algebra $\sf A$ is a Lie algebra under the graded bracket $[d_1, d_2]\defeq  d_1 \circ d_2 - (-)^{|d_1|\cdot|d_2|}d_2 \circ d_1$.}
 It satisfies $[\bs{L_X}, \iota_{\bs Y}]=\iota_{[\bs X, \bs Y]}$, and it is a Lie algebra morphism $[\bs{L_X}, \bs{L_Y}]=\bs L_{[\bs X, \bs Y]}$.

An exterior product $\w$ is defined as usual on the space $\Omega^\bullet(\Phi, \sf A)$ of variational differential forms with values in an algebra $(\sf A, \cdot)$ -- and kept tacit throughout for convenience -- so that  $\big( \Omega^\bullet(\Phi, \sf A), \w, d \big)$ is a differential graded algebra. The exterior product is not defined on $\Omega^\bullet(\Phi, \bs V)$ where $\bs V$ is merely a vector space. But if $(\bs V, \rho)$ is a representation for $\H$, 
  one  defines the vector space of \emph{equivariant} forms as $\Omega_\text{eq}^\bullet(\Phi, \bs V)=\left\{ \, \bs\alpha \in \Omega^\bullet(\Phi, \bs V)\,|\, R^\star_\gamma\bs\alpha=\rho(\gamma)\-\bs\alpha\, \right\}$. 
  The infinitesimal version of the equivariance property is $\bs L_{\chi^v}\bs\alpha=-\rho_*(\chi) \bs\alpha$.
  The subspace of \emph{invariant} forms is $\Omega_\text{inv}^\bullet(\Phi, \bs V)=\left\{ \, \bs\alpha \in \Omega^\bullet(\Phi, \bs V)\,|\, R^\star_\gamma \bs\alpha=\bs\alpha, \,\right\}$, infinitesimally $\bs L_{\chi^v}\bs\alpha=0$. 

The space of of \emph{horizontal} forms is $\Omega_\text{hor}^\bullet(\Phi)=\left\{ \, \bs\alpha \in \Omega^\bullet(\Phi)\,|\, \iota_{\chi^v}\bs\alpha=0 \right\}$. A form which is both horizontal and equivariant is said \emph{tensorial}: $\Omega_\text{tens}^\bullet(\Phi, \bs V)=\left\{ \, \bs\alpha \in \Omega^\bullet(\Phi, \bs V)\,|\, R^\star_\gamma\bs\alpha=\rho(\gamma)\-\bs\alpha,\, \text{ \& }\ \iota_{\chi^v}\bs\alpha=0\, \right\}$. 
Clearly, $\Omega_\text{tens}^0(\Phi, \bs V)=\Omega_\text{eq}^0(\Phi, \bs V)$. 
%
Finally, \emph{basic} forms are both horizontal and invariant: $\Omega_\text{basic}^\bullet(\Phi)=\left\{ \, \bs\alpha \in \Omega^\bullet(\Phi)\,|\, R^\star_\gamma\bs\alpha=\bs\alpha,\, \text{ \& }\ \iota_{\chi^v}\bs\alpha=0\, \right\}$. Alternatively, basic forms are defined as  Im$(\pi^\star)$, that is: $\Omega_\text{basic}^\bullet(\Phi)=\left\{ \, \bs\alpha \in \Omega^\bullet(\Phi)\,|\, \exists\, \bs\beta \in \Omega^\bullet(\M) \text{ s.t. } \bs\alpha=\pi^\star\bs\beta \,  \right\}$. 

\mbox{Remark} that the form analogue of $\Gamma_\H(T\Phi)$ -- that projects to well defined vector fields in $\Gamma(M)$ -- is not $\Omega^\bullet_\text{inv}(\Phi)$ but $\Omega_\text{basic}^\bullet(\Phi)$, only the latter projects to well-defined forms in $\Omega^\bullet(\M)$.
From $[\bs d, \pi^\star]=0$ follows that  $\big( \Omega^\bullet_\text{basic}(\Phi), \bs d \big)$ is a subcomplex of the de Rham complex: the basic subcomplex. The associated cohomology, isomorphic to the cohomology of $\big(\Omega^\bullet(\M), \bs d \big)$, is by definition the equivariant cohomology of $\Phi$. 
 As we advertised, the main preoccupation of this paper, in relation to the boundary problem in gauge theory, will  be to consider ways to construct basic forms.

\medskip

Notice  that, as in the finite dimensional case, $\Omega_\text{eq}^0(\Phi, \bs V)\simeq \Gamma(\bs E)$ where $\bs E \rarrow \M$ is an associated bundle to $\Phi$ built via the representation $(\bs V, \rho)$ in the usual way: Defining a right action of $\H$ on $\Phi \times \bs V$ by $(\phi, \gamma) \mapsto (\phi^\gamma, \rho(\gamma)\- \bs v)$, and considering it is an equivalence relation $\sim$ between pairs, the associated bundle $\bs E$ is defined as the space of equivalence classes; $\bs E= \Phi \times_\H \bs V \defeq \Phi \times \bs V /\!\sim$. The 1:1 correspondance between $\bs s \in \Gamma(\bs E)$ and $\bs \vphi \in \Omega_\text{eq}^0(\Phi, \bs V)$ is, $\bs s([\phi])=[ \phi, \bs\vphi(\phi)]$.

It is perhaps interesting to remark that, in view of this standard construction,  the base space $\M$ of $\Phi$ is itself a bundle associated with the $\H$-principal bundle $\A \rarrow \A/\H$. Indeed $\Gamma(E)$ is a $\H$-space, and considering the classes under the right action of $\H$ on $\Phi=\A \times \Gamma(E)$ one obtains the quotient space $\M=  \A \times_\H \Gamma(E) \defeq  \A \times \Gamma(E)/\!\sim$. 
Therefore, the field space is a doubly fibered space, $\Phi \rarrow \M \rarrow \A/\H$. The space of sections of $\M$, $\Gamma(\M)\defeq \big\{ \bs s: \A/\H \rarrow \M, [A] \mapsto \bs s([A])=[A, \vphi]\big\}$, is then isomorphic to $\Omega_\text{eq}^0\left(\A, \Gamma(E)\right)\defeq \left\{ \hat{\bs s}: \A \rarrow \Gamma(E)\, |\, R^\star_\gamma \hat{\bs s} =\rho(\gamma)\- \hat{\bs s} \right\}$. 
\mbox{One would} remark that the covariant derivative map of a section $\vphi \in \Gamma(E)$, $A \mapsto \left\{\iota_X D\vphi\right\}(A)=\iota_XD^A\vphi$ -- for any vector field $X \in \Gamma(T\P)$\footnote{The covariant derivative map would be  $D:\A \times \Gamma(E) \rarrow \Omega^1_\text{tens}(\P, V)$, $(A, \vphi)\mapsto D^A\vphi$. So that, for any vector field $X \in \Gamma(T\P)$, we have the map $\iota_X D:\A \times \Gamma(E) \rarrow  \Gamma(E)$, $(A, \vphi) \mapsto \iota_XD^A \vphi$.  } --
 is an element of this space since $R^\star_\gamma D^A\vphi = \rho(\gamma)\-D^A\vphi$.
This shows the centrality of the role payed by the connection space $\A$, which is reminiscent of the fact that a $H$-principal bundle $\P$ `controls' all associated bundles built via representations of $H$ (or $H$-spaces more generally). 
\medskip

In the construction above one can replace representations $(\rho, \bs V)$ by 1-cocycles for the action of $\H$, i.e. a smooth map  $C: \Phi \times \H \rarrow G$ , $(\phi, \gamma) \mapsto C(\phi, \gamma)$, satisfying the relation $C(\phi, \gamma\gamma')= C(\phi, \gamma)C(\phi^\gamma, \gamma')$, and $\bs V$ a $G$-space. Then we have a well-defined right action of $\H$ on $\Phi \times \bs V$ twisted by a cocycle $(\phi, \bs v) \mapsto (\phi^\gamma, C(\phi, \gamma)\- \bs v)$. One thus defined \emph{twisted} associated bundles as the spaces of equivalence classes under this action, $\tilde{\bs{E}} \defeq \Phi \times_\C \bs V$, whose space of sections is isomorphic to the space of $C$-equivariant $\bs V$-valued maps on $\Phi$, $\Gamma(\tilde{\bs{E}})\simeq \Omega_\text{eq}^0(\Phi, C)\defeq \left\{  \bs\vphi : \Phi \rarrow \bs V\, |\, \bs\vphi(\phi^\gamma)=C(\phi, \gamma)\- \bs\vphi(\phi) \right\}$. More generally, one has the well-defined spaces of $C$-equivariant forms $\Omega^\bullet_\text{eq}(\Phi, C) \defeq \left\{ \bs\alpha \in \Omega^\bullet(\Phi, \bs V)\, | \, R^\star_\gamma \bs\alpha = C(\ , \gamma)\- \bs\alpha \right\}$, 
and  of $C$-tensorial forms $\Omega^\bullet_\text{tens}(\Phi, C) \defeq \left\{ \bs\alpha \in \Omega^\bullet_\text{eq}(\Phi, C)\, | \, \iota_{\chi^v} \bs \alpha = 0 \right\}$.  

If the covariant differentiation of standard equivariant forms necessitates to endow $\Phi$ with variational Ehresmann (principal) connections, the covariant differentiation of $C$-equivariant forms requires the introduction of \emph{twisted} connections which are a generalisation of the latter. See section \ref{Variational connections on field space} below. We refer to \cite{Francois2019_II} for an in depth exposition of the basics of the twisted geometry on bundles, and to section 2.3 of \cite{Francois2021} for a nutshell presentation.

\paragraph{Gauge transformations} 

The local representative of a form $\bs\alpha \in \Omega^\bullet(\Phi)$ as seen through a section \mbox{$\bs\s :\U \rarrow \Phi$}  is $\bs a \defeq \bs\s^\star \bs\alpha$, so that $\bs a_{|[\phi]}\big(\bs X_{[\phi]}\big) =\bs\alpha_{|\bs\s([\phi])}\big( \bs \s_\star \bs X_{[\phi]} \big)$. In view of \eqref{Push-X-local}, two local representatives $\bs a=\bs \s^\star \bs\alpha$ and \mbox{$\bs a'={\bs \s'}^\star \bs\alpha$}  
 satisfy gluing relations on  $\U\cap \U' \neq \emptyset$ -- also called \emph{passive} gauge transformations -- controlled by the $\H$-equivariance and verticality properties of $\bs\alpha$. 
 In particular, it is clear that if $\bs\alpha$ is tensorial, the gluing relations of its local \mbox{representatives} are simply $\bs a'=\rho(\bs h)\- \bs a$. Similarly, if $\bs\alpha$ is $C$-tensorial, the gluing relations of its local representatives are simply $\bs a'=C(\bs\s , \bs h)\- \bs a$ .
 The local representatives of a basic form have trivial gluing relations: $\bs a'=\bs a$. Hence again the interest of such forms. We will not make use of this notion of passive, or $[\phi]$-dependent, gauge transformation, as we aim to work primarily on $\Phi$ rather than on $\M$. But we articulate it if only to clearly distinguish it from the notion of gauge transformation on $\Phi$ that will be of interest from now on. 
 
 The \emph{active}, $\phi$-dependent, gauge transformation of a form $\bs\alpha \in \Omega^\bullet(\Phi)$ is defined by the action of $\bs{\Aut}_v(\A)\simeq \bs{\H}$ via $\bs\alpha^{\bs\gamma}\defeq \bs\Psi^\star \bs \alpha$. So that, $\bs\alpha^{\bs\gamma}_{|\phi}\big(\bs X_\phi\big)=\bs\alpha^{\phantom .}_{|\bs\Psi(\phi)}\bs( \bs\Psi_\star \bs X_\phi \big)$. 
By \eqref{Pushforward-X-inf},  the $\bs\H$-gauge transformation of a variational form is thus controlled by its $\H$-equivariance and verticality properties. It  follows immediately that the $\bs\H$-gauge transformation of a tensorial  form is $\bs\alpha^{\bs\gamma}=\rho(\bs \gamma)\- \bs\alpha$, or infinitesimally $\bs L_{\bs \chi^v} \bs \alpha = -\rho_*(\bs\chi) \bs \alpha$, with $\bs\chi \in \text{Lie}\bs\H$. 
In the same way, $C$-tensorial forms $\bs\H$-transform as $\bs\alpha^{\bs\gamma}= C(\ , \bs \gamma)\- \bs\alpha$, or infinitesimally $\bs L_{\bs \chi^v} \bs \alpha = -\tfrac{d}{d\tau} C(\ \, , \exp{ \tau \bs\chi}) \big|_{\tau=0} \bs \alpha$.
 Basic  forms are $\bs \H$-invariant, $\bs\alpha^{\bs\gamma}= \bs\alpha$ or $\bs L_{\bs \chi^v} \bs \alpha = 0$. 

\paragraph{Basis 1-forms} 

As a relevant illustration, consider $\bs d \phi=\left\{ \bs d A, \bs d \vphi\right\} \in \Omega^1(\phi)$ the basis for variational forms on $\Phi$. Its verticality property reproduces, by definition, the infinitesimal gauge transformations of the fields,
\begin{align}
\label{Vert-dphi}
\bs d \phi_{|\phi}(\chi^v_\phi)= \left( \bs d A_{|A}(\chi^v_A)\,,\  \bs d \vphi_{|\vphi} (\chi^v_\phi)\right)= \left(D^A \chi\,, \ -\rho_*(\chi)\vphi \right)\rdefeq \delta_\chi\phi.
\end{align}
The last equality introduces a convenient notation. We thus find that on $\Gamma(V\Phi)$, its $\H$-equivariance property is, 
$R^\star_\gamma \bs d\phi_{|\phi^\gamma}(\chi^v)=  \bs d\phi_{|A^\gamma}(R_{\gamma \star} \chi^v_\phi) = \bs d\phi_{|\phi^\gamma} (\gamma\- \chi \gamma)^v_{\phi^\gamma}= \left( D^{A^\gamma}(\gamma\-\chi \gamma)\,,\ -\rho_*(\gamma\- \chi \gamma) \vphi^\gamma \right)=  \left( \gamma\- (D^A\chi) \gamma\,,\ -\rho(\gamma)\-\rho_*(\chi)\vphi \right) = \left(\gamma\- \bs{d}A_{|A}(\chi^v_A) \gamma\,, \ \rho(\gamma)\- \bs d\vphi_{|\phi}(\chi^v_\vphi) \right)$, by \eqref{Vert-dphi}. Introducing the notation  $\uprho \defeq (\Ad\,, \ \rho)$, we write the equivariance compactly as
\begin{align}
\label{Equiv-dphi}
R^\star_\gamma \bs d \phi = \uprho(\gamma)\- \bs d \phi \defeq \left( \gamma\- \bs d A \gamma\, ,\ \rho(\gamma)\- \bs d \vphi \right). 
\end{align}
We require this to hold on $\Gamma(T\Phi)$, so that $\bs d \phi \in \Omega_\text{eq}^1(\Phi, T\Phi)$. 
By  \eqref{Pushforward-X-inf}, it is now easy to find the $\bs \H$-gauge transformation of $\bs{d}\phi$~to~be, 
\begin{align*}
\bs d \phi^{\bs \gamma}_{|\phi}(\bs X_\phi)&\defeq (\bs\Psi^\star \bs d \phi)_{|\phi}(\bs X_\phi)=\bs d \phi_{|\phi^{\bs\gamma}} \left( \bs\Psi_\star \bs X_\phi \right),   \notag\\
						        & = \bs d \phi_{|\phi^{\bs\gamma}} \left( R_{\bs \gamma(\phi)\star} \left[ \bs X_\phi + \left\{ \bs{d \gamma\gamma}\-_{|\phi}(\bs X_\phi)\right\}^v_\phi \right] \right)
						         = (R_{\bs\gamma(\phi)}^\star  \bs d\phi_{|\phi^{\bs\gamma}})  \left( \bs{X}_\phi + \left\{ \bs{d\gamma\gamma}\-_{|\phi}(\bs X_\phi)\right\}^v_\phi \right), \notag \\
						        & = \uprho \big(\bs\gamma(\phi)\big)\-   \bs d\phi_{|\phi}  \left( \bs{X}_\phi + \left\{ \bs{d\gamma\gamma}\-_{|\phi}(\bs X_\phi)\right\}^v_\phi \right) 	
						        =    \uprho \big(\bs\gamma(\phi)\big)\-  \left(    \bs d\phi_{|\phi}(\bs X_\phi) + \delta_{ \bs{d\gamma\gamma}\-_{|\phi}(\bs X_\phi)} \phi \right). 
 \end{align*}
That is, for $\bs\gamma \in \bs\H$,
\begin{align}
\label{GT-dphi}
\bs d \phi^{\bs \gamma} &= \uprho (\bs\gamma)\-    \left( \bs d \phi  + \delta_{ \bs{d\gamma\gamma}\-} \phi \right)
		 = \left\{  \begin{array}{c}   \bs d A^{\bs \gamma} = \bs{\gamma}\-   \left( \bs{d}A  + D^A\big\{ \bs{d}\bs{\gamma} {\bs{\gamma}\- }\big\} \right)    \bs{\gamma}\ \\[1mm] 
		                                          \bs d \vphi^{\bs \gamma} =  \rho(\bs \gamma)\- \left( \bs d \vphi -\rho_*(\bs{d\gamma\gamma}\-) \vphi\right) 
		               \end{array} \right.
 \end{align}
The infinitesimal version is easily read from the above, by $\bs\gamma \rarrow \bs\chi \in \text{Lie}\bs\H$ and keeping only the linear terms, and of course seen to match the result obtained by, 
\begin{align}
\label{linearized-GT-dphi}
\bs L_{\bs \chi^v} \bs d \phi = \left( \iota_{\bs \chi^v} \bs d + \bs d \iota_{\bs \chi^v} \right) \bs d\phi = \bs d \delta_{\bs \chi}\phi  
				& = \bs d \left( D^A \bs \chi\, ,\ -\rho_*(\bs \chi) \vphi\right), \notag \\
				& = \left(   [\bs d A, \bs \chi] + D^A \bs (\bs d\bs\chi)   \, , \ -\rho_*(\bs\chi)\bs d \vphi - \rho_*(\bs d \bs\chi)\vphi   \right).
\end{align}

In term of the basis ${\bs d\phi}$, a generic variational form $\bs \alpha \in \Omega^\bullet(\Phi)$ will be written $\bs \alpha_{|\phi} =\alpha\big (\!\w^\bullet\! \bs d \phi_{|\phi} \, ;\,  \phi \big)$, 
or simply $\bs \alpha =\alpha\big (\!\w^\bullet\! \bs d \phi \, ;\,  \phi \big)$, where $\alpha(\ ;\ )$ is alternating multilinear in the first arguments and its second argument denotes the  functional dependence of $\bs\alpha$ on $\phi$. 
So, in concrete situations, given such an expression for $\bs\alpha$ and eq.\eqref{GT-dphi}, it is possible to compute algebraically its $\bs\H$-gauge transformation by $\bs\alpha^{\bs\gamma} =\alpha\big (\!\w^\bullet\! \bs d \phi^{\bs\gamma}\, ;\,  \phi^{\bs\gamma} \big)$, thereby cross-checking the geometric result obtained via $\bs{\alpha^\gamma} = \bs\Psi^\star \bs\alpha$.

\paragraph{Curvature and covariant derivative maps} 

We provide further illustrations by iterating the process on two quantities relevant for what follows, that we consider in turn. 
\smallskip

$\bullet$ {\bf Curvature map:} Consider the $\H$-equivariant curvature map $F:\Phi \rarrow \Omega^2_\text{tens}(\P, \text{Lie}H)$, $(A, \vphi) \mapsto F(A)=dA+\tfrac{1}{2}[A, A] $, s.t. $R^\star_\gamma F =\gamma\- F \gamma$. 
 Given  $\bs{X} \in \Gamma(T\A)$ with flow $f_\tau$, we have:
 \begin{align}
\label{dF}
 \bs{d}F_{|\phi} (\bs X_\phi)&
 =\bs X \big(F\big)(\phi)=\tfrac{d}{d\tau} F\big( f_\tau(\phi) \big) \big|_{\tau=0} = \tfrac{d}{d\tau} df_\tau(A) + \sfrac{1}{2}[f_\tau(A), f_\tau(A)] \big|_{\tau=0}= D^A\big( \tfrac{d}{d\tau}f_\tau(A)\big|_{\tau=0} \big),\notag \\
        				    &= D^A\big( \bs{d}A_{|\phi}(\bs{X}_\phi)\big), \notag\\[1mm]
	  \text{which is simply} \quad \bs{d} F&=D\big( \bs{d}A\big).
 \end{align}
Then, the verticality property of $\bs{d}F$ reproduces the infinitesimal $\H$-transformation~of~$F$:
\begin{align}
\label{Vert-dF}
\bs{d}F_{|\phi}(\chi^v_\phi)=D^A\big( \bs{d}A_{|\phi}(\chi^v_\phi) \big)= D^A\big( D^A\chi \big)= [F(A), \chi],      \quad \text{or} \quad       \iota_{\chi^v}\bs{d}F =[F, \chi].
\end{align}
Its $\H$-equivariance is  
\begin{align}
\label{Equiv-dF}
R^\star_\gamma \bs{d}F_{|\phi^\gamma}(\bs{X}_\phi)&=  \bs{d}F_{|\phi^\gamma}(R_{\gamma\star}\bs{X}_\phi) = D^{A^\gamma}\big( \bs{d}\phi_{|\phi^\gamma}(R_{\gamma\star}\bs{X}_\phi) \big) =D^{A^\gamma}\big( R^\star_\gamma \bs{d}A_{|\phi^\gamma}(\bs{X}_\phi) \big),  \notag\\
									& = D^{A^\gamma}\big( \gamma\- \bs{d}A_{|\phi}(\bs{X}_\phi) \gamma \big) = \gamma\- D^A\big( \bs{d}A_{|\phi}(\bs{X}_\phi) \big) \gamma = \gamma\- \bs{d}F_{|\phi} (\bs{X}_\phi) \gamma, \qquad \text{or}\quad R^\star_\gamma \bs{d}F =\gamma \- \bs{d}F \gamma.
\end{align}
This result is also found simply from the fact that pullbacks and $\bs d$ commute, so that $R^\star_\gamma \bs{d} F= \bs{d} R^\star_\gamma F = \gamma\- \bs{d}F \gamma$. 
From  \eqref{Vert-dF}-\eqref{Equiv-dF} and \eqref{Pushforward-X-inf},    the $\bs \H$-gauge transformation of $\bs{d}F$ is found to be 
\begin{align}
\label{GT-dF}
\bs{d}F^{\bs \gamma}_{|\phi}(\bs{X}_\phi)&\defeq (\bs{\Psi}^\star \bs{d}F)_{|\phi}(\bs{X}_\phi)=\bs{d}F_{|\phi^{\bs\gamma}} \left( \bs{\Psi}_\star \bs{X}_\phi \right),   \notag\\
						        & = \bs{d}F_{|\phi^{\bs\gamma}} \left(R_{\bs{\gamma}(\phi)\star} \left[ \bs{X}_\phi + \left\{ \bs{d}\bs{\gamma} {\bs{\gamma}\- }_{|\phi}(\bs{X}_\phi)\right\}^v_\phi \right] \right)
						         = (R_{\bs{\gamma}(\phi)}^\star  \bs{d}F_{|\phi^{\bs\gamma}})  \left( \bs{X}_\phi + \left\{ \bs{d}\bs{\gamma} {\bs{\gamma}\- }_{|\phi}(\bs{X}_\phi)\right\}^v_\phi \right), \notag \\
						        & = \bs{\gamma}(\phi)\-   	\bs{d}F_{|\phi}  \left( \bs{X}_\phi + \left\{ \bs{d}\bs{\gamma} {\bs{\gamma}\- }_{|\phi}(\bs{X}_\phi)\right\}^v_\phi \right) 		\bs{\gamma}(\phi) 
						        =    \bs{\gamma}(\phi)\-  \left(    \bs{d}F_{|\phi}(\bs{X}_\phi) + \big[F(\phi), \bs{d}\bs{\gamma} {\bs{\gamma}\- }_{|\phi}(\bs{X}_\phi)\big] \right)  \bs{\gamma}(\phi),   \notag\\
 \text{that is} \quad    \bs{d}F^{\bs \gamma} &=  \bs{\gamma}\-   \left( \bs{d}A  + \big[F, \bs{d}\bs{\gamma} {\bs{\gamma}\- }\big] \right)    \bs{\gamma}
\end{align}
This can also be checked algebraically, using \eqref{GT-dphi} in $\bs d F^{\bs\gamma}=D^{A^{\bs\gamma}}(\bs d A^{\bs\gamma})$. 
\smallskip

$\bullet$ {\bf Covariant derivative map:}  Let us consider the $\H$-equivariant covariant derivative map $D: \Phi \rarrow \Omega^1_\text{tens}(\P, V)$, $\phi=(A, \vphi) \mapsto D^A \vphi=dA +\rho_*(A)\vphi$, s.t. $R^\star_\gamma D = \rho(\gamma)\-D$. 
Given  $\bs{X} \in \Gamma(T\Phi)$ with flow $f_\tau$, we get:
\begin{align}
\label{d-Dphi}
\bs d D_{|\phi} (\bs X_\phi)=\bs X \big(D\big)(\phi)=\tfrac{d}{d\tau}\ D^{f_\tau(A)} f_\tau(\vphi)\ \big|_{\tau=0} &= \tfrac{d}{d\tau}\  df_\tau(\vphi)    +\rho_*\big(f_\tau(A)\big) f_\tau(\vphi)\  \big|_{\tau=0}, \notag \\
														 &=D^A\left(  \tfrac{d}{d\tau}\ f_\tau(\vphi) \ \big|_{\tau=0}\right) + \rho_*\left( \tfrac{d}{d\tau}\ f_\tau(A)  \big|_{\tau=0}\right) \vphi, \notag \\
														 &= D^A \left( \bs d \vphi_{|\phi} (\bs X_\phi)\right) +  \rho_*\left( \bs d A_{|\phi}(\bs X_\phi) \right) \vphi, \notag \\[1mm]
					\text{which we can write} \quad \bs d \,\!D^A\vphi &= D^A(\bs d \vphi) + \rho_*(\bs d A)\vphi,
\end{align}
by  a slight abuse of notation (on the left-hand side) to make the expression more transparent. 
As one expects, the verticality property of $ \bs d \,\!D^A\vphi $ reproduces the infinitesimal $\H$-transformation of $ D^A\vphi$:
\begin{align}
\label{Vert-dDphi}
 \bs d \,\!D^A\vphi_{|\phi}(\chi^v_\phi)=D^A\big( \bs d \vphi_{|\phi}(\chi^v_\phi) \big) + \rho_*\big(\bs d A_{|\phi}(\chi^v_\phi)\big)\vphi = D^A\big( -\rho_*(\chi)\vphi \big) + \rho_*\big(D^A\chi \big)\vphi= -\rho_*(\chi) D^A \vphi.
\end{align}
As for the $\H$-equivariance, we have
\begin{align}
\label{Equiv-dDphi}
R^\star_\gamma \bs d \,\!D^A\vphi_{|\phi^\gamma}(\bs{X}_\phi)&= \bs d \,\!D^A\vphi_{|\phi^\gamma}(R_{\gamma\star}\bs{X}_\phi) 
										= D^{A^\gamma}\big(  \bs d \vphi_{|\phi^\gamma} (R_{\gamma\star}\bs{X}_\phi)\big) + \rho_*\big( \bs d A_{|\phi^\gamma} (R_{\gamma\star}\bs{X}_\phi) \big) \vphi^\gamma, \notag \\
										&= D^{A^\gamma}\big(  R_{\gamma}^\star \bs d \vphi_{|\phi^\gamma} (\bs{X}_\phi)\big) + \rho_*\big( R_{\gamma}^\star \bs d A_{|\phi^\gamma} (\bs{X}_\phi) \big) \vphi^\gamma,  \notag \\
										&= D^{A^\gamma}\big(  \rho(\gamma)\- \bs d \vphi_{|\phi} (\bs{X}_\phi)\big) + \rho_*\big( \gamma\- \bs d A_{|\phi}  (\bs{X}_\phi)\,\gamma \big) \vphi^\gamma,  \notag \\
										&=\rho(\gamma)\- \left(  D^{A}\big(  \bs d \vphi_{|\phi} (\bs{X}_\phi)\big) + \rho_*\big(  \bs d A_{|\phi}  (\bs{X}_\phi) \big) \vphi  \right), \notag \\
           \text{which is } \quad   R^\star_\gamma \bs d \,\!D^A\vphi  &= \rho(\gamma)\-  \bs d \,\!D^A\vphi. 
\end{align}
Which also results from $[R^\star_\gamma, \bs d]=0$. Again, by \eqref{Vert-dDphi}-\eqref{Equiv-dDphi} and  \eqref{Pushforward-X-inf} we get the $\bs \H$-gauge transformation: 
\begin{align}
\label{GT-dPhi}
(\bs d\,\!D^A\vphi)^{\bs \gamma} (\bs X) &= \left( \bs\Psi^\star \bs d\, D^A\vphi \right)(\bs X) =  \bs d\,\!D^A\vphi \, (\bs\Psi_\star \bs X) = \ldots \notag \\
 \Rightarrow 
(\bs d\,\!D^A\vphi)^{\bs \gamma} &= \rho(\bs\gamma)\ \- \left( \bs d\,\!D^A\vphi - \rho_*(\bs{d\gamma\gamma}\-) D^A\vphi \right).
\end{align}

\paragraph{Lagrangians}

A  gauge theory is specified by a choice of Lagrangian functional $L:\Phi \rarrow \Omega^n(\P)$, $\phi\defeq(A, \vphi) \mapsto L(\phi)=L(A, \vphi)$, with $n=\text{dim}M$, whose $\H$-equivariance is usually prescribed. 
Said otherwise $L \in \Omega_\text{eq}^0(\Phi)$, so that a Lagrangian can be seen as a section of some bundle associated with $\Phi$. There are three possible cases.
\medskip

$\bullet$ {\bf Case 1: invariant gauge theories} The Lagrangian strictly satisfies the gauge principle, $R^\star_\gamma L =L$, so $L \in \Omega_\text{basic}^0(\Phi)$. This means that $\exists \, \b L \in  \Omega^0(\M)$ s.t. $L=\pi^\star \b L$. It also means that $L$ is the section of the bundle associated with $\Phi$ via the trivial representation. So is $Z \defeq \exp{i\int L} \in \Omega_\text{basic}(\Phi, \CC)$. 
Thus, they are both $\bs\H$-invariant: $L^{\bs\gamma}=L$ and $Z^{\bs\gamma}=Z$.
\medskip

$\bullet$ {\bf Case 2: non-invariant gauge theories transforming via representations} The Lagrangian has $\H$-equivariance $R^\star_\gamma L = L + \t\rho(\gamma)$, so that $R^\star_\gamma Z = \rho(\gamma)\- Z$, where $\rho(\gamma)\defeq \exp{-i \int \t\rho(\gamma)}$ is a representation  of $\H$ on $\CC$. We have  $L, Z \in \Omega^0_\text{eq}(\Phi)$ (by a slight abuse of notation in the case of $L$), they are sections of standard bundles associated with $\Phi$. So, their $\bs\H$-gauge transformations are $L^{\bs\gamma}\defeq \bs\Psi^\star L = L + \t\rho(\bs\gamma)$ and  $Z^{\bs\gamma}=\rho(\bs\gamma)\-Z$. 
\medskip

$\bullet$ {\bf Case 3: non-invariant gauge theories transforming via cocycles} The Lagrangian has  $\H$-equivariance $R^\star_\gamma L = L + c(\ , \gamma)$, so that $R^\star_\gamma Z = C(\, ,\gamma)\- Z$, where $C(\ ,\gamma)\defeq \exp{-i \int c(\ ,\gamma)}$ is a 1-cocycle for the action of $\H$ on $\CC$.  We have $L, Z \in \Omega^0_\text{eq}(\Phi, C)$ (by a slight abuse of notation in the case of $L$), they are sections of twisted bundles associated  to $\Phi$. So their  $\bs\H$-gauge transformations are $L^{\bs\gamma}\defeq \bs\Psi^\star L = L + c(\ , \bs\gamma)$ and $Z^{\bs\gamma}=C(\, ,\bs\gamma)\-Z$. 

\bigskip

In the next two sections, we discuss structures that are a priori non-canonical on the bundle $\Phi$.  In the first, we elaborate around the standard notion of Ehresmann -- or principal -- variational connections, and then  briefly comment on a generalisation called twisted variational connections. In the next, we give a nutshell presentation of a tool to build basic forms known as the dressing field method. The material exposed in both these sections will be used to address the problem of boundaries in the presymplectic structure of invariant gauge theories (case 1 above) in section \ref{Basic presymplectic structures}, where we will only make passing comments on the case of non-invariant theories (cases 2 and 3).

 \subsection{Variational connections on field space}
\label{Variational connections on field space}

\subsubsection{Variational Ehresmann connections}
\label{Variational Ehresmann connections}

As mentioned above, to split the SES \eqref{SESLieAlg-inf} and to define a notion of horizontality, one needs to introduce a choice of Ehresmann connection on $\Phi$. Such a variational connection $\bs\omega \in \Omega^1_\text{eq}(\Phi, \text{Lie}\H)$ is by definition s.t. 
\begin{align}
\label{var-connection1}
R^\star_\gamma \bs\omega &= \gamma\- \bs\omega\, \gamma, \\
\label{var-connection2}
\bs\omega (\chi^v)&=\chi\  \in \text{Lie}\H. 
\end{align}
It is not unique, and from the definition follows that the space of variational connections is affine and modeled on the vector space $\Omega^1_\text{tens}(\Phi, \text{Lie}\H)$. That is, for $\bs\beta \in \Omega^1_\text{tens}(\Phi, \text{Lie}\H)$, we have that $\bs\omega'=\bs\omega+\bs\beta$ is another connection. 
In view of the above defining axioms,  by \eqref{Pushforward-X-inf} the $\bs\H$-gauge transformation of a connection is  $\bs\omega^{\bs\gamma} \defeq \bs\Psi^\star \bs\omega=\bs\gamma\- \bs\omega \bs\gamma+ \bs\gamma\- \bs d \bs \gamma$. 
At any point $\phi \in \Phi$, the horizontal complement to $V_\phi \Phi \subset T_\phi\Phi$ is $H_\phi\Phi \defeq \ker \bs\omega_{|\phi}$. We have then the horizontal subbundle s.t. $T\Phi=H\Phi\oplus V\Phi$. This allows to define the horizontal projection, 
\begin{align}
\label{horiz-proj-map}
|^h: \Gamma(T\Phi) &\rarrow \Gamma(H\Phi), \notag \\
\bs X &\mapsto \bs X^h \defeq \bs X - [\bs\omega(\bs X)]^v,
\end{align}
as it is clear by \eqref{var-connection2} that $\bs X^h \in \ker\bs\omega$. One easily shows that, by \eqref{var-connection1}, $R_{\gamma\star}\bs X^h=(R_{\gamma\star} \bs X)^h$, proving that the horizontal distribution thus defined is $\H$-equivariant, $R_{\gamma\star} H_\phi\P = H_{\phi^\gamma}\Phi$, as it must. 

The connection thus allows to define a notion of covariant derivation, $\bs{D^\omega} \defeq \bs d \circ |^h : \Omega^\bullet_\text{eq}(\Phi, \bs V) \rarrow \Omega^{\bullet+1}_\text{tens}(\Phi, \bs V)$, that is $\bs{D^\omega}\bs\alpha(\bs X_1, \bs X_2, \ldots) \defeq \bs d \bs\alpha (\bs X^h_1, \bs X^h_2, \ldots)$. The horizontality of the resulting form is obvious, and the preservation of the $\H$-equivariance is shown by 
$R^\star_\gamma \bs{D^\omega}\bs\alpha(\bs X,  \ldots) = \bs{D^\omega}\bs\alpha( R_{\gamma\star}\bs X, \ldots) \defeq \bs d \bs \alpha \big((R_{\gamma\star}\bs X)^h, \ldots\big) =  \bs d \bs \alpha \big(R_{\gamma\star}\bs X^h, \ldots\big)=\bs d R^\star_\gamma \bs\alpha (\bs X^h, \ldots)= \rho(\gamma)\- \bs d \bs\alpha (\bs X^h, \ldots) \rdefeq \rho(\gamma)\- \bs{D^\omega \alpha}(\bs X, \ldots)$. 
On $\Omega^\bullet_\text{tens}(\Phi, \bs V)$, the covariant derivative has the alternative expression $\bs{D^\omega}=\bs d \, + \rho_*(\bs\omega)$. 
And on $\Omega^\bullet_\text{basic}(\Phi, \bs V)$, for which $\rho$ is trivial, it clearly reduces to the exterior derivative $\bs{D^\omega}=\bs d$. That is, $\bs d$ is a canonical covariant derivative on basic forms of $\Phi$.\footnote{Which we knew already from the observation in section \ref{Field space as a principal bundle}  that the basic complex $\big(\Omega^\bullet_\text{basic}(\Phi), \bs d\big)$ is a subcomplex of the de Rham complex of $\Phi$.}

 The curvature of the connection is $\bs\Omega\defeq \bs{D^\omega \omega}=\bs d\bs\omega +\tfrac{1}{2}[\bs\omega, \bs\omega] \in \Omega^2_\text{tens}(\Phi, \text{Lie}\H)$,\footnote{It is the covariant derivative of the connection, but as the latter isn't tensorial, the second equality \emph{is not} due to the alternative expression just mentioned with $\rho_*=\ad$, as the factor $\sfrac{1}{2}$ should tell. It is rather known as the \emph{Cartan structure equation} for the curvature. } and its $\bs\H$-transformation is thus $\bs\Omega^{\bs\gamma}=\bs\gamma\- \bs\Omega \bs \gamma$. It is easily shown that on  $\Omega^\bullet_\text{tens}(\Phi, \bs V)$,  $\bs{D^\omega} \circ \bs{D^\omega} = \rho_*(\bs\Omega)$. 
 The curvature satisfies the Bianchi identity $\bs{D^\omega} \bs\Omega=\bs d\bs\Omega + [\bs \omega, \bs \Omega] \equiv 0$.
 As we have $\bs\Omega(\bs X, \bs Y)= \bs d \bs \omega (\bs X^h, \bs Y^h) = -\bs\omega([\bs X^h, \bs Y^h])$, if $[\bs X^h, \bs Y^h]=[\bs X, \bs Y]^h$ then $\bs\Omega \equiv 0$. The curvature thus mesures the failure of \eqref{horiz-proj-map} to be a Lie algebra morphism.

 \paragraph{Connection associated with an equivariant metric} 
 
As verticality is canonically defined on $\Phi$, if it is endowed with a Riemannian metric $\bs g: \Gamma(T\Phi)\times \Gamma(T\Phi)\rarrow \C^\infty(\Phi)$, the horizontal bundle $H\Phi$ can be defined as the $\bs g$-orthogonal complement of $V\Phi$. But for the horizontal distribution to be $\H$-equivariant, so must be the metric, $R^\star_\gamma \bs g = \bs g$ (the bundle structure of $\Phi$ must be respected). Thus, given a $\H$-equivariant metric $\bs g$ there is  an associated connection $\bs{\omega^g}$ s.t. $\ker \bs{\omega^g} = \Gamma(V\Phi)^\bot \defeq \left\{ \ker \bs g \big(\chi^v, \  \big), \forall \chi^v \in \Gamma(V\Phi) \right\} $, and defined implicitly using \eqref{horiz-proj-map} by $\bs g\big( \chi^v,  \bs X - [\bs{\omega^g}(\bs X)]^v\big)=0$, $\forall \chi \in \text{Lie}\H$. 
Of course a metric is a priori no more canonical than a connection, and it is an even ``richer" structure. 

In the case of  the $\H$-subbundle of connections (on $\P$) $\A \subset \Phi$, we have $T_A\A \simeq  \Omega^1_\text{tens}(\P, \text{Lie}H)$. So, assuming the existence of a Hodge dual (thus of a metric) on $\P$, and for Lie$H$  semisimple, there is  a `natural' metric built via the Killing form on Lie$H$, i.e. the trace: $\mathring{\bs g}_{|A}\!:~\!\!T_A\A \times T_A\A \rarrow \RR$, $(\alpha, \beta) \mapsto \mathring{\bs g}_{|A}\big(\alpha, \beta \big)\defeq \int \Tr\left( \alpha \w  *\beta \right)$ (it is tacit that the arguments are compactly supported). The connection associated with this metric on $\A$ is \mbox{introduced} by Singer \cite{Singer1978, Singer1981}, and termed the Singer-de Witt connection in \cite{Gomes-Riello2018, Gomes-et-al2018, Gomes2019}. Singer indeed defines the horizontal subspace as $H_A\A\defeq \left\{ A+\alpha \ |\  D^{A\dagger}\alpha =0 \right\}$ where $D^{A\dagger}\defeq(-)^{k}*^{-1}\!D^A\, *: \Omega^k_\text{eq}(\P, \text{Lie}H) \rarrow \Omega^{k-1}_\text{tens}(\P, \text{Lie}H)$ it the adjoint of the covariant derivative $D^A$ w.r.t. the inner product $\langle \zeta,\eta \rangle \defeq \int\Tr(\zeta\w * \eta)$, with $\zeta, \eta \in \Omega^k_\text{eq}(\P, \text{Lie}H)$,\footnote{This generalises the construction of the codifferential  $\delta\defeq (-)^{k}*^{-1}\!d\,*: \Omega^k(\P, \RR) \rarrow \Omega^{k-1}(\P, \RR)$ as the adjoint of the exterior derivative $d$ w.r.t. the inner product $\langle \zeta, \eta \rangle\defeq \int \zeta \w * \eta$ for $\zeta, \eta \in \Omega^k(\P, \RR)$. } 
  the orthogonality condition being $0=\mathring{\bs g} \big(D^A\chi, \alpha \big)= \int\Tr\big( D^A \chi \w  * \alpha \big)=\int\Tr\big( \chi \w  * D^{A\dagger}\alpha \big)$, $\forall \chi \in \text{Lie}\H=\Omega^0_\text{eq}(\P, \text{Lie}H)$ and the metric being non-degenerate.
  

 \paragraph{Dual horizontalisation of forms}
 
Since a connection allows to define the horizontal projection of vector fields, by duality it allows to define the ``horizontalisation'' of forms. Given $\bs\alpha \in \Omega^\bullet(\Phi)$, one defines the corresponding $\bs\alpha^h \defeq \bs\alpha \circ |^h \in \Omega^\bullet_\text{hor}(\Phi)$. 
 In  view of the expression for the horizontal projection \eqref{horiz-proj-map}, the explicit form of $\bs \alpha^h$ in terms of $\bs\alpha$ and $\bs\omega$  depends on the verticality property of $\bs \alpha$.

It is thus clear that if $\bs\alpha \in \Omega_\text{eq}^\bullet(\Phi, \bs V)$, then $\bs\alpha^h \in \Omega_\text{tens}^\bullet(\Phi, \bs V)$.
  Indeed, 
  much like the computation of the equivariance of $\bs{D^\omega \alpha}$ above, we have
$
  R^\star_\gamma \bs \alpha^h(\bs X, \ldots) =  \bs \alpha^h\big(R_{\gamma\star}\bs X, \ldots \big) \defeq \bs\alpha \big( (R_{\gamma\star}\bs X)^h, \ldots\big) =\bs\alpha \big( R_{\gamma\star}\bs X^h, \ldots\big) = R^\star_\gamma \bs \alpha(\bs X^h, \ldots)
  								= \rho(\gamma)\- \bs \alpha(\bs X^h, \ldots) \rdefeq \rho(\gamma)\- \bs \alpha^h(\bs X, \ldots).
  $ 
Idem, then, for twisted forms: for $\bs\alpha \in \Omega_\text{eq}^\bullet(\Phi, C)$, we have $\bs\alpha^h \in \Omega_\text{tens}^\bullet(\Phi, C)$.

As illustrative applications, consider first the basis $\bs d\phi=\left\{ \bs d A, \bs d \vphi \right\} \in \Omega^1_\text{eq}(\Phi)$, to which is associated $\bs d\phi^h=\left\{ \bs d A^h, \bs d \vphi^h \right\} \in \Omega^1_\text{tens}(\Phi)$, whose explicit form is,
 by \eqref{Vert-dphi} and \eqref{horiz-proj-map},
\begin{align}
\label{dphi-horiz}
\bs d \phi^h_{|\phi}= \bs d \phi_{|\phi} - \delta_{\bs \omega} \phi \quad \Rightarrow \quad \left\{  \begin{array}{c} \bs d A^h_{|\phi} = \bs dA_{|\phi} - D^A\bs\omega,  \\[1mm]  \bs d \vphi^h_{|\phi} = \bs d\vphi_{|\phi} + \rho_*(\bs\omega) \vphi.  \end{array} \right.
\end{align}
Remark that despite a superficial resemblance, these are not covariant derivative formulae ($\phi$ is a point of $\Phi$ not an equivariant function on it). 
We find that applying the covariant derivative on $\bs d \phi^h$ results in $\bs{D^\omega} \bs d \phi^h = -\delta_{\bs\Omega}\phi$. Indeed,
\begin{align}
\bs{D^\omega} \bs d A^h &= \bs d (\bs d A^h) + \rho_*(\bs \omega) \bs dA^h = \bs d(-D^A \bs\omega) + [\bs\omega, \bs d A - D^A\bs\omega], \notag \\
				       &= -D^A(\bs{d\omega}) - [\bs d A, \bs \omega] + [\bs\omega, \bs dA] - D^A\big(\sfrac{1}{2}[\bs\omega, \bs \omega] \big) = -D^A \bs \Omega. \\[1mm]
\bs{D^\omega} \bs d \vphi^h &= \bs d (\bs d \vphi^h) + \rho_*(\bs \omega) \bs d\vphi^h = \bs d\big(\rho_*(\bs\omega)\vphi \big) + \rho_*(\bs\omega)\big( \bs d\vphi + \rho_*(\bs\omega)\vphi \big), \notag\\
					    &= \rho_*(\bs{d\omega}) \vphi + \rho_*(\bs\omega \w \bs \omega)\vphi= \rho_*(\bs\Omega)\vphi.
\end{align}
Equation \eqref{dphi-horiz} gives, for example, a general expression for 1-forms: Given $\bs \alpha = \alpha \big(\bs d \phi; \phi \big) \in \Omega^1(\Phi)$, we have  
\begin{align}
\label{horiz-1-form}
\bs\alpha^h =\alpha\big(\bs{d}\phi^h ; \phi\big)= \bs\alpha -  \iota_{[\bs\omega]^v} \bs\alpha =\bs\alpha -  \alpha\big( \delta_{\bs\omega}\phi; \phi \big) \ \ \in \Omega^1_\text{hor}(\Phi).
\end{align}

Consider then $\bs dF, \bs d\,\!D^A\vphi \in \Omega^1_\text{eq}(\Phi)$ to which, by \eqref{Vert-dF}, \eqref{Vert-dDphi} and  \eqref{horiz-proj-map},  one associates  $\bs dF^h, (\bs d\,\!D^A\vphi)^h \in \Omega^1_\text{tens}(\Phi)$ given respectively by
\begin{align}
\label{dF-dDphi-hor}
\bs dF^h = \bs d F - [F, \bs\omega] \quad \text{and} \quad (\bs d\,\!D^A\vphi)^h = \bs d\,\!D^A\vphi + \rho_*(\bs\omega) D^A\vphi. 
\end{align} 
But remark that since $F, D^A\vphi \in \Omega^0_\text{eq}(\Phi)=\Omega^0_\text{tens}(\Phi)$ we have $\bs{D^\omega}F, \bs{D^\omega}(D^A\vphi) \in \Omega^1_\text{tens}(\Phi)$ with explicit expressions,
\begin{align}
\bs{D^\omega}F = \bs d F + [\bs\omega, F] \quad \text{and} \quad \bs{D^\omega}(D^A\vphi) = \bs d\,\!D^A\vphi + \rho_*(\bs\omega) D^A\vphi. 
\end{align} 
In this occasion then, \eqref{dF-dDphi-hor} are indeed covariant derivative expressions. Which is not surprising as, by definition, $\bs{D^\omega}\defeq \bs d \circ |^h$ on $\Omega^\bullet_\text{eq}(\Phi)$, so that the `horizontalisation' of $\bs dF$ and $\bs d\,\!D^A\vphi$ is precisely the definition of the covariant derivative of $F$ and $D^A\vphi$. It is then immediate that,
\begin{align}
\bs{D^\omega} (\bs dF^h) = \bs{D^\omega} \circ \bs{D^\omega} F = [\bs\Omega, F] \qquad \text{and} \qquad \bs{D^\omega} (\bs d\,\!D^A\vphi)^h = \bs{D^\omega} \circ \bs{D^\omega} (D^A\vphi) = \rho_*(\bs\Omega)D^A\vphi.  
\end{align}

 A  case of special relevance to our general purpose is that to  $\bs\alpha \in \Omega_\text{inv}^\bullet(\Phi)$ one then associates $\bs\alpha^h \in \Omega_\text{basic}^\bullet(\Phi)$. 
 In~this case only, one may observe that both $\bs{D^\omega}\bs\alpha\defeq(\bs d\bs \alpha)^h$ and $\bs d \bs\alpha^h$ are  basic forms, yet in general $\bs d\bs\alpha^h \neq \bs{D^\omega \alpha}$.   
 For~$\bs\alpha=\alpha\big(\bs d\phi; \phi \big) \in \Omega^1_\text{inv}(\Phi)$ in particular, we prove in Appendix \ref{Proof of formula 1} that  the following formula holds, 
 \begin{align}
 \label{Formula1}
 \bs{D^\omega\alpha}= \bs d \bs\alpha^h +  \iota_{[\bs\Omega]^v}\bs\alpha
 				=  \bs d \bs\alpha^h + \alpha\big(  \delta_{\bs\Omega} \phi; \phi \big), 
 \end{align}
 using the notation introduced in equation  \eqref{Vert-dphi}. So that, only in the special case of a  flat $\mathring{\bs\omega}$ do we get $ \bs D^{\mathring{\bs\omega}} \bs \alpha= \bs d \bs\alpha^h$.
 
 \paragraph{Ambiguity in the choice of connection} 

As already seen, connections are non-canonical and form  an affine space. So,~the~above `horizontalisation' and `basification' of forms suffers from an ambiguity due to an a priori \mbox{arbitrariness} in the choice of connection.
  Given any $\bs \beta \in \Omega^1_\text{tens}(\Phi, \text{lie}\H)$, both $\bs\omega$  and $\bs\omega' = \bs\omega +\bs\beta$ are valid choices of \mbox{connections}. Given $\bs\alpha \in \Omega^\bullet(\Phi)$, denote $\bs\alpha^h_{\bs\omega}$ and $\bs\alpha^h_{\bs\omega'}$  the corresponding horizontal forms obtain by the above procedure respectively through $\bs\omega$ and $\bs\omega'$. One may ask how they are related. The answer is easily found for 1-forms via a well known~trick (familiar e.g. from the proof of the Chern-Weil homomorphism theorem).
  
  Consider the affine curve in the space of connections $\bs\omega_\tau \defeq \bs\omega + \tau \bs \beta$, with $\tau \in [0,1]$, s.t. $\bs\omega_0=\bs\omega$ and  $\bs\omega_1=\bs\omega'$. Given $\bs\alpha \in \Omega^1(\Phi)$, we have by definition 
  $\bs \alpha^h_{\bs\omega_\tau} = \bs\alpha - \iota_{[\bs\omega_\tau]^v} \bs\alpha \in \Omega^1_\text{hor}(\Phi)$. We have then the generic result, 
 \begin{align}
 \label{Ambiguity-hor}
 \textstyle\int_0^1d\tau\ \tfrac{d}{d\tau}  \bs \alpha^h_{\bs\omega_\tau}& = \bs\alpha^h_{\bs\omega'} - \bs\alpha^h_{\bs\omega} = - \iota_{[\bs\beta]^v} \bs\alpha, \notag \\[1mm]
& \hookrightarrow \quad  \bs\alpha^h_{\bs\omega'} =  \bs\alpha^h_{\bs\omega} - \iota_{[\bs\beta]^v} \bs\alpha \quad  \in \Omega^1_\text{hor}(\Phi).
 \end{align} 
 In particular, considering $\bs d \phi \in \Omega^1_\text{eq}(\Phi)$ we have
  \begin{align} 
  \label{Ambiguity-hor-dphi}
 \bs d\phi^h_{\bs\omega'} =  \bs d\phi ^h_{\bs\omega} - \iota_{[\bs\beta]^v} \bs d \phi = \bs d\phi ^h_{\bs\omega} - \delta_{\bs\beta} \phi \quad \in \Omega^1_\text{tens}(\Phi).
\end{align} 
From this, one can in principle work out the precise relation between $\bs\alpha^h_{\bs\omega'}=\alpha\big(\!\w^\bullet\! \bs d\phi^h_{\bs\omega'}\, ;\,  \phi \big)$ and $\bs\alpha^h_{\bs\omega}=\alpha\big(\!\w^\bullet\! \bs d\phi^h_{\bs\omega}\, ;\,  \phi \big)$. We do not do so  as we will not make use of such a general result. 

 In the case of special interest to us $\bs\alpha = \alpha\big(\bs d \phi; \phi \big)\in \Omega^1_\text{inv}(\Phi)$, as the above procedure produces basic forms, we  alter the notation in eq.\eqref{Ambiguity-hor} to have more suggestively,
 \begin{align} 
 \label{Ambiguity-basic-1-form}
 \bs\alpha^b_{\bs\omega'} &=  \bs\alpha^b_{\bs\omega} - \iota_{[\bs\beta]^v} \bs\alpha \quad  \in \Omega^1_\text{basic}(\Phi), \notag \\
 					&=\bs\alpha^b_{\bs\omega} - \alpha\big( \delta_{\bs\beta} \phi; \phi \big). 
\end{align} 
Consequently, the covariant derivatives are related by
 \begin{align} 
 \label{Ambiguity-basic-2-form}
\bs d \bs\alpha^b_{\bs\omega'} = \bs d\bs\alpha^b_{\bs\omega} - \bs d \alpha\big( \delta_{\bs\beta} \phi; \phi \big)   \quad  \in \Omega^2_\text{basic}(\Phi).  	
\end{align} 
These two results we will use when considering the basic presymplectic structure of gauge theories. 

The ambiguity in the procedure being worked out,  it may still be that there is a choice of connection more natural than others. Such would be the case,  as seen above, if it is associated with a natural bundle metric.

\subsubsection{Variational twisted connections}
\label{Variational twisted connections}

We briefly review a notion of connection which, relatively to the aim of covariant differentiation, is a slight generalisation of Ehresmann's. 
A variational twisted connection is by definition $\t{\bs\omega} \in \Omega^1_\text{eq}(\Phi, \text{Lie}G)^C$ satisfying:
\begin{align}
\label{var-twisted-connection1}
R^\star_\gamma \t{\bs\omega}_{|\phi^\gamma} &= C(\phi, \gamma)\- \t{\bs\omega}_{|\phi}\, C(\phi, \gamma) +C(\phi, \gamma)\- \bs d C(\ , \gamma)_{|\phi}, \\[1mm]
\label{var-twisted-connection2}
\t{\bs\omega}_{|\phi} (\chi^v_\phi)&=\tfrac{d}{d\tau} C\big(\phi, \exp(\tau \chi)\big) \big|_{\tau=0}, \ \ \chi\  \in \text{Lie}\H. 
\end{align}
It is non-canonical, and from the above defining axioms follows that the space of twisted connections is affine and modelled on the vector space $\Omega^1_\text{tens}(\Phi, \text{Lie}G)^C$: for $\bs\beta \in \Omega^1_\text{tens}(\Phi, \text{Lie}G)^C$,  we have that $\t{\bs\omega}'=\t{\bs\omega}+\bs\beta$ is another twisted connection. 
In view of \eqref{var-twisted-connection1}-\eqref{var-twisted-connection2}, and  by \eqref{Pushforward-X-inf}, it takes some doing to show that  the $\bs\H$-gauge transformation of a twisted connection is  
$\bs\omega^{\bs\gamma} \defeq \bs\Psi^\star \bs\omega=C(\bs\gamma)\- \bs\omega \, C(\bs\gamma)+ C(\bs\gamma)\- \bs d C(\bs \gamma)$, where we introduce the convenient notation $C\big(\bs\gamma(\phi)\big)\defeq C\big(\phi, \bs\gamma(\phi) \big)$. 

A covariant derivation is defined on twisted forms by $\bs D^{\t{\bs\omega}} \defeq \bs d \ + \t{\bs\omega} : \Omega^\bullet_\text{tens}(\Phi, \bs V)^C \rarrow \Omega^{\bullet+1}_\text{tens}(\Phi, \bs V)^C$, and of course  $\bs D^{\t{\bs\omega}} =\bs d$ on $\Omega^\bullet_\text{basic}(\Phi, \bs V)$. 
 The  curvature of a twisted connection  is defined via Cartan's structure equation, $\t{\bs\Omega}\defeq \bs d\t{\bs\omega} +\tfrac{1}{2}[ \t{\bs\omega}, \t{\bs\omega}] \in \Omega^2_\text{tens}(\Phi, \text{Lie}G)^C$,   thus it satisfies a Bianchi identity $\bs D^{\t{\bs\omega}}  \t{\bs\Omega}\equiv0$, and has $\bs\H$-transformation  $\t{\bs\Omega}^{\bs\gamma}=C(\bs\gamma)\- \t{\bs\Omega} C(\bs \gamma)$. As usual $\bs D^{\t{\bs\omega}} \circ \bs D^{\t{\bs\omega}} = \t{\bs \Omega}$. 
\bigskip

Twisted connection are the adequate tool to tackle general non-invariant gauge theories (case 3, just above \ref{Variational connections on field space}). As we have seen, the Lagrangian of such a theory has $\H$-equivariance $R^\star_\gamma L = L + c(\ , \gamma)$, whose infinitesimal version gives the \emph{classical gauge anomaly} $\bs L_{\chi^v} L=\tfrac{d}{d\tau} c(\ \, , \exp{\tau \chi})\big|_{\tau=0} \rdefeq \alpha(\chi, \ )$. 
Considering the associated functional $Z \defeq \exp{i\int L} \in \Omega^0_\text{tens}(\Phi, \CC)^C$, s.t.  $R^\star_\gamma Z = C(\, ,\gamma)\- Z$ with $C(\ ,\gamma)\defeq \exp{-i \int c(\ ,\gamma)} \in U(1)$, we see that a non-invariant theory defines a section of the twisted associated line bundle $\t{\bs E} \defeq \Phi\times_C \CC$. 
The adapted abelian twisted connection is  $\t{\bs\omega} \in \Omega^1_\text{eq}(\Phi, i\RR)^C$ s.t. 
\begin{align}
\label{var-twisted-connection1bis}
R^\star_\gamma \t{\bs\omega}_{|\phi^\gamma}  &=  \t{\bs\omega}_{|\phi}   - i \textstyle\int \bs d c(\, , \gamma)_{|\phi}, \\[1mm]
\label{var-twisted-connection2bis}
\t{\bs\omega}_{|\phi}(\chi^v_\phi)  &= -i \textstyle\int \alpha(\chi, \phi) \rdefeq -i\,a(\chi, \phi).
\end{align}
The verticality property of the twisted connection encodes the gauge anomaly, and consequently the tensoriality of the curvature $\t{\bs \Omega} = \bs d \t{\bs \omega}$ encodes the Wess-Zumino consistency condition:
 \begin{align}
 \t{\bs \Omega}(\chi^v, \eta^v) = \bs d \t{\bs \omega}(\chi^v, \eta^v) &= \chi^v \cdot \t{\bs \omega}(\eta^v) - \eta^v \cdot  \t{\bs \omega}(\chi^v) -  \t{\bs \omega}([\chi^v, \eta^v]) \equiv 0  \\
												    \hookrightarrow & \quad  \chi^v \cdot a(\eta, \phi) -  \eta^v \cdot a(\chi, \phi)  - a([\chi, \eta], \phi) = 0.  \notag
 \end{align}
 To be compared e.g. to  eq.(8.62) and eq.(10.76) in \cite{Bertlmann}, or eq.(12.25) in \cite{GockSchuck}.

 Interestingly, one find a `modified basic action' from the twisted covariant derivative of $Z$. Indeed,
  \begin{align}
  \bs D^{\t{\bs\omega}}Z = \bs d Z +\t{\bs\omega} Z = \big( i \bs d \textstyle\int L + \t{\bs\omega} \big)\, Z =  \big( i \bs dS + \t{\bs\omega} \big) \, Z,
  \end{align}
  where $S\defeq \int L$ is the classical action.
 Since $Z$ and $\bs D^{\t{\bs\omega}}Z \in \Omega^\bullet_\text{tens}(\Phi, \CC)^C$, we have that $i \bs d S + \t{\bs\omega} \in \Omega^1_\text{basic}(\Phi, i\RR)$. This is easily checked explicitly by \eqref{var-twisted-connection1bis}-\eqref{var-twisted-connection2bis},
 \begin{align}
& R^\star_\gamma \big(i\bs d S + \t{\bs\omega} \big)=i\bs d (\textstyle\int L+ c(\, \, ,\gamma)) + \t{\bs\omega} - i \textstyle\int\bs d c(\ ,\gamma)=i\bs d S + \t{\bs\omega}, \\
& \big(  i\bs d S + \t{\bs\omega} \big)(\chi^v) = i \textstyle\int \bs L_{\chi^v} L  + \t{\bs\omega} (\chi^v) = i \textstyle\int \alpha(\chi, \ ) - i \textstyle\int \alpha(\chi, \ )=0.
\end{align}
 The  quantity $\bs d S - i \t{\bs\omega}$ is a generalisation of the notion of Wess-Zumino `improved' action \cite{Manes-Stora-Zumino1985, Attard-Lazz2016}. Indeed, if the connection is locally exact, $\t{\bs\omega}=\bs d \t{\bs\varpi}$, then $S-i\t{\bs\varpi} \in \Omega^0_\text{basic}(\Phi, \RR)=\Omega^0_\text{inv}(\Phi, \RR)$, and $ \t{\bs\varpi}$ is a Wess-Zumino functional. 
 In  section \ref{Field-dependent dressing fields and variational connections} we will see how such a term can be built explicitly from the cocycle $c$ and a dressing field. 
  
What is done here for classical gauge anomalies holds also for quantum gauge anomalies, see  sections of 2.3 and 4.2 of \cite{Francois2021}, and \cite{Francois2019_II}  for a full exposition of the geometry of twisted connections

\medskip
The extraction of a basic action 0-form above is quite incidental. As a twisted connection is a priori not meant to split the  SES \eqref{SESLieAlg-inf}, it does not provide a notion of horizontality on $\Phi$ and thus would not be used to extract the horizontal part of any given form. A fortiori, it is unlikely to provide a general strategy to build basic forms. In the following section we discuss a general method designed to do just that.

\subsection{The dressing field method}
\label{The dressing field method}

The dressing field method (DFM) is a systematic way to build basic forms on a bundle, thus to obtain gauge-invariants in gauge theory. It has gradually developed  \cite{GaugeInvCompFields, Francois2014, Attard_et_al2017} in recent years, and its immediate implications regarding the philosophy of gauge theories -- relevant to the present paper --  as been first expound in \cite{Francois2018}. 
In relation to the topic of the presymplectic structure of gauge theories, the DFM is the geometric underpinning of the so-called \emph{edge modes} \cite{DonnellyFreidel2016, Geiller2017, Speranza2018, Geiller2018, Geiller2019, Freidel-et-al2020-1, Freidel-et-al2020-2, Freidel-et-al2020-3}. 
A relatively complete and self-contained exposition can be found in sections 3 and 4.3 of \cite{Francois2021}. But in the interest of the reader, we give here a summary of the essentials of the method. 
\medskip

Consider a $H$-principal bundle $\P$ with connection $A$, and $\alpha \in \Omega^\bullet_\text{tens}(\P, V)$. Let $K \triangleleft H$ be a normal subgroup of the structure group, so that the quotient $H/K=J$ is a group. Correspondingly we have the gauge subgroups $\K, \J \subset \H$. Also, let $G$ be a group s.t.  $G \subset H$. 

We define the space of $K$-dressing fields as 
$\D r[K,G] \defeq \big\{ u : \P \rarrow G\ | \ R^*_k u = k\- u, \forall k\in K\big\}$, a definition that implies that the $\K$-gauge transformation on $\P$ of a dressing is $u^\gamma =\gamma\- u$. By means of the dressing field, one defines the map $f_u: \P \rarrow \P/K=\P'$, $p\mapsto f_u(p)\defeq pu(p)$, where $\P'$ is a $J$-principal subbundle of $\P$. In other word, the existence of a $K$-dressing field implies that the bundle $\P$ is trivial along $K$: $\P \simeq \P' \times K$. This map satisfies $f_u \circ R_k = f_u$, so $R^*_k \circ f^*_u=f^*_u$ and $f_{u*} X^v =0$ for $X \in $ Lie$K$ (it kills vertical vector fields generated by the action of $K$). Therefore, the following \emph{dressed fields}
\begin{align} 
\label{dressed-fields} 
A^u\defeq f^*_u A = u\- Au +u\-du, \quad \text{ and } \quad \alpha^u \defeq f^*_u \alpha = \rho(u)\-\alpha,  
\end{align}
are $K$-basic on $\P$, thus descend on $\P'$. This implies that both are $\K$-invariant: $(A^u)^\gamma=A^u$ and $(\alpha^u)^\gamma=\alpha^u$ for $\gamma \in \K$. 
In particular, the curvature $F \in \Omega^2_\text{tens}(\P, \text{Lie}H)$ and $\vphi \in  \Omega^0_\text{tens}(\P, V)\simeq \Gamma(E)$ dress as $F^u=u\- F u = dA^u+\sfrac{1}{2}[A^u, A^u]$ (which is then the curvature of $A^u$) and $\vphi^u=\rho(u)\-\vphi$.

Let us emphasize an important fact: It is clear from the definition that $u \notin \K$, so that despite the formal resemblance \eqref{dressed-fields} are \emph{not} gauge transformations. This means, in particular, that the dressed connection is not a $H$-connection, $A^u \notin \A$, and a fortiori is not a point in the gauge $\K$-orbit $\O_\K[A]$ of $A$, so that $A^u$ must not be confused with a gauge-fixing of $A$. 

We also point out that,  with minimal adjustments,  the above results can be extended to the case $G \supset H$: One needs only to assume that $G$ is (a~subgroup~of) the structure group of a bigger principal bundle of which $\P$ is a subbundle/a reduction. This is typically the case for Cartan geometries, on which gauge gravity theories are based. See section 2.2, and footnote 12, in  \cite{Francois2021}.

Finally, remark that if $K=H$ the bundle is trivial, $\P \simeq M \times H$, and $A^u, \alpha^u \in \Omega^\bullet_\text{basic}(\P)$. This means in particular that in this case there is a 1:1 association between the dressed fields $\phi^u=(A^u, \vphi^u)$ and the $\H$-gauge orbit $\O_\H[\phi]$. So,  $\phi^u$ can be thought of as a `coordinatisation' for the gauge class $[\phi]$ such that $\Phi^u \simeq \Phi/\H=\M$. 
This must be qualified, as the dressed fields may exhibit residual transformations.

\paragraph{Residual gauge symmetry} 

 If $K \triangleleft H$, one expects \eqref{dressed-fields} to display a residual $\J$-gauge transformations which will depend on the $\J$-transformation of $u$ (that of $A, \alpha$ being known already), the latter in turn given by  its $J$-equivariance. We will not linger on the details here, referring  to section 3.2 of \cite{Francois2021} for details, only to mention one  interesting case which is when the $K$-dressing field has $J$-equivariance $R^*_j u = j\- u j$ so that   its $\J$-gauge transformation is $u^\eta=\eta\- u \eta$ and that of the dressed fields is then, 
 \begin{align} 
\label{residual-1} 
(A^u)^\eta\defeq \Psi^* A^u  = \eta\- A^u \eta +\eta\-d\eta, \quad \text{ and } \quad (\alpha^u)^\eta \defeq \Psi^* \alpha^u = \rho(\eta)\-\alpha^u,  
\end{align}
for $\Psi \in \Aut_v(\P, J) \simeq \Aut_v(\P')$ ($J$-automorphism) s.t. $\Psi(p)=p\eta(p)$ with $\eta \in \J$. 
This  can be checked algebraically from $(A^u)^\eta= (A^\eta)^{u^\eta}$. 
 
 In the rest of this paper we will consider the case of dressing fields $u \in \D r[H, G]$ for simplicity.\footnote{Although, as we have remarked, it is a strong constraint on the topology of $\P$ -- it is trivial -- which in turn as  interpretive implications \cite{Francois2018} quite significant for the `edge mode' strategy, as  argued in  \cite{Francois2021} and as we will further comment ahead.} Yet, even in this case the dressed fields   may display another form of residual transformations worth stressing, stemming from an ambiguity in the choice of dressing field. 
 
 \paragraph{Ambiguity in the choice of a dressing field}

Given their defining property, two dressing fields $u, u' \in \D r[H, G]$ may a priori be related by $u'=u\xi$, with $\xi \in  \G \defeq \left\{\,   \xi:\P \rarrow G\, |\,  R^*_h \xi = \xi  \, \right\}$.  By analogy with the notation for the action of the gauge group $\H$, let us denote the action of $\G$ on $\D r[H, G]$  as $u^\xi=u\xi$. 
By definition $\G$ acts trivially  on  $A$ and $\alpha$, a fact we denote by $A^\xi=A$ and $\alpha^\xi=\alpha$. On the other hand, it is clear how $\G$ must act on  dressed fields: 
\begin{align}
\label{Residual-2}
(A^u)^\xi \defeq A^{u^\xi}= A^{u\xi}=\xi\- A^u \xi + \xi\-d\xi, \quad \text{ and } \quad (\alpha^u)^\xi \defeq \alpha^{u^\xi}= \alpha^{u\xi} = \rho(\xi)\-\alpha^u, 
\end{align}
which implies  in particular $(F^u)^\xi=\xi\- F^u \xi$ and $(\vphi^u)^\xi=\rho(\xi)\- \vphi^u$.

This invites to think of the space of dressed fields $\Phi^u$ as fibered by the right action of $\G$, noted then $R_\xi \phi^u \defeq (\phi^u)^\xi$, so that $\Phi^u$ is a $\G$-principal bundle over $\Phi^u/\G\defeq \M^u$.
As such, and in complete analogy with $\Phi$, it gives rise to a SES
 \begin{align}
 \label{SESgroups-dressed}
\makebox[\displaywidth]{
\hspace{-18mm}\begin{tikzcd}[column sep=large, ampersand replacement=\&]
\&0     \arrow [r]         \& \bs{\Aut}_v(\Phi^u) \simeq \bs{\G}     \arrow[r, "\iota"  ]          \& \bs{\Aut}(\Phi^u)       \arrow[r, "\t\pi"]      \&  \bs{\Diff}(\M^u)        \arrow[r]      \& 0,
\end{tikzcd}}  \raisetag{3.4ex}
\end{align}
 where $\bs{\Aut}(\Phi^u)$ is the automorphism group defined as usual, and the subgroup of vertical automorphisms is isomorphic to the gauge group $\bs{\G}\defeq \left\{ \bs{\xi}: \Phi^u \rarrow \G\, | \ R^\star_\xi\bs{\xi}=\xi\-\bs{\xi} \xi \right\}$. Its infinitesimal version is,
\begin{align}
\label{SESLieAlg-dressed}
\makebox[\displaywidth]{
\hspace{-18mm}\begin{tikzcd}[column sep=large, ampersand replacement=\&]
\&0     \arrow [r]         \& \Gamma_{\!\G}(V\Phi^u) \simeq \text{Lie}\bs{\G}     \arrow[r, "\iota"  ]          \&  \Gamma_{\!\G}(T\Phi^u)     \arrow[r, "\pi_\star"]      \&  \Gamma(T\M^u)         \arrow[r]      \& 0,
\end{tikzcd}}  \raisetag{3.4ex}
\end{align}
As usual, the action by pullback of $\bs{\Aut}_v(\Phi^u)$ on elements of $\Omega^\bullet(\Phi^u)$ defines their $\bs\G$-gauge transformations which is thus determined by their $\G$-equivariance and verticality properties, i.e. the result of their evaluation on  $\zeta^v \in \Gamma(V\Phi^u)$ for $\zeta \in \text{Lie}\G$. We will have more to say on this shortly. 

Or course, $(\phi^u)^\xi$ is  $H$-basic (and $\H$-invariant)  $\forall \xi \in \G$, so any given representative in the $\G$-orbit $\O_{\G}[\phi^u]$ is as good a coordinatisation for $[\phi] \in \M$ as any other. Said otherwise there is a a $1:1$ correspondence $\O_\H[\phi] \sim \O_{\G}[\phi^u]$. Which means that, contrary to a first analysis, it is $\Phi^u/\G \defeq \M^u$ that is isomorphic to $\Phi/\H =\M$ (not $\Phi^u$). 
As~the latter is the physical state space, it follows that $\G$ is not a permutation group of physical states. Rather, as we know, $\bs{\Diff}(\Phi/\H) \simeq \bs{\Diff}(\Phi^u/\G)$ is, with infinitesimal counterpart $\Gamma(\Phi/\H) \simeq \Gamma(\Phi^u/\G)$.

This being clarified, there are only two relevant options regarding the physical status of the group $\G$: Either it is an interesting new gauge symmetry, as is the case in gauge gravity theories where $\G=\GL(n)$ is the group of local coordinate changes (see, section 5.3.1.b in \cite{Francois2021}), 
 and as such the associated Noether charges may be observables when `measured' against background field configurations for whom elements of $\G$ are Killing symmetries (a topic we address first for $\H$ in section \ref{Presymplectic structure of invariant matter  coupled gauge theories}, then for $\G$ in section \ref{Via dressing fields}).  
Or  there are compelling reasons as to why the group $\G$ must be `small' compared to $\H$ (perhaps even reduced to a global/rigid or discrete group). 
For either options to stand a chance of being realised, a dressing field must be introduced not by hand as new degrees of freedom, but built from elements of  the initial field space. 

This~suggests to consider \emph{field-dependent} dressing fields which,  as it turns out,  also permit to build basic forms on the $\H$-bundle $\Phi$.

 \paragraph{Field-dependent dressing fields}  
 
 A $\Phi$-dependent dressing field is a map $\bs u : \Phi \rarrow \D r[H, G]$, $\phi \mapsto \bs u(\phi)$, thus satisfying $R^\star_\gamma \bs{u}=\gamma\- \bs{u}$ -- i.e. $\bs u(\phi^\gamma)=\gamma\-\bs u(\phi)$ -- for $\gamma\in \H$.\footnote{Again we here work with the simplifying assumption of a $H$-dressing. But the following can be adapted with minor adjustments to $\Phi$-dependent $K$-dressing fields $\bs u : \Phi \rarrow \D r[K, G]$, s.t.  $R^\star_\gamma \bs{u}=\gamma^{-1} \bs{u}$ for $\gamma \in \K \subset \H$, leaving then  residual $\J$-gauge transformations (called residual transformations of the first kind in \cite{Francois2021}).}
  Given the above considerations, it allows to define   
  \begin{align}
 \label{Dressing-map}
 \text{F}_{\bs{u}} : \Phi &  \rarrow  \M    \notag\\
 			     \phi & \mapsto \text{F}_{\bs{u}}(\phi) \defeq \phi^{\bs u}=(A^{\bs u}, \vphi^{\bs u}), \quad \text{s.t.} \quad  \text{F}_{\bs u}\circ R_\gamma=\text{F}_{\bs u}.
 \end{align}
 This map in a sense \emph{realises} the projection map $\pi$ of $\Phi$.
So clearly, $\Gamma(V\Phi) \in \ker \text{F}_{\bs u}$: for $\chi^v  \in \Gamma(V\Phi)$ generated by $\chi \in$ Lie$\H$, we have $\text{F}_{\bs{u}\star} \, \chi^v_\phi = \tfrac{d}{d\tau} \big( \text{F}_{\bs u} \circ R_{e^{\tau \chi}}\big)(\phi) \big|_{\tau =0} = \tfrac{d}{d\tau} \text{F}_{\bs u}(\phi)\big|_{\tau=0}=0$. 
Despite the formal resemblance with a vertical automorphism, $\text{F}_{\bs u} \notin \Aut_v(\Phi)$, as is clear from the fact that  $\bs{u} \notin \bs\H$. 
Yet, in exact analogy with the computation of  $\bs\Psi_\star \bs X$ for $\bs X \in \Gamma(T\Phi)$, eq. \eqref{Pushforward-X-inf}, (which owes nothing to the $\H$-equivariance of $\bs \gamma \in \bs\H$), we have:
  \begin{alignat}{2}
  \label{Dressed-vector}
   \text{F}_{\bs{u}\star} : T_\phi \Phi & \rarrow T_{\phi^{\bs{u}}} \M   \notag\\
 			         \bs{X}_\phi & \mapsto  \text{F}_{\bs{u}\star} \bs{X}_\phi  
			         										     = \uprho (\bs u)\-    \left( X(\phi)  +  \delta_{ \bs{d uu}\-_{|\phi}(\bs X_\phi)} \phi \right) \tfrac{\delta}{\delta [\phi]},
  \end{alignat}
Dually, the pullback application allows to realise \emph{basic} forms on $\Phi$,
 \begin{align}
 \label{Dressed-basic-forms}
 \text{F}^\star_{\bs{u}} : \Omega^\bullet (\M) &  \rarrow   \Omega_\text{basic}^\bullet(\Phi )  \notag\\*
 			    \b{\bs{\alpha}}_{|[\phi]} & \mapsto \text{F}_{\bs{u}}^\star  \b{\bs{\alpha}}_{\, |\phi}  \rdefeq    {\bs{\alpha}^{\bs u}}_{|\phi}.
 \end{align}
 Indeed, as $\text{F}_{\bs u}\sim \pi$, $ \Omega_\text{basic}^\bullet(\Phi )= \im \pi^\star \simeq \im \text{F}_{\bs u}^\star$. The $\H$-basicity of $\bs{\alpha}^{\bs u}$ is easily proven:
 $R^\star_\gamma \bs{\alpha}^{\bs u} = R^\star_\gamma \text{F}_{\bs u}^\star \b{\bs{\alpha}} = \text{F}_{\bs u}^\star \b{\bs{\alpha}} = \bs{\alpha}^{\bs u}$, with $\gamma \in \H$, and $\bs{\alpha}^{\bs u}\big( \chi^v \big) = \big(  \text{F}_{\bs u}^\star  \b{\bs{\alpha}} \big) (\chi^v)= \b{\bs{\alpha}}\big(  \text{F}_{\bs{u} \star}\, \chi^v\big) =0$. The $\bs \H$-invariance ensues, $(\bs{\alpha}^{\bs u})^{\bs \gamma} = \bs{\alpha}^{\bs u}$ for $\bs{\gamma} \in \bs\H$. 
 
\medskip
 
As it stands, $\bs{\alpha}^{\bs u}=\text{F}_{\bs u}^\star \b{\bs{\alpha}}$ is the basic counterpart of $\b{\bs{\alpha}} \in \Omega^\bullet(\M)$.
But as the notation suggests, we would rather like to see $\bs{\alpha}^{\bs u}$ as the basic version of some given form $\bs\alpha \in \Omega^\bullet(\Phi)$ whose functional expression could presumably be given in in terms of $\bs\alpha$ and  $\bs u$. This is indeed possible via a shift of viewpoint: 
One may notice that  to a given $\bs \alpha_{|\phi} =\alpha\big (\!\w^\bullet\! \bs d \phi \, ;\,  \phi \big) \in \Omega^\bullet(\Phi)$  corresponds  $\b{\bs \alpha}_{|[\phi]} =\alpha\big (\!\w^\bullet\! \bs d [\phi] \, ;\,  [\phi] \big) \in \Omega^\bullet(\M)$ built via the same functional $\alpha(\ \, ; \ )$ (the~only difference being the type of arguments the latter takes in). Now we can define the \emph{dressed version}, or  \emph{dressing}, of $\bs\alpha$ via  \eqref{Dressed-basic-forms} as being
 \begin{align}
 \label{Dressed-variational-form}
  {\bs{\alpha}^{\bs u}}_{|\phi} \defeq \text{F}^\star_{\bs u} \b{\bs{\alpha}}_{\, |\phi} = \alpha\big( \!\w^\bullet\!  \text{F}^\star_{\bs u} \bs{d}[\phi]; \text{F}_{\bs u}(\phi) \big)=\alpha \big( \!\w^\bullet\!\bs{d}\phi^{\bs u} ; \phi^{\bs u}   \big)
  \ \  \in \Omega^\bullet_\text{basic}(\Phi),
 \end{align} 
 where we have defined the basic basis 1-form $\bs d \phi^{\bs u} \defeq \text{F}^\star_{\bs u} \bs d[\phi] \in \Omega^1_\text{basic}(\Phi)$, with $ \bs d[\phi] \in \Omega^1(\M)$  basis of forms on $\M$. As per our stated desiderata, the latter can be written explicitly,  via  \eqref{Dressed-vector},  in terms of $\bs d\phi$ and $\bs u$:
\begin{align}
\label{Dressed-dphi}
{\bs{d}\phi^{\bs{u}}}_{|\phi} (\bs{X}_\phi)  \defeq &\ \big( \text{F}_{\bs u}^\star \bs{d}[\phi]_{|[\phi]} \big) (\bs{X}_\phi)= \bs{d}[\phi]_{|[\phi]} \big(\text{F}_{\bs u\star} \bs{X}_\phi \big)
																			   		= \uprho (\bs u)\-    \left( X(\phi)  +  \delta_{ \bs{d uu}\-_{|\phi}(\bs X_\phi)} \phi \right)  \notag\\
						  = & \uprho (\bs u)\-    \left(  \bs{d}\phi_{|\phi}(\bs{X}_\phi)   +  \delta_{ \bs{d uu}\-_{|\phi}(\bs X_\phi)} \phi \right)
						  =  \left[  \uprho (\bs u)\-    \left(  \bs{d}\phi   +  \delta_{ \bs{d uu}\-} \phi \right) \right]_{|\phi} (\bs{X}_\phi), \notag \\[2mm]
\text{ that is } \quad  \bs d\phi^{\bs u} &= \uprho (\bs u)\-    \left(  \bs d \phi   +  \delta_{ \bs{d uu}\-} \phi \right) 
							   = \left\{  \begin{array}{c}    \bs d A^{\bs u}= \bs u\-\!\left( \bs dA+ D\left\{    \bs{du}\bs{u}\- \right\}  \right) \bs{u} \\[1mm]
							   				           \bs d \vphi^{\bs u}= \rho(\bs u) \-\!\left( \bs d\vphi -  \rho_*( \bs{du}\bs{u}\-) \vphi  \right) 
							                  \end{array} \right.
\end{align} 
Comparing this to \eqref{GT-dphi}, we see that due to the formal similarity between $\text{F}_{\bs u}$ and $\bs \Psi$, their actions are formally alike. 
This generalises to $\bs \alpha$ above. Indeed, given that its $\bs \H$-gauge transformation is 
 \begin{align}
 \label{GT-var-form}
  {\bs{\alpha}^{\bs \gamma}}_{|\phi} \defeq \bs{\Psi}^\star {\bs{\alpha}}_{\, |\phi} = \alpha\big(  \!\w^\bullet\!  \bs{\Psi}^\star\bs d\phi; \bs{\Psi}(\phi) \big)=\alpha \big( \!\w^\bullet\!\bs d\phi^{\bs \gamma} ; \phi^{\bs \gamma}   \big),
 \end{align}  
 by comparison with \eqref{Dressed-variational-form} we see that the general rule of thumb to obtain the dressed version $\bs\alpha^{\bs u} \in \Omega^\bullet_\text{basic}(\Phi)$ of $\bs \alpha \in \Omega^\bullet(\Phi)$ is to replace $\bs \gamma \rarrow \bs u$ in $\bs\alpha^{\bs \gamma}$.\footnote{The latter being obtained, as we've seen, either geometrically via the equivariance and verticality properties of $\bs \alpha$, or algebraically by inserting \eqref{GT-dphi} in its functional expression as suggested by \eqref{GT-var-form}.}

 \medskip

Seing now \eqref{Dressed-variational-form} as a form on the $\G$-bundle of dressed fields $\Phi^{\bs u}$, the $\bs\G$-transformation of $\bs\alpha^{\bs u}$ is obtained in exactly the same way as the  $\bs\H$-transformation of $\bs \alpha$ on $\Phi$. We thus obtain $(\bs\alpha^{\bs u})^{\bs \xi}$ by replacing $\bs\alpha \rarrow \bs\alpha^{\bs u}$ and $\bs\gamma\rarrow \bs \xi$ in the formula for $\bs\alpha^{\bs\gamma}$. Both $\bs\alpha^{\bs u}$ and $(\bs\alpha^{\bs u})^{\bs \xi}$ are basic forms (corresponding to $\alpha$) on the $\H$-bundle $\Phi$. 

 Taking the example of $\bs d\phi^{\bs u}$, in analogy with \eqref{Vert-dphi}-\eqref{Equiv-dphi},we have  $R^\star_\xi \bs d \phi^{\bs u} =\uprho(\xi)\-   \bs d \phi^{\bs u}$ and  $\bs d\phi^{\bs u}_{|\phi^{\bs u}}\big( \zeta^v_{\phi^{\bs u}} \big)=\delta_\zeta \phi^{\bs u}$ (the infinitesimal version of eq.\eqref{Residual-2}), where $\zeta^v \in \Gamma(V\Phi^{\bs u})$ and $\zeta \in \text{Lie}\G$.%
\footnote{These  can be taken as axiomatic on $\Phi^{\bs u}$, or can motivated by reference to $\Phi$: Since by assumption $\phi^\xi=\phi$,  we can formally admit $\bs d \phi(\zeta^v)=0$, and since by definition $\bs u^\xi=\bs u \xi$, infinitesimally we can formally admit $\bs{du}(\zeta^v)=\zeta^v(\bs{u})=\bs{u}\zeta$. By the definition \eqref{Dressed-dphi} of $\bs d\phi^{\bs u}$, its $\G$-equivariance is clear, while it is easy to show explicitly that $\bs d\phi^{\bs u}_{|\phi^{\bs u}}\big( \zeta^v_{\phi^{\bs u}} \big)=\uprho(\bs u)^{-1}  \delta_{\bs u  \zeta \bs u ^{-1} } \phi = \delta_\zeta \phi^{\bs u}$.} 
So, by the same computation leading to \eqref{GT-dphi}, the $\bs\G$-transformation of $\bs d\phi^{\bs u}$ is 
 \begin{align}
 \label{dphi-dressed-ambiguity}
(\bs d \phi^{\bs u})^{\bs \xi} &= \uprho (\bs\xi)\-    \left( \bs d \phi^{\bs u}  + \delta_{ \bs{d\xi\xi}\-} \phi^{\bs u} \right)
		 = \left\{  \begin{array}{c}   (\bs d A^{\bs u})^{\bs \xi} = \bs{\xi}\-   \left( \bs{d}A^{\bs u}  + D^{A^{\bs u}} \big\{ \bs{d}\bs{\xi} {\bs{\xi}\- }\big\} \right)  \bs\xi\ \\[1mm] 
		                                          (\bs d \vphi^{\bs u})^{\bs \xi} =  \rho(\bs \xi)\- \left( \bs d \vphi^{\bs u}  -\rho_*(\bs{d\xi\xi}\-) \vphi^{\bs u}\right) 
		               \end{array} \right.
 \end{align}
This result, together with eq.\eqref{Residual-2}, allows to cross-check algebraically the  $\bs\G$-transformation of $\bs{\alpha^u}$, which  is
\begin{align}
\label{Dressed-alpha-residual-2nd-kind}
 \big( \bs{\alpha}^{\bs u} \big)^{\bs \xi} = \alpha \left( \w^\bullet \big(\bs{d}\phi^{\bs u}\big)^{\bs \xi}; (\phi^{\bs u})^{\bs \xi} \right).
\end{align}
To repeat, given the functional properties of $\alpha$, the latter is formally identical to the $\bs\H$-transformation of $\bs \alpha$. 
In particular, for $\bs\alpha^{\bs u} \in \Omega^1_\text{inv}(\Phi^{\bs u})$, eq.\eqref{Dressed-alpha-residual-2nd-kind} specialises as
\begin{align}
\label{Dressed-alpha-res-2nd-kind-inv}
 \big( \bs{\alpha}^{\bs u} \big)^{\bs \xi} &= \alpha \left( \bs d \phi^{\bs u}  + \delta_{ \bs{d\xi\xi}\-} \phi^{\bs u}; \phi^{\bs u} \right), \notag\\
 							  &= \bs\alpha^{\bs u} + \alpha\big( \delta_{\bs{d\xi\xi\-}} \phi^{\bs u}; \phi^{\bs u} \big).
\end{align}

Reverting back to the original viewpoint, equations \eqref{dphi-dressed-ambiguity}-\eqref{Dressed-alpha-res-2nd-kind-inv} can also be seen as relations between forms on the initial $\H$-bundle $\Phi$.  Seing that indeed  \eqref{dphi-dressed-ambiguity} is rewritten as
\begin{align}
 \label{dphi-dressed-ambiguity2}
(\bs d \phi^{\bs u})^{\bs \xi} &= \uprho (\bs\xi)\-    \left( \bs d \phi^{\bs u}  +  \uprho(u)\-\delta_{ \bs u\bs{d\xi\xi}\-\bs u\-} \phi \right) \quad \in \Omega^1_\text{basic}(\Phi),
\end{align}
 so that for $\bs\alpha = \alpha\big(\bs d \phi; \phi \big)\in \Omega^1_\text{inv}(\Phi)$, \eqref{Dressed-alpha-res-2nd-kind-inv}  is also
 \begin{align}
 \label{Dressed-alpha-res-2nd-kind-inv2}
 \big( \bs{\alpha}^{\bs u} \big)^{\bs \xi} = \bs\alpha^{\bs u} + \alpha\big( \delta_{ \bs u \bs{d\xi\xi\-} \bs u\-} \phi; \phi \big) \quad \in \Omega^1_\text{basic}(\Phi).
\end{align}
One may then notice the striking similarity between 
\eqref{Dressed-alpha-res-2nd-kind-inv2} and eq.\eqref{Ambiguity-basic-1-form} reflecting the ambiguity in building basic forms from variational Ehresmann connections. 
This is no coincidence, as we clarify in the following section concluding this review of the DFM. 

 \subsubsection{Field-dependent dressing fields and  variational connections}  
  \label{Field-dependent dressing fields and  variational connections}  

  \paragraph{Flat Ehresmann variational connections}  

By definition of a field-dependent dressing field, we have for $\gamma\in \H$ and $\chi^v_\phi \in V_\phi\Phi$:
\begin{align*}
R^\star_\gamma (-\bs{duu}\-)&= -\bs d (R^\star_\gamma \bs u)\  R^\star_\gamma \bs u\-= \gamma\- (-\bs{duu})\- \gamma, \\
-\bs{duu}\-_{|\phi}(\chi^v_\phi)&=-(\chi^v \bs u )(\phi)\bs u(\phi)\- = +\chi \bs u(\phi)\bs u(\phi)\- =\chi \ \in \text{Lie}\H.
\end{align*}
As it is furthermore clear that $\bs d (-\bs{duu}\-) +\tfrac{1}{2}[\bs{duu}\-, \bs{duu}\-]\equiv 0$, 
the quantity $\mathring{\bs\omega}\defeq -\bs{duu}\-=\bs{udu}\-$ is thus a flat variational Ehresmann connection, $\mathring\Omega=0$.

 The same is true of $\mathring{\bs\omega}'\defeq -\bs{du}'{\bs u'}\-=\bs{u}'{\bs{du}'}\-$ with $\bs u ' =\bs u^{\bs \xi}= \bs u \bs\xi $, since by definition $R^\star_\gamma \bs \xi=\bs\xi$. From this also follows that $\mathring{\bs\omega}' = \mathring{\bs\omega} + \mathring{\bs\beta}$, where $\mathring{\bs\beta}\defeq - \bs u \bs{d\xi\xi\-} \bs u\- \in \Omega^1_\text{tens}(\Phi)$ since 
\begin{align*}
R^\star_\gamma  \mathring{\bs\beta} &= \gamma\- (- \bs u \bs{d\xi\xi\-} \bs u\-) \gamma = \gamma\- \mathring{\bs\beta} \gamma, \\
\mathring{\bs\beta}(\chi^v) &=  -\bs u \bs{d\xi} (\chi^v)\, \bs\xi\- \bs u\- =0. 
\end{align*}
Therefore, the existence of a field-dependent dressing field $\bs u$ is equivalent to the existence of a flat variational Ehresmann connection $\mathring{\bs\omega}$ on field space $\Phi$, and the a priori ambiguity in choosing/building  such a dressing field ($\bs u'$/$\bs u$) translates as an ambiguity (assuming a specific form, $\mathring{\bs\beta}$ having the form it has) in picking a choice within the affine space of flat connections.  It is thus not surprising to find some similarities in the way basic form are built via dressing and via non-flat connections. Yet, the differences are also worth stressing.
\medskip

Using $\mathring{\bs\omega}$ we could perform the horizontalisation procedure seen in section \ref{Variational Ehresmann connections}, with in particular 
\begin{align}
 \bs d \phi^h_{\mathring{\bs\omega}} =  \bs d \phi   -  \delta_{\mathring{\bs\omega}} \phi \in \Omega^1_\text{hor}(\Phi)
\end{align}
as a special case of eq.\eqref{dphi-horiz}. 
Then, we have  that eq.\eqref{Dressed-dphi} is rewritten as 
\begin{align}
 \bs d\phi^{\bs u} &=  \uprho (\bs u)\-  \bs d \phi^h_{\mathring{\bs\omega}}  = \uprho (\bs u)\-    \left(  \bs d \phi   -  \delta_{\mathring{\bs\omega}} \phi \right)  \quad \in \Omega^1_\text{basic}(\Phi),
\end{align}
Through this simplest example, we see on display the crucial difference between the DFM and the horizontalisation via connection: The dressing operation takes care not only of horizontalisation, via the term  $\delta_{\mathring{\bs\omega}} \phi$, but also of trivialising the equivariance via the term $\uprho (\bs u)\- $. Hence, while one can use Ehresmann connections to produce basic forms out of invariant forms only, one can use the DFM to produce basic forms out of forms of \emph{any} equivariance.\footnote{ Twisted equivariant forms included.} 

 As to the matter of ambiguity, eq.\eqref{dphi-dressed-ambiguity2} is
\begin{align}
(\bs d \phi^{\bs u})^{\bs \xi} &= \uprho (\bs\xi)\-    \left( \bs d \phi^{\bs u}  +  \uprho(u)\-\delta_{ \mathring{\bs\beta}} \phi \right) \quad \in \Omega^1_\text{basic}(\Phi),
\end{align}
which is close to eq.\eqref{Ambiguity-hor-dphi} but not quite the same. The difference being again only in the equivariance, i.e. the presence of $\uprho(\bs \xi)$ and  $\uprho(\bs u)$. But in the case of  $\bs\alpha = \alpha\big(\bs d \phi; \phi \big)\in \Omega^1_\text{inv}(\Phi)$,  the latter disappear so  eq.\eqref{Dressed-alpha-res-2nd-kind-inv2}  rewritten  as
\begin{align}
\label{Ambiguity-basic-1-form-dressing}
 \big( \bs{\alpha}^{\bs u} \big)^{\bs \xi} = \bs\alpha^{\bs u} - \alpha\big( \delta_{  \mathring{\bs\beta}} \phi; \phi \big) \quad \in \Omega^1_\text{basic}(\Phi)
\end{align}
 is indeed seen to be special case of \eqref{Ambiguity-basic-1-form}, 
 with $\big( \bs{\alpha}^{\bs u} \big)^{\bs \xi}  =\bs{\alpha}^b_{\mathring{\bs\omega}'} $ and $\bs\alpha^{\bs u} =\bs{\alpha}^b_{\mathring{\bs\omega}}$.
 \medskip
 
 When considering the construction of basic forms out of invariant forms, as will be our main concerns ahead w.r.t. invariant gauge theories, one may thus expect convergence between the formal results one obtains from either variational Ehresmann connections or the DFM. But the respective merits of both approaches must be parsed and kept in mind.
 
 One the one hand,  the existence of a field-dependent dressing field, thus of a flat connection, is likely a strong constraint on the topology of field space $\Phi$.\footnote{In the finite dimensional case, a principal bundle over a connected manifold can have a flat connection only if it is trivial! } There is then no guarantee that it will always be possible to find/build such a dressing field globally defined across $\Phi$ (Gribov-Singer-like obstructions may exist). On the contrary, $\Phi$ can always be endowed with a non-flat Ehresmann connection, which imposes no such topological constraint.  
 
 On the other hand, the ambiguity of choice among the affine space of connections generally cannot be associated with some underlying useful symmetry transformations. Especially so if a connection comes from a natural bundle metric (as e.g. on $\A$), so that this ambiguity could arguably be discarded. Whereas, as we have remarked already, the group $\G$ controlling the ambiguity in the choice/building of dressing fields may be a physically relevant symmetry to which potentially observable charges can be associated. 
 
 There is no telling in advance which method to choose, as much will depend on the specific examples under consideration. If a bundle metric exists, then it is natural to use the associated connection. If a dressing field is readily identified, one should use the full power of the DFM. In section \ref{Basic presymplectic structures}, we will nonetheless give the most general form, according to both scheme, of the basic presymplectic structure for invariant gauge theories.  Before that, we say a final word about how dressing fields may also give rise to twisted variational connections.

  \paragraph{Flat twisted variational connections}  

Considering again a field-dependent dressing field $\bs u: \Phi \rarrow \D r[H, G]$ and a 1-cocycle $C$, we define a \emph{twisted} dressing field $C(\bs u)$ by $C\big(\bs u \big)(\phi) \defeq C\big(\phi, \bs u(\phi)\big)$.  Due to the cocycle defining property we indeed have, for $\gamma \in\H$,
\begin{align}
\big[R^\star_\gamma C\big(\bs u\big)\big] (\phi) &=C\big(\phi^\gamma, \bs u(\phi^\gamma) \big) = C\big(\phi^\gamma, \gamma\- \bs u(\phi) \big)  = C\big(\phi^\gamma, \gamma \big)  C\big(\phi, \gamma \bs u(\phi) \big) = C\big(\phi, \gamma \big) \- C\big(\phi, \gamma \bs u(\phi) \big), \notag \\
									&= \big[C(\ \, ,\gamma)\- C\big(\bs u\big)\big] (\phi) .
\end{align}
The infinitesimal version of which is 
\begin{align}
\label{inf-twisted-dress-field}
\bs L_{\chi^v} C\big(\bs u\big) =\iota_{\chi^v} \bs d C\big(\bs u\big) = -\tfrac{d}{d\tau} C\big(\ \, ,\exp{\tau \chi}\big)\big|_{\tau=0} C\big(\bs u\big),
 \end{align}
with $\chi^v \in \Gamma(V\Phi)$.
 
In the same manner that $\bs u$ defines a flat variational Ehresmann connection $\mathring\omega$, via the cocycle it defines a flat twisted variational connection
 $\mathring{\bs\varpi}=-\bs d C\big(\bs u\big) C\big(\bs u\big)\-=C\big(\bs u\big) \bs d C\big(\bs u\big)\-$. From the above follows indeed easily, 
\begin{align*}
R^\star_\gamma \mathring{\bs\varpi}_{|\phi^\gamma} &= C(\phi, \gamma)\- \mathring{\bs\varpi}_{|\phi}\, C(\phi, \gamma) +C(\phi, \gamma)\- \bs d C(\ , \gamma)_{|\phi}, \\[1mm]
\mathring{\varpi}_{|\phi}(\chi^v_\phi)&=\bs d C\big(\bs u\big) C\big(\bs u\big)\-(\chi^v)= +\tfrac{d}{d\tau} C\big(\ \, ,\exp{\tau \chi}\big)\big|_{\tau=0},
\end{align*}
the defining axioms \eqref{var-twisted-connection1}-\eqref{var-twisted-connection2} of a twisted connection. Clearly, $\bs d \mathring{\bs\varpi} + \tfrac{1}{2}[ \mathring{\bs\varpi},  \mathring{\bs\varpi}]=0$. 
\medskip

In section \ref{Variational twisted connections}, we saw that a general non-invariant Lagrangian is $R^\star_\gamma L = L + c(\ , \gamma)$, with \emph{classical gauge anomaly} $\bs L_{\chi^v} L=\tfrac{d}{d\tau} c(\ \, , \exp{\tau \chi})\big|_{\tau=0} \rdefeq \alpha(\chi, \ )$.
 It is associated with the twisted functional   $Z \defeq \exp{i\int L} \in \Omega^0_\text{tens}(\Phi, \CC)^C$ s.t.  $R^\star_\gamma Z = C(\, ,\gamma)\- Z$ with $C(\ ,\gamma)\defeq \exp{-i \int c(\ ,\gamma)} \in U(1)$. 
 
Admitting a dressing field exists, the corresponding flat twisted connection is $\mathring{\bs\varpi}=i\bs d \int c(\bs u)\defeq i\bs d\int c(\  ,\bs u)$. 
The~associated twisted covariant derivative of $Z$ is thus, 
  \begin{align}
  \bs D^{\mathring{\bs\varpi}}Z = \bs d Z +\mathring{\bs\varpi} Z =  \left( i \bs d \textstyle\int L +c(\bs u) \right) \, Z \quad  \in \Omega^1_\text{tens}(\Phi, \CC)^C.
  \end{align}
We have then that $ L +c(\bs u) \in  \Omega^0_\text{basic}(\Phi)$. This is none other than a Wess-Zumino improved (i.e. $\H$-invariant) Lagrangian, and $c(\bs u)$ is a Wess-Zumino functional which (usually by special design) satisfies $\bs L_{\chi^v} c\big(\bs u\big) = - \alpha(\chi,\ \,)$. 

A WZ functional is thus  seen to be the pre-potential of a flat twisted connection. 
Any twisted connection $\tilde{\bs\omega}$ adapted to this context (i.e. fit to induce a twisted covariant derivative) satisfies $\tilde{\bs\omega}(\chi^v)= -i \int \alpha(\chi,\ \,)$ -- see eq. \eqref{var-twisted-connection2bis}. 
Here, we have indeed $\mathring{\bs\varpi}(\chi^v) = i \int  \bs d c(\bs u)(\chi^v) = i \int \bs L_{\chi^v} c(\bs u) =  -i \int  \alpha(\chi,\ \,)$ as a special case of \eqref{inf-twisted-dress-field}. 
\smallskip

Let us finally remark that the above improved Lagrangian is precisely what is immediately given by application of the DFM: The dressed version of the non-invariant Lagrangian $L \in\Omega^0_\text{eq}(\Phi)^C$ is 
\begin{align}
L^{\bs u}(\phi) \defeq&\, \big(\text{F}^\star_{\bs u} L \big) (\phi) =L \big(\text{F}_{\bs u}(\phi)\big)  = L(\phi^{\bs u})= L(\phi)+ c\big(\phi, \bs u(\phi)\big), \notag \\[1mm]
\hookrightarrow \quad L^{\bs u} =&\, L + c(\bs u)  \ \  \in \Omega^0_\text{basic}(\Phi).
\end{align}
This illustrates a commentary we made in the previous subsection, to the effect that the DFM can associate basic forms to forms of any equivariance, including twisted equivariance. This was extensively used in \cite{Francois2021}, section 5.3.2, to produce the basic (dressed) presymplectic structure of  non-invariant pure gauge theories.  

\medskip

After these long technical preliminaries, we are ready for our main application concerning the presymplectic structure of invariant matter coupled gauge theories.

\section{Presymplectic structures of matter coupled gauge theories over bounded regions}
\label{Presymplectic structures of matter coupled gauge theories over bounded regions}

 In this section we first briefly remind one original aim of the covariant phase space formalism for gauge field theory, and the impediment posed by the presence of boundaries: that is the boundary problem. 
 Then we provide results of some generality about the presymplectic structure of invariant matter coupled gauge theories. Some of these are necessary to show how one may try to answer the boundary problem by constructing  \emph{basic} presymplectic structures obtained either via the DFM or via  variational connections. 
 
 \subsection{Covariant phase space formalism}
\label{Covariant phase space formalism}

 As  already evoked just before section \ref{Variational connections on field space}, a gauge theory is specified by a Lagrangian functional $L:\Phi \rarrow \Omega^n(\P, \RR)$, $\phi \mapsto L(\phi)=L(A, \vphi)$, $n$ the dimension of spacetime~$M$. 
 For the associated action $S=\int_U L$\footnote{Since we took the viewpoint that $L$ is a n-form on $\P$,  actually  $S=\int_{\s(U)} L=\int_U \s^*L$ with $\s:U\rarrow \P$ a local section and $\s^*L$ written in terms of the gauge potential $\s^*A$. For convenience, we shall omit $\s$ in writing integration domains.}
 to be finite, one usually assumes the region  $U\subset M$ to be compact or closed, or that  the fields are either compactly supported or satisfy sufficiently fast fall-off conditions at infinity (which amounts to an effective compactification of $\M$). 

The variational principle stipulates that the field equations are found from requiring $S$ to be stationary, $\delta S=0$ $\forall \delta \phi$, under well-defined boundary conditions. In the formulation adopted here this translates as $\bs{d}S(\bs{X})=0$ $\forall \bs{X} \in \Gamma(T\Phi)$ , i.e the functional $S:\Phi\rarrow \RR$ is closed, $\bs{d}S=0$. Admitting that $\bs d$ and $\int$ commute, this gives 
\begin{align}
\label{variational_principle}
\bs{d}S=\int_U \bs{d}L= \int_U \bs{E} + d\bs{\theta}=\int_U \bs{E}\ + \int_{\d U} \bs{\theta} =0
\end{align}
where $\bs{E}_{|\phi}=E(\bs{d}\phi; \phi)$ is the field equations 1-form and $\bs{\theta}_{|\phi}=\theta(\bs{d}\phi; \phi)$ is the presymplectic potential current \mbox{1-form}. Here $E$ and $\theta$ are different functionals of $\phi$,  both linear in $\bs{d}\phi$ (nonetheless based on the same functional as $L$). 
\medskip

The point of the covariant phase space approach, or covariant Hamiltonian formalism, is to associate a phase space equipped with a symplectic form to a field theory over a region $U \subseteq \M$, and doing so while keeping all spacetimes symmetries manifest. Such a symplectic structure would be the starting point for a canonical or geometric quantization procedure, or so was one original motivation. 
 Some attribute the inception of the idea to \cite{Zuckerman1986,CrnkovicWitten1986, Crnkovic1987}, but it actually goes further back and has close ties to the  multisymplectic formalism as shown in  \cite{Helein2012}, which we recommand. Classical references are \cite{Lee-Wald1990, Ashtekar-et-al1990}, and modern introductions are \cite{Compere-Fiorucci2018, Harlow-Wu2020}  (see also \cite{Farajollahi-Luckock2002} for a compact  summary).

The configuration space is the field space, here the $\H$-bundle $\Phi$.
The covariant phase space is the solution space $\S$ -- the \emph{shell} -- defined by $\boldsymbol E=0$. 
The physical, or reduced, phase space is  $\S/\H\rdefeq \M_\S$  if it can be endowed with a well-defined symplectic 2-form. Notice then that the Hamiltonian flow belongs to the physical transformation group $\bs\Diff(\M)$ in the SES \eqref{SESgroups-inf}, and the corresponding Hamiltonian vector field thus belongs to $\Gamma(T\M)$ in the SES~\eqref{SESLieAlg-inf}. 

The presymplectic potential $\bs\theta$ allows to define the Noether currents and charges associated with the action of $\H$, and a 
 natural candidate symplectic form is derived  from it. Since $[\bs d, d]=0$,  we have $0\equiv \bs{d}^2 L = \bs{dE}+d(\bs{d\theta})$. So, the 2-form $\bs \Theta \defeq \bs{d\theta}$ is $d$-closed on-shell, $d \bs\Theta =0_{\, |\S}$.
 Given  a codimension 1 submanifold $\Sigma \subset U$,  we have  $\bs\Theta_\Sigma \defeq \int_\Sigma \bs\Theta \in \Omega^2(\Phi, \RR)$. The presymplectic potential is  $\bs\theta_\Sigma\defeq \int_\Sigma \bs\theta \in \Omega^1(\Phi, \RR)$, so that $\bs{\Theta}_\Sigma=\bs{d} \bs{\theta}_\Sigma$. Since $\bs{d} \bs{\Theta}_\Sigma=0$, $\bs{\Theta}_\Sigma$ is a presymplectic 2-form (hence the name given to $\bs\theta$ and $\bs{\theta}_\Sigma$). It allows to define a Poisson bracket between charges. 
 
 For $\bs\theta_\Sigma$ and $\bs\Theta_\Sigma$ to induce a symplectic structure on  $\S/\H=\M_\S$, they must be basic on $\Phi$. This requires that on-shell the $\H$-equivariance and verticality properties of $\bs \theta$ and $\bs \Theta$ are right,i.e. that adequate boundary conditions are specified. This last requirement is jeopardised when considering bounded regions, or when one considers the partitioning  a region into subregions sharing a fictitious boundary. This is what we call the boundary problem: the obstruction to the basicity of $\bs\theta_\Sigma$ and $\bs\Theta_\Sigma$ due a boundary $\d \Sigma$. 
 
 Before considering potential answers to the boundary problem in section \ref{Basic presymplectic structures}, we want to give general results on the presymplectic structure of invariant gauge theories. Precisely,  we are interested in identifying the Noether currents and charges, the Poisson bracket of charges, and most importantly the $\bs\H$-gauge transformations of $\bs \theta_\Sigma$ and $\bs \Theta_\Sigma$. Whenever possible we will give these results as functions of the field equations $\bs E$ so that on-shell restrictions are read-off immediately.

  \subsection{Presymplectic structure of invariant matter coupled gauge theories}
\label{Presymplectic structure of invariant matter coupled gauge theories}

We are concerned with gauge theories that strictly respect the gauge principle, so that $L\in \Omega^0_\text{basic}(\Phi)$ (case 1 mentioned just before section \ref{Variational connections on field space}). 
As $\bs d$ is a covariant derivative on basic forms, we know immediately that $\bs d L \in \Omega^1_\text{basic}(\Phi)$. 

In the pure gauge case, $L\in \Omega^0_\text{basic}(\A)$, the Utiyama theorem \cite{Castrillon-Lopez-et-al_2019, Bruzzo1987} states that $L$ must factorise through the curvature map, $L=\t L \circ F$, with $\t L$ a $\Ad(H)$-invariant functional on $\Omega^\bullet_\text{tens}(\P, \text{Lie}H)$. By extension, in the coupled case $L$ must depend on tensorial quantities and otherwise factors through the curvature and covariant derivative maps, 
\begin{align}
\label{inv-Lagrangian}
L(\phi)=\t L(\vphi, F, D^A\vphi)\rdefeq \t L(\{\phi\}),
\end{align}
with $\t L$ a $H$-invariant multilinear fonctional on $\Omega^\bullet_\text{tens}(\P)$: $\t L\big(\uprho(h)\{\phi\}\big)=\t L(\{\phi\})$, $h \in H$ and we remind that $\uprho \defeq (\Ad, \rho)$. We notice that the set $\{\phi\}=(F, \vphi, D^A\vphi)$ is closed under $D$, as on the one hand by Bianchi $DF=0$, and on the other hand $D^2\vphi=\rho_*(F)\vphi$, $D^3\vphi=\rho_*(F) D\vphi$ and {\color{gray} $D^{2n}\vphi =\rho_*(F^{n})\vphi$,  $D^{2n+1}\vphi =\rho_*(F^{n})D\vphi$}. 
We may denotes this $D\{\phi \} \subset \{\phi \}$. 
Then it comes that $\bs d L$ will be linear in $\bs d \phi$ or $D^A(\bs d \phi)$, $\bs d L_{|\phi} = \t L \left( \bs d \phi, D^A(\bs d \phi) ; \{\phi \}\right)$. By  \eqref{dF}-\eqref{d-Dphi} and using the $H$-invariance of $\t L$, we have
\begin{align}
\label{def-E-theta}
\bs d L_{|\phi} 
 		       &= \t L \left(\bs d \vphi,   \bs d F, \bs d\, D^A(\vphi);  \{\phi \} \right), \notag \\
		       &=  \t L \big( \bs d \vphi;  \{\phi \} \big) +   \t L \left(  D^A(\bs d A); \{\phi \}\right) + \t L\left( D^A(\bs d\vphi)+ \rho_*(\bs d A)\vphi; \{\phi \}\right),  \notag \\
		       &=  \t L \big( \bs d \vphi;  \{\phi \} \big) + d \t L\big( \bs d A; \{\phi \} \big) + \t L \left( \bs dA ; D^A \{\phi \} \right)  \notag \\
		       & \hspace{2.3cm} + d \t L \big( \bs d \vphi ; \{\phi \} \big) - \t L \left( \bs d \vphi ; D^A \{\phi \} \right) + \t L \big( \rho_*(\bs dA)\vphi ; \{\phi \} \big),     \notag \\
		       &=  \t L \left( \bs dA ; D^A \{\phi \} \right) + \t L \big( \rho_*(\bs dA)\vphi ; \{\phi \} \big) +  \t L \big( \bs d \vphi;  \{\phi \} -D^A  \{\phi \} \big)  \ \  + \ \ d \t L\big(\bs d \phi; \{\phi\} \big), \notag\\
		       &\rdefeq \t E\big( \bs d \phi; \{\phi\} \big) + d \t\theta\big(  \bs d \phi; \{\phi\} \big) =E\big(\bs d \phi; \phi \big) + d\theta\big(\bs d \phi; \phi \big) = \bs E + d \bs \theta. 
\end{align}
The last three equalities will help keep track of what the notations of the functionals $E$ and $\theta$ means here. 
It is clear in particular that by \eqref{Equiv-dphi}, since $\{\phi\} \in \Omega^\bullet_\text{tens}(\P)$, and due to the $H$-invariance of $\t L$, the $\H$-equivariance of $\bs E$ and $\bs \theta$ are trivial
\begin{alignat}{5}
\label{trivial-eq-E-theta}
R^\star_\gamma \bs E &= \bs E\quad &&\text{and}&& \, \ \quad\qquad R^\star_\gamma \bs \theta &&=  \bs \theta, \quad\qquad \text{i.e.} \quad \bs E, \bs \theta \in \Omega^1_\text{inv}(\Phi). \\
\hookrightarrow \quad  E(\bs d\phi^\gamma; \phi^\gamma)&=E(\bs d\phi; \phi) \quad &&\text{and}&& \quad \theta(\bs d\phi^\gamma; \phi^\gamma)&&= \theta(\bs d\phi; \phi).  \notag
\end{alignat}
We can already say that their $\bs \H$-gauge transformations are thus controlled only by their respective verticality properties. The latter are also related to the definition of conserved currents and charges associated with the action of~$\H$. 

\paragraph{Noether currents and charges:}  

As $L$ is basic, the infinitesimal version of its trivial equivariance is, for $\chi^v \in \Gamma(V\Phi)$, $\bs L_{\chi^v}L= \iota_{\chi^v} \bs d L = \iota_{\chi^v} \bs E + d \iota_{\chi^v} \bs \theta =0$. 
The quantity $J(\chi; \phi)\defeq \iota_{\chi^v} \bs \theta$ is thus conserved on-shell, $dJ(\chi; \phi)=0_{|\S}$. This is the Noether current associated with $\H$. We might be interested in further determining its general form. Using the above definition \eqref{def-E-theta} of $\bs\theta$ and $\bs E$ we have, 
\begin{align}
\label{Noether-current}
J(\chi; \phi)\defeq \iota_{\chi^v} \bs \theta &= \t L \big( \iota_{\chi^v} \bs d\phi ;  \{\phi\}  \big), \notag \\
							       &= \t L\big( D^A\chi; \{\phi\} \big) +  \t L\big( -\rho_*(\chi)\vphi; \{\phi\} \big), \notag \\
							       &= d \t L \big(  \chi; \{\phi\}  \big) - \t L \big( \chi; D^A\{\phi\} \big) -  \t L\big( \rho_*(\chi)\vphi; \{\phi\} \big), \notag \\
							       &= d \theta\big( \chi; \phi \big) - E\big(\chi; \phi \big). 
\end{align}
The on-shell $d$-exacteness of $J(\chi; \phi)$ is  manifest in this form.
In the last equality the notation means that the current is (of course) written in terms of the Lie$H$-linear pieces of $\bs\theta$ and $\bs E$. Notice it implies that the details of the contribution of the matter field to the presymplectic potential is irrelevant to the on-shell form  of the current!

The Noether charge is defined as $Q_\Sigma(\chi; \phi)\defeq \int_\Sigma J(\chi; \phi)$, and is also written $Q_\Sigma(\chi; \phi)=\iota_{\chi^v} \bs\theta_\Sigma$. Given \eqref{Noether-current}, it is explicitly
\begin{align}
\label{Noether-charge}
Q_\Sigma(\chi; \phi)= \int_{\d \Sigma} \theta\big( \chi; \phi \big) - \int_\Sigma E\big(\chi; \phi \big). 
\end{align}
On-shell, it is a purely boundary term.
To reiterate the previous point: the above result proves that the contribution of the matter field is irrelevant to the on-shell form of the charge, which depends only on the contribution of the connection/gauge potential. 
 Now, the presymplectic 2-form $\bs\Theta =\bs{d\theta}$ 
 induces a Poisson bracket for these charges.

\paragraph{Poisson bracket of charges:} 

To see this, let us first notice that the infinitesimal version of the trivial $\H$-equivariance of $\bs \theta$, eq. \eqref{trivial-eq-E-theta},  gives  a relation between the Noether charge and the presymplectic 2-form
\begin{align}
\label{presympl-form-Noether-charge}
&\bs L_{\chi^v} \bs\theta= \iota_{\chi^v} \bs{d \theta} + \bs d \iota_{\chi^v} \bs \theta \rdefeq  \iota_{\chi^v} \bs \Theta + \bs dJ(\chi; \phi) =0. \notag\\[1mm]
&\hookrightarrow  \text{ so that } \quad \iota_{\chi^v} \bs \Theta_\Sigma = - \bs d Q_\Sigma(\chi; \phi) 
																			= - \int_{\d\Sigma} \bs d \theta(\chi; \phi) + \int_\Sigma \bs d E(\chi; \phi). 
\end{align}
From this,  using $[\bs L_{\bs Y}, \iota_{\bs X}]= \iota_{[\bs Y, \bs X]}$, we  obtain that  for $\chi^v, \eta^v \in \Gamma(V\A)$
\begin{align}
\label{verticality-Theta}
\bs\Theta(\chi^v, \eta^v)=\iota_{\eta^v} \big(\iota_{\chi^v} \bs \Theta \big) = - \iota_{\eta^v} \bs d \iota_{\chi^v}\bs \theta = -\bs L_{\eta^v} \iota_{\chi^v} \bs \theta= -\iota_{\chi^v} \bs L_{\eta^v}\bs\theta - \iota_{[\eta^v, \chi^v]} \bs \theta 
					 = \iota_{[\chi, \eta]^v} \bs \theta . 
\end{align}
where in the last step we use the fact that the map Lie$\H \rarrow \Gamma(V\A)$ is a isomorphism. 
The  Poisson bracket of charges defined by the presymplectic 2-form is thus.
\begin{align}
\label{Poisson-bracket}
\big\{ Q_\Sigma(\chi; \phi) ,  Q_\Sigma(\eta; \phi)\big\}\defeq\, \bs\Theta_\Sigma(\chi^v, \eta^v) = \int_\Sigma \iota_{[\chi, \eta]^v} \bs \theta= \int_\Sigma J([\chi. \eta]; \phi) 
										    = Q_\Sigma([\chi, \eta]; \phi). 
\end{align}
It is clearly antisymmetric, and the Jacobi identity is satisfied for the Poisson bracket because it holds in Lie$\H$. So the map Lie$\H \rarrow \big( Q_\Sigma(\ ;\phi), \big\{\, ,\, \big\})$ is a Lie algebra morphism.
Written functionally, \eqref{Poisson-bracket} reproduces the Peierls-DeWitt bracket (see \cite{Forger-VieraRomero2005} Theorem 4, also \cite{Harlow-Wu2020}).

Through this Poisson bracket, the Noether charges are also generators of Lie$\H$-transformations. 
Consider indeed a functional $f : \Phi \rarrow \Omega^{n-1}(\P)$, $\phi \mapsto f(\phi)$. Define  its associated variational Hamiltonian vector field $V^f$, as one does, via $\iota_{V^f} \bs \Theta_\Sigma = -\int_\Sigma \bs d f$. The action of Lie$\H$ on $f$ is usually given by the Lie derivative along a vertical vector field, 
\begin{align}
\int_\Sigma \bs L_{\chi^v} f = \int_\Sigma \iota_{\chi^v}\bs d f + \bs d  \cancel{\iota_{\chi^v} f} =  \iota_{\chi^v} \left(- \iota_{V^f} \bs \Theta_\Sigma  \right) = \iota_{V^f} \iota_{\chi^v} \bs \Theta_\Sigma  \rdefeq&\ \big\{ Q_\Sigma(\chi; \phi), \ f \big\},   \\*
																					\hookrightarrow\  =&\ -  \iota_{V^f} \bs d Q_\Sigma(\chi; \phi). \notag
\end{align}
The first line shows why Noether charges generate  Lie$\H$-transformations via the Poisson bracket, the second line gives the explicit mean of computation: One must first determine the Hamiltonian vector field of $f$ via the symplectic 2-form $ \bs \Theta_\Sigma$, then feed it to the variational 1-form $\bs d Q_\Sigma(\phi; \chi)$.

\paragraph{Field-dependent gauge transformations: }  

We are interested in finding the general form of the field-dependent $\bs\H$-gauge transformations of $\bs\theta$ and $\bs\Theta$. But let us first, as a warm-up, show that an invariant theory is well-behaved by finding the $\bs\H$-transformation of the field equation $\bs E$. 

As already pointed out in section \ref{Field space as a principal bundle}, given eq.\eqref{Pushforward-X-inf} the $\bs\H$-transformation of a form depends on its $\H$-equivariance and its verticality property. For the field equations we have $R^\star_\gamma \bs E=\bs E$ by \eqref{trivial-eq-E-theta}, and $\iota_{\chi^v}\bs E= d E(\chi; \phi)$ by $\bs L_{\chi^v} L=0$ and \eqref{Noether-current} above. So, for $\bs\gamma \in \bs\H$ corresponding to $\bs\Psi \in \bs\Aut_v(\A)$, we get
 \begin{align}
\label{Field-depGT-FieldEq}
{\bs E^{\bs\gamma}}_{|\phi}(\bs X_\phi) &\defeq \big( \bs\Psi^\star \bs E \big)_{|\phi}(\bs X_\phi) = \bs E_{\phi^{\bs\gamma}}\big(\bs\Psi_\star \bs X_\phi\big)
					=   \bs E_{|\phi^{\bs\gamma}} \left(R_{\bs{\gamma}(\phi)\star} \left( \bs{X}_\phi + \left\{ \bs{d}\bs{\gamma} {\bs{\gamma}\- }_{|\phi}(\bs{X}_\phi)\right\}^v_\phi \right) \right), \notag \\				                                                  	&= R^\star_{\bs{\gamma}(\phi)} \bs E_{|A^{\bs\gamma}}  \left( \bs{X}_\phi + \left\{ \bs{d}\bs{\gamma} {\bs{\gamma}\- }_{|\phi}(\bs{X}_\phi)\right\}^v_\phi \right)
					= \bs E_{|\phi} \left( \bs{X}_\phi + \left\{ \bs{d}\bs{\gamma} {\bs{\gamma}\- }_{|v}(\bs{X}_\phi)\right\}^v_\phi \right),  \notag \\[1mm]
			                 		&=  \bs E_{|\phi} \big( \bs{X}_\phi \big) + d E \big( \{ \bs{d\gamma\gamma}_{|\phi}\-(\bs X_A) \}; \phi \big),  \notag \\[1mm]
		\text{that is \quad} \bs E^{\bs \gamma}&= \bs E + dE \big(  \bs{d\gamma\gamma}\- ;  \phi \big).
\end{align}
The action of $\bs\H$ does not take us off-shell. Which is to be expected if the bundle $\S \xrightarrow{\H} \M_\S$, and the reduced phase space $ \M_\S$, are to be well-defined.\footnote{But this is by no means automatic, as some non-invariant theories are not well behaved in this way (e.g. massive Yang-Mills theory, see Appendix F in \cite{Francois2021}). }

The $\bs\H$-transformation of $\bs\theta$ goes similarly. We have $R^\star_\gamma\bs\theta=\bs\theta$ by \eqref{trivial-eq-E-theta}, and its verticality property is the very definition of the Noether current \eqref{Noether-current}. Thus, 
\begin{align}
\label{Field-depGT-presymp-pot-current}
{\bs\theta^{\bs\gamma}}_{|\phi}(\bs X_\phi) &\defeq \big( \bs\Psi^\star \bs \theta \big)_{|A}(\bs X_\phi) = \bs \theta_{A^{\bs\gamma}}\big(\bs\Psi_\star \bs X_\phi\big)
					=   \bs \theta_{|\phi^{\bs\gamma}} \left(R_{\bs{\gamma}(\phi)\star} \left( \bs{X}_\phi + \left\{ \bs{d}\bs{\gamma} {\bs{\gamma}\- }_{|\phi}(\bs{X}_A)\right\}^v_\phi \right) \right),  \notag \\[1mm]
					&= R^\star_{\bs{\gamma}(\phi)} \bs \theta_{|\phi^{\bs\gamma}}  \left( \bs{X}_\phi + \left\{ \bs{d}\bs{\gamma} {\bs{\gamma}\- }_{|\phi}(\bs{X}_\phi)\right\}^v_\phi \right)
					= \bs \theta_{|\phi} \left( \bs{X}_\phi + \left\{ \bs{d}\bs{\gamma} {\bs{\gamma}\- }_{|\phi}(\bs{X}_\phi)\right\}^v_\phi \right),  \notag \\[1mm]
					&=  \bs \theta_{|\phi} \big( \bs{X}_\phi \big) +  J\big( \{ \bs{d\gamma\gamma}_{|\phi}\-(\bs X_\phi) \}; \phi \big), \notag \\[1mm]
		\text{that is \quad} \bs\theta^{\bs \gamma}&= \bs \theta + d\theta \big(  \bs{d\gamma\gamma}\- ;  \phi \big) - E \big( \bs{d\gamma\gamma}\- ; \phi \big). 
\end{align} 
The $\bs\H$-transformation of  presymplectic potential is then obviously,
\begin{align}
\label{Field-depGT-presymp-pot}
\bs\theta_\Sigma^{\bs \gamma} = \bs \theta_\Sigma + Q_\Sigma(\bs{d\gamma\gamma}\-; \phi) =  \bs \theta_\Sigma + \int_{\d\Sigma} \theta( \bs{d\gamma\gamma}\- ; \phi) - \int_\Sigma E(\bs{d\gamma\gamma}\-; \phi),
\end{align}
From this, or from \eqref{trivial-eq-E-theta}-\eqref{Noether-current}, it is clear that the presymplectic potential is $\bs\H$-invariant, basic, if we are on-shell and if either $\d\Sigma=\emptyset$ or $\phi \rarrow 0$ and/or  $\bs\gamma \rarrow 1$  
 at $\d\Sigma$ or at infinity.

We finally turn our attention to the $\bs\H$-gauge transformation of $\bs\Theta\defeq\bs{d\theta}$. It can be guessed from  \eqref{Field-depGT-presymp-pot-current} above using  the naturality of pullbacks, i.e. $[\bs\Psi^\star, \bs d ]=0$ here. But we might want an explicit check using the method above. For this, as the verticality of $\bs\Theta$ is given by \eqref{presympl-form-Noether-charge}-\eqref{verticality-Theta} in therm of the Noether current, we only need to determine its $\H$-equivariance. But then again, we must appeal to the naturality of pullbacks, in this case $[R^\star_\gamma, \bs d ]=0$, so that $R^\star_\gamma \bs\Theta= R^\star_\gamma \bs{d\theta}=\bs{d}R^\star_\gamma\bs\theta=\bs{d\theta}=\bs\Theta$. Then,
\begin{align}
\label{1}
{\bs\Theta^{\bs\gamma}}_{|\phi}\big(\bs X_\phi, \bs Y_\phi \big)\defeq&\, \big(  \bs\Psi^\star \bs\Theta \big)_{|\phi}(\bs X_\phi, \bs Y_\phi) = \bs\Theta_{|\phi^{\bs\gamma}} \left(\bs\Psi_\star \bs X_\phi, \bs\Psi_\star\bs Y_\phi  \right), \notag\\
			=&\, \bs\Theta_{|\phi^{\bs\gamma}} \left( R_{\bs{\gamma}(\phi)\star} \left( \bs{X}_\phi + \left\{ \bs{d}\bs{\gamma} {\bs{\gamma}\- }_{|\phi}(\bs{X}_A)\right\}^v_\phi   \right) ,  R_{\bs{\gamma}(\phi)\star} \left( \bs Y_\phi + \left\{ \bs{d}\bs{\gamma} {\bs{\gamma}\- }_{|\phi}(\bs Y_\phi)\right\}^v_\phi   \right)  \right), \notag \\[1mm]
			=&\, R^\star_{\bs\gamma(\phi)} \bs \Theta_{|\phi^{\bs\gamma}} \left( \bs{X}_\phi + \left\{ \bs{d}\bs{\gamma} {\bs{\gamma}\- }_{|\phi}(\bs{X}_\phi)\right\}^v_\phi,     \bs{Y}_\phi + \left\{ \bs{d}\bs{\gamma} {\bs{\gamma}\- }_{|\phi}(\bs{Y}_\phi)\right\}^v_\phi \right) ,\notag\\[1mm]
			=&\, \bs\Theta_{|\phi} \big(\bs X_\phi, \bs Y_\phi \big) +  \bs\Theta_{|\phi} \left(  \left\{ \bs{d}\bs{\gamma} {\bs{\gamma}\- }_{|\phi}(\bs{X}_\phi)\right\}^v_\phi, \bs Y_\phi  \right) + \bs\Theta_{|\phi}\left( \bs X_\phi,  \left\{ \bs{d}\bs{\gamma} {\bs{\gamma}\- }_{|\phi}(\bs{Y}_\phi)\right\}^v_\phi\right)  \notag\\
			&\hspace{5cm} +  \bs\Theta_{|\phi} \left(  \left\{ \bs{d}\bs{\gamma} {\bs{\gamma}\- }_{|\phi}(\bs{X}_\phi)\right\}^v_\phi,  \left\{ \bs{d}\bs{\gamma} {\bs{\gamma}\- }_{|\phi}(\bs{Y}_\phi)\right\}^v_\phi \right), \notag\\[1mm]
			=&\,  \bs\Theta_{|\phi} \big(\bs X_v, \bs Y_\phi \big)  - \iota_{\bs Y} \bs d\, J\left(\big\{ \bs{d\gamma\gamma}\-_{|\phi}(\bs X_\phi) \big\}; \phi \right)  +  \iota_{\bs X} \bs d\, J\left( \big\{ \bs{d\gamma\gamma}\-_{|\phi}(\bs Y_\phi) \big\} ; \phi \right) \notag\\ 
			&\hspace{5cm}  +  \bs\theta_{|\phi} \left(\big[\bs{d\gamma\gamma}\-_{|\phi}(\bs X_\phi), \bs{d\gamma\gamma}\-_{|\phi}(\bs Y_A)\big]^v_\phi\right), \notag \\[1mm]
			=&\, \bs\Theta_{|\phi} \big(\bs X_\phi, \bs Y_\phi \big) - \bs Y \cdot J\left( \big\{ \bs{d\gamma\gamma}\-_{|\phi}(\bs X_\phi) \big\}; A \right)  + \bs X \cdot J\left( \big\{ \bs{d\gamma\gamma}\-_{|\phi}(\bs Y_\phi) \big\}; A \right) \notag \\
			&\hspace{5cm} +  J\left( \big\{ [\bs{d\gamma\gamma}\-_{|\phi}(\bs X_\phi), \bs{d\gamma\gamma}\-_{|\phi}(\bs Y_\phi)\big] \big\}; \phi \right). 
\end{align}
Notice that  in the step before last, we used \eqref{verticality-Theta}, and \eqref{presympl-form-Noether-charge} which is only valid for $\phi$-independent gauge parameters $\chi \in $ Lie$\H$. So
the quantity $ \bs{d\gamma\gamma}\-_{|\phi}(\bs Z_\phi)$ is considered $\phi$-independent, and only  the underlined $\phi$'s in $J\left(\big\{ \bs{d\gamma\gamma}\-_{|\phi}(\bs Z_\phi) \big\}; \munderline{blue}{\phi} \right)$ are acted upon by the variational vector fields in \eqref{1}.

Actually, the quantity $\bs d\, J\left( \big\{ \bs{d\gamma\gamma}\-_{|\phi} \big\}; \phi \right)$ is a 2-form on $\Phi$, and by the Kozsul formula, evaluated on two vectors it gives
 \begin{align}
 \label{2}
 \bs d\, J\left(\big\{ \bs{d\gamma\gamma}\-_{|\phi} \big\}; \phi \right) \big( \bs X_\phi, \bs Y_\phi \big)&= \bs X \cdot J\left( \big\{ \bs{d\gamma\gamma}\-_{|\phi}(\bs Y_\phi) \big\}; \phi \right)
 																			- \bs Y \cdot J\left( \big\{ \bs{d\gamma\gamma}\-_{|\phi}(\bs X_\phi) \big\}; \phi \right) \notag\\*
															  & \hspace{5cm} - J\left(    \big\{ \bs{d\gamma\gamma}\-_{|\phi}([\bs X, \bs Y]_\phi) \big\}; \phi  \right),
 \end{align}
where all the $\phi$'s in the terms $J\left( \big\{ \bs{d\gamma\gamma}\-_{|\phi}(\bs Z_\phi) \big\}; \phi \right)$ are acted upon. Observe also that 
\begin{align*}
\big[ \bs{d\gamma\gamma}\-(\bs X), \bs{d\gamma\gamma}\-(\bs Y) \big] &= \bs{d\gamma\gamma}\-(\bs X) \bs{d\gamma\gamma}\-(\bs Y) - \bs{d\gamma\gamma}\-(\bs Y) \bs{d\gamma\gamma}\-(\bs X) 
														= -\bs{d\gamma}(\bs X)\bs{d\gamma}\-(\bs Y) + \bs{d\gamma}(\bs Y)\bs{d\gamma}\-(\bs X), \\
														&= \big(\! -\bs{d\gamma}\bs{d\gamma}\- \big)(\bs X, \bs Y) 
														= \bs d \big( \bs{d\gamma\gamma}\-\big) (\bs X, \bs Y),
\end{align*}
which is simply a ``flatness", or Maurer-Cartan type, condition  $\bs d \big(\bs{d\gamma\gamma}\-\big) -\sfrac{1}{2}\big[\bs{d\gamma\gamma}\-, \bs{d\gamma\gamma}\- \big]=0$
But then, again by Kozsul we have, 
\begin{align*}
\big[ \bs{d\gamma\gamma}\-_{|\phi}(\bs X_\phi), \bs{d\gamma\gamma}\-_{|\phi}(\bs Y_\phi) \big] = \bs d \big( \bs{d\gamma\gamma}\-\big)_{|\phi} (\bs X_\phi, \bs Y_\phi) 
																  = \bs X \cdot \big\{  \bs{d\gamma\gamma}\-_{|\munderline{blue}{\phi}} (\bs Y_{\munderline{blue}{\phi}})\big\} 
																  - \bs Y \cdot \big\{  \bs{d\gamma\gamma}\-_{|\munderline{blue}{\phi}} (\bs X_{\munderline{blue}{\phi}})\big\}   
																  -  \bs{d\gamma\gamma}\-_\phi\big( \big[\bs X, \bs Y\big]_\phi \big),
\end{align*}
where we stressed that the underlined $\phi$'s are acted upon.
Inserting this in the last term of  \eqref{1},  remembering that $J(\ ; \phi)$ is linear in the first argument and using \eqref{2}, we have 
\begin{align*}
{\bs\Theta^{\bs\gamma}}_{|\phi}\big(\bs X_\phi, \bs Y_\phi \big)   =&\,   \bs\Theta_{|\phi} \big(\bs X_\phi, \bs Y_\phi \big)   + \bs X \cdot J\left( \big\{ \bs{d\gamma\gamma}\-_{|\phi}(\bs Y_\phi) \big\}; \munderline{blue}{\phi} \right) - \bs Y \cdot J\left(\big\{ \bs{d\gamma\gamma}\-_{|\phi}(\bs X_\phi) \big\}; \munderline{blue}{\phi} \right)  \notag \\
			&\hspace{3cm}  + J\left( \big\{ \bs X \cdot \big\{  \bs{d\gamma\gamma}\-_{|\munderline{blue}{\phi}} (\bs Y_{\munderline{blue}{\phi}})\big\} 
																  - \bs Y \cdot \big\{  \bs{d\gamma\gamma}\-_{|\munderline{blue}{\phi}} (\bs X_{\munderline{blue}{\phi}})\big\}   
																  -  \bs{d\gamma\gamma}\-_\phi\big( \big[\bs X, \bs Y\big]_\phi \big) \big\}; \phi \right), \notag\\
										=&\,  \bs\Theta_{|\phi} \big(\bs X_\phi, \bs Y_\phi \big)  + \bs X \cdot J\left( \big\{ \bs{d\gamma\gamma}\-_{|\munderline{blue}{\phi}}(\bs Y_{\munderline{blue}{\phi}}) \big\}; \munderline{blue}{\phi} \right)
 																			- \bs Y \cdot J\left( \big\{ \bs{d\gamma\gamma}\-_{|\munderline{blue}{\phi}}(\bs X_{\munderline{blue}{\phi}}) \big\}; \munderline{blue}{\phi} \right) 
															  - J\left(    \big\{ \bs{d\gamma\gamma}\-_{|\phi}([\bs X, \bs Y]_\phi) \big\}; \phi  \right), \notag\\
										=&\,  \bs\Theta_{|\phi} \big(\bs X_\phi, \bs Y_\phi \big)  +  \bs d\, J\left( \big\{ \bs{d\gamma\gamma}\-_{|\phi} \big\}; \phi \right) \big( \bs X_\phi, \bs Y_\phi \big).
\end{align*}
Which is finally, using \eqref{Noether-current},
\begin{align}
\label{Field-depGT-Theta}
\bs\Theta^{\bs\gamma} = \bs\Theta +  \bs d\, J\left(\big\{ \bs{d\gamma\gamma}\- \big\}; \phi \right) = \bs\Theta + \bs d \left( d\theta\big(\bs{d\gamma\gamma}\-; \phi\big) - E\big(\bs{d\gamma\gamma}\-; \phi\big) \right), 
\end{align}
consistent with \eqref{Field-depGT-presymp-pot-current}.
The $\bs \H$-gauge transformation of the presymplectic 2-form is then, 
\begin{align}
\label{Field-depGT-presymp-form}
\bs\Theta_\Sigma^{\bs\gamma} =\bs\Theta_\Sigma + \bs d Q_\Sigma( \bs{d\gamma\gamma}\-; \phi)   =\bs\Theta_\Sigma + \int_{\d\Sigma}   \bs d \theta\big(\bs{d\gamma\gamma}\-; \phi\big) - \int_\Sigma \bs d E\big(\bs{d\gamma\gamma}\-; \phi\big),
\end{align}
consistent with  \eqref{Field-depGT-presymp-pot}. As for  $\bs\theta_\Sigma$, the presymplectic 2-form is $\bs\H$-invariant, basic,
if we are on-shell and if either $\d\Sigma=\emptyset$ or $\phi \rarrow 0$ and/or  $\bs\gamma \rarrow 1$ at $\d\Sigma$ or at infinity.
In which case it induces a symplectic 2-form on $\M_\S$. 
\smallskip

We can use the results just derived to say a word about the charges associated with field-dependent gauge parameters $\bs\chi \in $ Lie$\bs\H$ and their Poisson bracket. 

\paragraph{On charges and bracket associated with field-dependent gauge parameters: }  
As $L \in \Omega^0_\text{inv}(\Phi)$, we have $L^{\bs\gamma}=L$ for $\bs\gamma \in \bs\H$, whose linearisation is $\bs L_{\bs\chi^v} L =0$ for $\bs\chi \in $ Lie$\bs\H$. As the latter relation is $\iota_{\bs\chi^v} \bs E + d \iota_{\bs\chi^v} \bs\theta =0$, it is still true that the current associated with a field-dependent gauge parameter defined by
\begin{align}
\label{Current-field-dep-parameter}
J(\bs\chi; \phi)\defeq \iota_{\bs\chi^v} \bs\theta =  d \theta\big( \bs\chi; \phi \big) - E\big(\bs\chi; \phi \big) 
\end{align}
is conserved on-shell. Integration over a codimension 1 submanifold $\Sigma$ defines the associated Noether charge,
\begin{align}
\label{Noether-charge-fied-dep-parameter}
Q_\Sigma(\bs\chi; \phi)= \int_{\d \Sigma} \theta\big( \bs\chi; \phi \big) - \int_\Sigma E\big(\bs\chi; \phi \big). 
\end{align}
 The question is now to find the relation of this current/charge to the presymplectic potential and 2-form. 
 
 This will be easily done after we have clarified a few technical points. First, let us define the bracket 
 \begin{align}
 \label{extended-bracket}
 \{\bs\chi, \bs\eta\}\defeq [\bs\chi, \bs\eta] + {\bs\chi}^v(\bs\eta) - {\bs\eta}^v(\bs\chi),
 \end{align}
 where the first term on the right-hand side is the standard Lie bracket in Lie$H$. 
 Of course, for $\bs\chi,\bs\eta \in$ Lie$\bs\H$ we have that ${\bs\chi}^v(\bs\eta)=[\bs\eta, \bs \chi]$,\footnote{Which is the linearisation of the $\bs\H$-gauge transformation law $\bs\eta^{\bs\gamma}=\bs\gamma^{-1} \bs\eta \bs\gamma$, itself resulting from the defining $\H$-equivariance of $\bs\eta \in$ Lie$\bs\H$, $R^\star_\gamma \bs\eta =\gamma^{-1} \bs\eta \gamma$.} so \eqref{extended-bracket} reduces to  $\{\bs\chi, \bs\eta\}= - [\bs\chi, \bs\eta]$. While for  $\chi,\eta \in$ Lie$\H$ this bracket reduces to the standard one $\{\chi, \eta\}= [\chi, \eta]$. We will keep the general bracket because we have in mind  applications of this formalism where the relevant transformation group is not a gauge group $\H$, but $\Diff(M)$. In this case, the linearised gauge parameters are vector fields $\chi, \eta \in$ Lie$\Diff(M)\simeq \Gamma(TM)$, with $[\,,]$ the Lie bracket of vector fields. Then, $\bs\chi, \bs\eta : \Phi \rarrow \Gamma(TM)$ are field-dependent vector fields s.t.  ${\bs\chi}^v(\bs\eta) \neq [\bs\eta, \bs \chi]$ and is  generally  left unspecified. 
 
 The bracket \eqref{extended-bracket}  enters the following commutation relations, proven in appendix \ref{Commutation relations with the extended bracket},
   \begin{equation}
   \label{Commutation-relations}
  \begin{aligned}
  [\iota_{\bs\chi^v}, \iota_{\bs{d\eta}^v}] &= \iota_{[\bs\chi^v(\bs \eta)]^v}, \\
  [  \bs L_{\bs\chi^v}, \iota_{\bs\eta^v}] &=\iota_{\{\bs\chi, \bs \eta\}^v},\\
   [  \bs L_{\bs\chi^v}, \bs L_{\bs\eta^v}] &= \bs L_{\{\bs\chi, \bs \eta\}^v}.
  \end{aligned}
  \end{equation}
 The third is a consequence of the second. Notice that in the  case $\chi,\eta \in$ Lie$\H$ we recover  $[\bs L_{\chi^v}, \bs L_{\eta^v}] = \bs L_{[\chi^v, \eta^v]}$, since $[\chi, \eta]^v=[\chi^v, \eta^v]$. While in the the  case $\bs\chi,\bs\eta \in$ Lie$\bs\H$ we also recover $[\bs L_{\bs\chi^v}, \bs L_{\bs\eta^v}] = \bs L_{[\bs\chi^v, \bs \eta^v]}$, since $-[\bs\chi, \bs\eta]^v=[\bs\chi^v, \bs\eta^v]$. In line with the general formula $[\bs L_{\bs X}, \bs L_{\bs Y}] = \bs L_{[\bs X, \bs Y]}$   valid $\forall \bs X, \bs Y \in \Gamma(T\Phi)$. 
 
 With the above relation at hand, we proceed. The linearisation version of the $\bs\H$-transformation of $\bs\theta$, eq.\eqref{Field-depGT-presymp-pot-current}, is $ \bs L_{\bs\chi^v} \bs \theta = J(\bs{d\chi}; \phi) =\iota_{\bs{d\chi}^v} \bs\theta$. That is,  
 \begin{align}
 \label{Non-integrable-current}
\quad  \iota_{\bs\chi^v}\bs\Theta &= -\bs d J(\bs\chi; \phi) + J(\bs{d\chi}; \phi), \\
						&= -\bs dJ( \munderline{red}{\bs\chi}; \phi). \notag
 \end{align}
 The notation in the second line is meant to suggest that, given the linearity of $J(\,;\phi)$ in its first argument,  the first line  gives the same result as eq.\eqref{presympl-form-Noether-charge} for a field-independent parameter $\chi$. That is, if $\bs\chi$ was held ($\phi$-) constant, the result would of course be integrable.\footnote{This is often denoted as $ \iota_{\bs\chi^v}\bs\Theta=  -\slashed{\delta} J( \bs\chi; \phi)$ in the extent literature.} 
 From  \eqref{Non-integrable-current} follows immediately the relation between the presymplectic 2-form and the charge for field-dependent parameters, 
  \begin{align}
 \label{Non-integrable-charge}
\quad  \iota_{\bs\chi^v}\bs\Theta_\Sigma &= -\bs d Q_\Sigma(\bs\chi; \phi) + Q_\Sigma(\bs{d\chi}; \phi), \\
						&= -\bs dQ_\Sigma( \munderline{red}{\bs\chi}; \phi). \notag
 \end{align}
 The non-integrable term $Q_\Sigma(\bs{d\chi}; \phi)$ is sometimes called `symplectic flux' -- e.g. in \cite{Freidel-et-al2021}. 
 
 In this circumstance, can we still define a Poisson Bracket for the charges \eqref{Noether-charge-fied-dep-parameter}? As it turns out yes, and it has the same expression as in the $\phi$-independent case. To see this, consider
 \begin{align}
 \bs\Theta(\bs\chi^v, \bs\eta^v)=\iota_{\bs\eta^v} \big(\iota_{\bs\chi^v} \bs \Theta \big) &= - \iota_{\bs\eta^v} \bs d \iota_{\bs\chi^v}\bs \theta + \iota_{\bs\eta^v} \iota_{\bs{d\chi}^v} \bs \theta \quad   \text{ by \eqref{Non-integrable-current}, } \notag\\
  									&= -\iota_{\bs\chi^v} \bs L_{\bs\eta^v}\bs\theta -\iota_{\{ \bs\eta,\, \bs \chi\}^v}\bs\theta - \cancel{\iota_{\bs{d\chi}^v}\iota_{\bs\eta^v} \bs\theta} + \iota_{[\bs\eta^v(\chi)]^v}\bs\theta \quad  \text{ by \eqref{Commutation-relations}, } \notag\\
									&= -\iota_{\bs\chi^v} \left(\iota_{\bs{d\eta}^v}\bs\theta  \right) + \iota_{\{\bs\chi, \bs\eta\}^v} \bs \theta  + \iota_{[\bs\eta^v(\chi)]^v} \bs\theta,  \notag \\
							& = -\cancel{\iota_{\bs{d\eta}^v} \iota_{\bs\chi^v}\bs\theta}  -\iota_{[\bs\chi^v(\bs\eta)]^v}\bs\theta + \iota_{\{\bs\chi, \bs\eta\}^v} \bs \theta  + \iota_{[\bs\eta^v(\chi)]^v} \bs\theta 
							= \iota_{[\bs\chi, \bs \eta]^v} \bs\theta,
 \end{align}
 by definition of the extended bracket  \eqref{extended-bracket} in the last step. If we define the Poisson bracket the usual way, we get 
 \begin{align}
\label{Poisson-bracket-field-dep}
\big\{ Q_\Sigma(\bs\chi; \phi) ,  Q_\Sigma(\bs\eta; \phi)\big\}\defeq\, \bs\Theta_\Sigma(\bs\chi^v, \bs\eta^v) = \int_\Sigma \iota_{[\bs\chi, \bs\eta]^v} \bs \theta \rdefeq
										     Q_\Sigma([\bs\chi, \bs\eta]; \phi)
\end{align}
by \eqref{Current-field-dep-parameter}-\eqref{Noether-charge-fied-dep-parameter}. This matches eq.\eqref{Poisson-bracket}, and obviously reduces to it when $\bs\chi \rarrow \chi$. 
The Noether charges \eqref{Noether-charge-fied-dep-parameter} for $\phi$-dependent gauge parameters  $Q_\Sigma(\bs\chi; \phi)$ generate Lie$\bs\H$-transformation via the PB \eqref{Poisson-bracket-field-dep}.

In \cite{Francois2021} a centrally extended PB of charges for $\phi$-independent parameter was defined for anomalous (non-invariant) gauge theories. We will show elsewhere that non-integrable charges for $\phi$-dependent parameters are defined similarly as above and that their PB is the same as the centrally extended one just mentioned.  
\medskip

In the final paragraph of this subsection, we address the question of the physical status, and observability, of the Noether charges derived in this framework.

\paragraph{On the observability of charges}
\label{On the observability of charges}

Physically interpretable charges are computed against a background field configuration enjoying special symmetries,  in asymptotic domains of spacetime where the theory can be approximated by its free regime \cite{Barnich-et-al2000, Barnich-Brandt2002}. Let us see how to make contact with this viewpoint in the above framework. 
Using the affine structure of $\A$, we suppose that the connexion (the gauge potential) is written,
\begin{align}
\label{Perturbation-ansatz}
A=A_0 + \alpha, 
\end{align}
with $A_0$ a background connection, s.t. by definition of the term $\bs d A_0 =0$, and $\alpha \in \Omega_\text{tens}^1(\P, \text{Lie}H)$. We have then, 
\begin{equation}
\label{Perturb-identities}
\begin{aligned}
\bs d A &=\bs d\alpha, \\
F&=F_0 + f + \tfrac{1}{2}[\alpha, \alpha], \quad \text{and}\quad \bs d F= D^A(\bs d\alpha),
\end{aligned}
\end{equation}
where $F_0$ is the background curvature ($\bs dF_0=0$) and $f \defeq  D^{A_0}\alpha = D_0\alpha$ is the field strength we associate to the potential $\alpha$. 
The field equation and presymplectic potential are then 
\begin{align}
\bs E=E(\bs d\phi; \phi)=E(\bs d \alpha, \bs d\vphi; \phi) \quad \text{and} \quad \bs \theta=\theta(\bs d\phi; \phi)=\theta(\bs d \alpha, \bs d\vphi; \phi),
\end{align}
 with functionnal expressions defined in \eqref{def-E-theta}. 
 %
  Plugging the ansatz \eqref{Perturbation-ansatz} in the Noether current  \eqref{Noether-current} and using \eqref{def-E-theta}, we have on the one hand
 \begin{align*}
 J(\chi; \phi)= \iota_{\chi^v} \bs\theta &= \t L \big( \iota_{\chi^v} \bs d\alpha; \{\phi\} \big) + \t L \big( \iota_{\chi^v} \bs d\vphi; \{\phi\} \big), \notag \\
 							&= \t L \big([\alpha, \chi] ;  \{\phi\} \big) -   \t L \big( \rho_*(\chi) \vphi;  \{\phi\} \big),  \notag \\
							&=  - \t L \big(\chi ;  [\alpha, \{\phi\}] \big) -   \t L \big( \rho_*(\chi) \vphi;  \{\phi\} \big),
 \end{align*} 
by the $H$-invariance of $\t L$ in the last step. On the other hand the field equations are, 
 \begin{align*}
\bs E=E(\bs d \alpha, \bs d\vphi; \phi) &=  \t L \left( \bs d \alpha ; D^A \{\phi \} \right) + \t L \big( \rho_*(\bs d\alpha)\vphi ; \{\phi \} \big) +  \t L \big( \bs d \vphi;  \{\phi \} -D^A  \{\phi \} \big), \notag \\
							&=  \t L \left( \bs d \alpha ; D_0 \{\phi \} + [\alpha,  \{\phi\} ] \right) + \t L \big( \rho_*(\bs d\alpha)\vphi ; \{\phi \} \big) +  \t L \big( \bs d \vphi;  \{\phi \} -D^A  \{\phi \} \big). 
 \end{align*} 
Therefore, we get to write the current above in terms of the Lie$H$-linear piece of $\bs E$, so that:
 \begin{align}
 \label{Pertub-current}
 J(\chi; \phi)&= \t L \big(\chi ;  D_0 \{\phi\} \big) -   E \big( \chi;  \{\phi\} \big), \notag \\
		  &=   d\t L \big(\chi, \{\phi\} \big)  - \t L\big( D_0 \chi;  \{\phi\}  \big)-   E \big( \chi ;  \{\phi\} \big), \notag \\
		  &=d \theta\big(\chi,\phi\big) - \t L\big( D_0 \chi;  \{\phi\}  \big)-   E \big( \chi ;  \phi \big). 
 \end{align} 
This result may be interpreted as meaning that the current is conserved if we are on-shell \emph{and} if $\chi$ is a Killing symmetry of the background connection $A_0$, $D_0 \chi=0$. The associated Noether charges $Q_\Sigma(\chi; \phi)= Q_\Sigma(\chi; \alpha, \vphi)$ can then be seen as  conserved charges associated with the symmetries of the background. Notice that the field equation for the matter field, the piece of $\bs E$ linear in $\bs d\vphi$, plays no role in this. 

We notice that  \eqref{Pertub-current} with the Killing condition is formally the same as eq.\eqref{Noether-current}. 
This means that the general expressions for the  currents and charges \eqref{Noether-current}-\eqref{Noether-charge} can be used from the onset, simply plugging in the ansatz \eqref{Perturbation-ansatz} (in concrete situations) and declaring  $\chi$ to be a Killing symmetry for $A_0$. What one typically gets from doing so, is that $Q_\Sigma(\chi; \alpha, \vphi)$ splits as a background contribution from $A_0$ -- treated as a renormalisation constant -- and a physically interpreted contribution from $\alpha$, considered as the dynamical gauge field. 

This in particular reproduces and generalises the treatment of Abbott and Deser \cite{Abbott-Deser1982bis, Abbott-Deser1982} who derive charges in classical Yang-Mills theory and in metric gravity. For comparison, we re-express in the language of differential forms their treatment of the YM case in appendix \ref{The Abbott-Deser derivation of charges in YM theory}. 
Finally, we remark that the above manifestly holds for $\phi$-dependent gauge parameters $\bs\chi$ as well.
 
 We now illustrate the  general results of the previous subsections with the classic examples of Yang-Mills theory and for 4D gravity.

\subsection{Applications}
\label{Applications}

In the following applications, we will consider that all the geometric objects are pulled-back on spacetime $M$. So that a point in field space $\phi=(A, \vphi) \in \Phi$ can now be understood as referring to local fields, i.e. $A$ is a local connection -- that is a gauge potential -- and $\vphi \in \Gamma(E)$ is indeed a section of the associated bundle $E$. 

\subsubsection{Coupled Yang-Mills theory}
\label{Coupled Yang-Mills theory}

 The Lagrangian describing the coupling of a Yang-Mills field to a scalar field (that we keep massless to avoid unnecessary complication) is  
\begin{align}
\label{YM-Lagrangian}
 L_\text{\tiny YM}(\phi)= L_\text{\tiny YM}(A, \vphi)=\tfrac{1}{2}\Tr(F *\!F)+ \tfrac{1}{2} \langle D\vphi, *D\vphi \rangle,
 \end{align}
 and it is clear that $R^\star_\gamma L_\text{\tiny YM} =L_\text{\tiny YM}$ for $\gamma \in \H=\SU(n)$, i.e. $L_\text{\tiny YM} \in \Omega^0_\text{basic}(\Phi)$. Thus,  using $\langle u, v \rangle=\Tr( | v \rangle \langle u | )$,
 \begin{align}
 \bs d L_\text{\tiny YM} &=   \bs E_\text{\tiny YM} + d\bs \theta_\text{\tiny YM}  \quad \in \Omega^1_\text{basic}(\Phi) \quad \text{with:}  \notag \\[1mm]
\label{YM-E}
 \bs E_\text{\tiny YM} &= E_\text{\tiny YM}(\bs d A; A) = \Tr\big( \bs d A \, \{D\!*\!F -  |\!*\!D \vphi\rangle \langle \vphi| \} \big) + \langle \bs d\vphi , D\!*\!D \vphi\rangle  \quad \in \Omega^1_\text{inv}(\Phi),  \\   \label{YM-theta}
\bs \theta_\text{\tiny YM} &=   \theta_\text{\tiny YM}(\bs d A; A) = \Tr\big( \bs d A *\!F\big) + \langle \bs d\vphi, *D\vphi \rangle \quad \in \Omega^1_\text{inv}(\Phi).
  \end{align}
  We denote  $J=  |\!*\!D \vphi\rangle \langle \vphi |$ the $(n-1)$-form current sourcing the YM field.
  By the general formula \eqref{Noether-charge}, the Noether charge associated with $\chi \in$ Lie$\SU(n)$ is then
  \begin{align}
\label{Noether-charge-YM}
Q^\text{\tiny YM}_\Sigma(\chi; \phi)&= \int_{\d\Sigma} \theta_\text{\tiny YM}(\chi; \phi) - \int_\Sigma E_\text{\tiny YM}(\chi; \phi) , \notag \\
			  &= \int_{\d\Sigma} \Tr\big(\chi *\!F \big) - \int_\Sigma \Tr\big( \chi\,  \{D\!*\!F -  J \}  \big).
\end{align}
A result that can be checked by direct computation from the definition $Q^\text{\tiny YM}_\Sigma(\chi; \phi)\defeq \iota_{\chi^v} \bs\theta^\text{\tiny YM}_\Sigma$, using \eqref{YM-theta} and  \eqref{Vert-dphi}.
Notice how only the Lie$H$-linear piece of $\bs E_\text{\tiny YM}$ contributes. 
On-shell, this charge is the same as in the pure YM case, $Q^\text{\tiny YM}_\Sigma(\chi; \phi)=Q^\text{\tiny YM}_\Sigma(\chi; A)_{\ |\S}$ (see \cite{Francois2021} section 5.1.1). 
The presymplectic 2-form is,
\begin{align}
\label{YM-Theta}
\bs\Theta^\text{\tiny YM}_\Sigma=\int_\Sigma \bs{d\theta}_\text{\tiny YM}=-\int_\Sigma \Tr\big( \bs d A *\!\bs d F  \big) - \langle \bs d \vphi, *\bs d D\vphi \rangle \quad \in \Omega^2_\text{inv}(\Phi).
\end{align}
 and  by \eqref{presympl-form-Noether-charge} relates to the charge as
\begin{align}
 \iota_{\chi^v} \bs \Theta^\text{\tiny YM}_\Sigma = - \bs d Q^\text{\tiny YM}_\Sigma(\chi; \phi) 
 								          = -\int_{\d\Sigma} \Tr\big(\chi *\!\bs dF \big) + \int_\Sigma \Tr\big( \chi\, \bs d  \{D\!*\!F -  J \}  \big).
\end{align} 
 By \eqref{Poisson-bracket} it induces the Poisson bracket of charges $\big\{ Q^\text{\tiny YM}_\Sigma(\chi; \phi) ,  Q^\text{\tiny YM}_\Sigma(\eta; \phi)\big\}=Q^\text{\tiny YM}_\Sigma([\chi, \eta]; \phi)$, as could be checked explicitly by computing $\bs \Theta^\text{\tiny YM}_\Sigma (\chi^v, \eta^v)$ from \eqref{YM-Theta} -- and using  \eqref{Vert-dphi}, \eqref{Vert-dF} and \eqref{Vert-dDphi}. 
 The map $\chi \rarrow Q^\text{\tiny YM}_\Sigma(\chi; \phi)$ is thus a morphism of Lie algebras. 

By \eqref{Noether-charge-fied-dep-parameter}, the charge associated with a field-dependent gauge parameter $\bs\chi \in$ Lie$\bs\H$ is 
  \begin{align}
\label{Noether-charge-YM-field-dep}
Q^\text{\tiny YM}_\Sigma(\bs\chi; \phi)&= \int_{\d\Sigma} \theta_\text{\tiny YM}(\bs\chi; \phi) - \int_\Sigma E_\text{\tiny YM}(\bs\chi; \phi) , \notag \\
			  &= \int_{\d\Sigma} \Tr\big(\bs\chi *\!F \big) - \int_\Sigma \Tr\big(\bs\chi\,  \{D\!*\!F -  J \}  \big).
\end{align}
This time it is non-integrable, as by  \eqref{Non-integrable-charge} we have
\begin{align}
 \iota_{\bs\chi^v} \bs \Theta^\text{\tiny YM}_\Sigma &= - \bs d Q^\text{\tiny YM}_\Sigma(\bs\chi; \phi) +  Q^\text{\tiny YM}_\Sigma(\bs{d\chi}; \phi) =-\bs dQ_\Sigma( \munderline{red}{\bs\chi}; \phi) , \notag \\			 									    &= -\int_{\d\Sigma} \Tr\big(\bs\chi *\!\bs dF \big) + \int_\Sigma \Tr\big( \bs\chi\, \bs d  \{D\!*\!F -  J \}  \big).
\end{align} 
Still, by \eqref{Poisson-bracket-field-dep} the non-integrable charges \eqref{Noether-charge-YM-field-dep} satisfy the Poisson algebra $\big\{ Q^\text{\tiny YM}_\Sigma(\bs\chi; \phi) ,  Q^\text{\tiny YM}_\Sigma(\bs\eta; \phi)\big\}=Q^\text{\tiny YM}_\Sigma([\bs\chi, \bs\eta]; \phi)$, with PB defined by $ \bs \Theta^\text{\tiny YM}_\Sigma$, thus representing  Lie$\bs\H$. 
\medskip

Considering the question of the physical interpretation of charges (integrable or not):  By the affine ansatz \eqref{Perturbation-ansatz} $A=A_0+ \alpha$ s.t. $D_0 \chi \equiv 0$, i.e. $\chi$ is a Killing symmetry of the background YM field $A_0$, one plugs the expansion \eqref{Perturb-identities} of the field strength in the charge \eqref{Noether-charge-YM} which is then  on-shell
 \begin{align}
\label{Noether-charge-YM-perturb}
Q^\text{\tiny YM}_\Sigma(\chi; \phi)= \int_{\d\Sigma} \theta_\text{\tiny YM}(\chi; \phi)_{\ |\S} 
			  &= \int_{\d\Sigma} \Tr\big(\chi *\! \{ F_0 + f + \tfrac{1}{2}[\alpha, \alpha] \} \big),  \notag \\ 
			  &= \int_{\d\Sigma} \Tr\big(\chi *\! F_0\big)  + \Tr\big(\chi *\! f\big) \rdefeq Q^\text{\tiny YM}_\Sigma(\chi; A_0) + Q^\text{\tiny YM}_\Sigma(\chi; \alpha),
\end{align}
where we used $\Tr\big(\chi  [\alpha, \alpha]  \big) = -\Tr\big( [\alpha, \chi] \alpha  \big) \equiv 0$ by the $\H$-invariance of  $\Tr(\alpha \w \alpha)$. The contribution $Q^\text{\tiny YM}_\Sigma(\chi; A_0)$ comes entirely from the background, while the second term  
\begin{align}
\label{Noether-charge-YM-AB}
Q^\text{\tiny YM}_\Sigma(\chi; \alpha)=\int_{\d\Sigma}  \Tr\big(\chi *\! f\big) = \int_{\d\Sigma}  * d \Tr\big(\chi  \alpha \big)
\end{align}
is the contribution from the perturbation $\alpha$ in the YM field, and reproduces the charge $Q_\chi$ of Abbott \& Deser \cite{Abbott-Deser1982} -- see \eqref{charge-AB}-\eqref{charge-AB-bis} in appendix \ref{The Abbott-Deser derivation of charges in YM theory}. All this holds still for field-dependent gauge parameters $\chi \rarrow \bs\chi$.

One may be interested in expressing this conserved charge in terms of the charged field $\vphi$ sourcing the YM equation. To do so, one first looks at the expansion of the field equation $D*F=J$,
\begin{align}
\label{YM-eq-perturb}
&D_0*\!F_0 + [\alpha, *F_0] + D_0*\!f + [\alpha, *\! f] + D^A *\! \tfrac{1}{2}[\alpha, \alpha] =J.
\end{align}
Collecting on one side the terms linear in $\alpha$ and redefining a new source $j\defeq J-  \{D*F\}^R$ where $ \{D*F\}^R$ are the remaining terms, we get
\begin{align}
\label{YM-eq-perturb-bis}
&D_0*\!f +[\alpha, *F_0] =  j.
\end{align}
Together with the Killing condition $D_0\chi=0$, this can be used to rewrite the charge \eqref{Noether-charge-YM-AB} as a bulk integral of $j$,
\begin{align}
\label{Noether-charge-YM-AB-current}
Q^\text{\tiny YM}_\Sigma(\chi; \alpha)&=\int_{\Sigma}  d\Tr\big(\chi *\! f\big) = \int_{\Sigma}  \Tr\big(D_0\chi *\! f + \chi\, D_0\!*\! f\big) = \int_{\Sigma}  \Tr\big(\chi \,\{ j - [\alpha, *F_0] \}\big), \notag\\
							 &=  \int_{\Sigma}  \Tr\big(\chi \, j\big).
\end{align}
Where we have used  $\Tr\big(\chi \, [\alpha, *F_0] \big)= - \Tr\big( [\alpha, \chi ]*\!F_0 \big) = \Tr\big( \alpha\, [*F_0, \chi] \big)\equiv 0$, by the $\H$-invariance of $\Tr$ in the second equality and the Killing condition in the last. 

This is the reverse of the logic of Abbott-Deser, who start with the expansion of the coupled YM equation to define the $d$-closed singlet current  $\Tr\big(\chi \, j\big)$, and work so as to obtain the charge \eqref{Noether-charge-YM-AB} above.  
As their procedure depends only on the field equation $\bs E_\text{\tiny YM}$, it is insensitive to $L_\text{\tiny YM}$, thus to $\bs\theta_\text{\tiny YM}$. One may take advantage of this and renormalise the charge $Q^\text{\tiny YM}_\Sigma(\chi; \phi)$ by the addition of a boundary term to $L_\text{\tiny YM}$, so as to eliminate the background contribution $Q^\text{\tiny YM}_\Sigma(\chi; A_0)$. The AB charge can thus be seen as a renormalisation of the Noether charge coming from the covariant phase space formalism. 
\medskip

For future reference, when we will consider the basic presymplectic structure of the theory, lets us finally write down the field-dependent $\bs\H$-gauge transformations of the field equations and presymplectic structure. 
By  \eqref{Field-depGT-FieldEq}, for $\bs\gamma \in\bs\H= \bs\SU (n)$ we get,
\begin{align}
\label{GT-E-YM}
 \bs E_\text{\tiny YM}^{\bs \gamma}= \bs E_\text{\tiny YM} + dE_\text{\tiny YM} \big(  \bs{d\gamma\gamma}\- ;  \phi \big) = \bs E + d\Tr\big(  \bs{d\gamma\gamma}\- \{D*\!F-J \}  \big).
\end{align}
 This can be verified algebraically by $\bs E_\text{\tiny YM}^{\bs \gamma}=E_\text{\tiny YM}\big( \bs d \phi^{\bs \gamma}; \phi^{\bs \gamma}\big)$  using \eqref{GT-dphi} in \eqref{YM-E}.
By \eqref{Field-depGT-presymp-pot}-\eqref{Field-depGT-presymp-form} we have immediately,
\begin{align}
(\bs\theta^\text{\tiny YM}_\Sigma)^{\bs \gamma}&= \bs \theta^\text{\tiny YM}_\Sigma + \int_{\d\Sigma}\theta_\text{\tiny YM} \big(  \bs{d\gamma\gamma}\- ;  \phi \big) - \int_\Sigma E_\text{\tiny YM} \big( \bs{d\gamma\gamma}\- ; \phi \big), \notag\\
              &= \bs \theta^\text{\tiny YM}_\Sigma + \int_{\d \Sigma }\Tr\big(\bs{d\gamma\gamma}\- *\!F \big) - \int_\Sigma\Tr\big( \bs{d\gamma\gamma}\-\, \{D\!*\!F-J\}\big),     \label{GT-thetaYM} \\[1mm]
(\bs\Theta^\text{\tiny YM}_\Sigma)^{\bs\gamma} &= \bs\Theta^\text{\tiny YM}_\Sigma +  \int_{\d \Sigma}\bs d \theta_\text{\tiny YM}\big(\bs{d\gamma\gamma}\-; \phi \big) - \int_\Sigma \bs d E_\text{\tiny YM}\big(\bs{d\gamma\gamma}\-; \phi \big), \notag\\
				                    &= \bs\Theta^\text{\tiny YM}_\Sigma + \int_{\d \Sigma} \bs d \Tr\big( \bs{d\gamma\gamma}\- *\!F \big)  - \int_\Sigma \bs d  \Tr\big(  \bs{d\gamma\gamma}\-\, \{D\!*\!F-J\} \big).   \label{GT-ThetaYM}
\end{align}
Notice how only the Lie$\H$-linear pieces of $\bs\theta$ and $\bs E$ contribute to the final results. This can be  verified by direct computation,  using \eqref{GT-dphi}, \eqref{GT-dF} and \eqref{GT-dPhi} in  \eqref{YM-theta} and \eqref{YM-Theta}.
Clearly, only on-shell and under proper boundary conditions are $\bs\theta^\text{\tiny YM}_\Sigma$ an $\bs\Theta^\text{\tiny YM}_\Sigma$ $\bs\H$-invariant, i.e. basic forms on $\Phi$,  and  induce a symplectic structure on  $\M_\S$.

\paragraph{Comments}
Considering spinor fields $\psi$ instead of scalar fields $\vphi$ would change nothing of substance in the above results. Indeed, as we have observed several times now, only the Lie$\H$-linear part of 
$\bs\theta_\text{\tiny YM}$ and $\bs E_\text{\tiny YM}$ 
 contributes to the expression of the Noether currents and charges, as well as to the $\bs\H$-gauge transformations, so that the matter field contribution plays little to  no role. Therefore, here the only change would be hidden in the current $J$ sourcing the gauge field, which would then be the Dirac current: $J$ can thus be treated as a black box wherein one can have any type of matter field.

For the same reason, the addition of a potential term  $V(\vphi)$ for the scalar field to $L_\text{\tiny YM}$ wouldn't affect the results presented in this section, as it  just modifies the $\bs d \vphi$-linear part of $\bs E_\text{\tiny YM}$.  In particular, with the typical potential $V(\vphi)= \mu^2 \langle \vphi, * \vphi\rangle + \lambda \langle \vphi, *\vphi \rangle^2$ where $ \mu^2 \in\RR$ and $\lambda >0$, the Lagrangian becomes the prototype of a Yang-Mills-Higgs model, encompassing the electroweak model as a special case, and the above gives its charge and presymplectic structure.

\subsubsection{Coupled 4D gauge gravity}
\label{Coupled 4D gauge gravity}

We employ here the language of Cartan geometry \cite{Sharpe, Cap-Slovak09}. Meaning that we consider the underlying kinematics as given by Cartan-deSitter geometry $(P, \b A)$, where $P$ is a principal bundle whose structure group is $H=S\!O(1,3)$ and whose gauge group is thus $\H=\SO(1,3)$, while the Cartan connection $\b A$ takes value in the de Sitter/anti-de Sitter Lie algebra, Lie$S\!O(1,4)$ or Lie$S\!O(2,3)$, according to the sign of $\Lambda$. 

The geometry is called reductive as we have the $H$-invariant splitting $\b A= A +  \tfrac{1}{\ell} e$, where $A={A^a}_b$ is the Ehresmann Lie$H$-valued connection (the Lorentz or spin connection) and $e=e^a$ is the $\RR^4$-valued soldering form. The Cartan curvature is thus $\b F=d\b A+\tfrac{1}{2}[\b A, \b A]=F+ \tfrac{1}{\ell} T = \big( R-\tfrac{\epsilon}{\ell^2}e e^t\big) + \tfrac{1}{\ell}D^A e $, where $e^t\defeq e^T \eta= e^a \eta_{ab}$ and $R=dA+\tfrac{1}{2}[A, A]$ is the Lie$H$-valued Riemann 2-form. In matrix form, 
\begin{align*}
\b A = \begin{pmatrix} A & \tfrac{1}{\ell}e \\ \tfrac{-\epsilon}{\ell}e^t & 0 \end{pmatrix}, \qquad \b F = d\b A + \b A ^2= \begin{pmatrix} F & \tfrac{1}{\ell}T \\ \tfrac{-\epsilon}{\ell}T^t & 0 \end{pmatrix}=\begin{pmatrix} R-\tfrac{\epsilon}{\ell^2}e e^t & \tfrac{1}{\ell}D^A e \\[1mm] \tfrac{-\epsilon}{\ell}(D^Ae)^t & 0 \end{pmatrix},
\end{align*}
 with $\tfrac{1}{\ell^2}= \tfrac{2|\Lambda|}{(n-1)(n-2)}=\tfrac{|\Lambda|}{3}$ for $n=4=$ dim$\M$, and $\epsilon=\pm$ is the sign of $\Lambda$. The so-called \emph{normal} Cartan connection $\b A_{|N}$ is the unique torsion-free connection, so that $A=A(e)$, meaning that the only d.o.f. in the normal connection are those of the soldering, $\b A_{|N}= \b A_{|N}(e)$.\footnote{This is generically what the normality condition implies for normal Cartan connections in more elaborate situations, such as conformal Cartan geometry or more general parabolic geometries.} 
 Cartan flatness, $\b F\equiv 0$, means (in addition to vanishing torsion)  $F=0 \rarrow R=\tfrac{\epsilon}{\ell}ee^t$, that is the base manifold (spacetime) is the homogeneous de Sitter or anti-de Sitter space, $M\simeq(A)dS$. 
 
 Given the bilinear form $\eta: \RR^4 \times \RR^4 \rarrow \RR$,  the Cartan connection induces via its soldering component a metric on $M$, $g\defeq \eta(e, e): \Gamma(TM) \times\Gamma(TM)  \rarrow \RR$. In components $e^a={e^a}_\mu dx^{\,\mu}$, where ${e^a}_\mu$  is the (co-) tetrad field, so we have the well-known relation $g_{\mu\nu}={e_\mu}^a \eta_{ab} {e^b}_\mu$. To introduce notations that will be useful latter on (in section \ref{4D gauge gravity}), let us rewrite this in the index-free fashion $\bs e \defeq {e^a}_\mu$ and $\bs g = \bs e^T \eta \bs e$. 
 
 Dirac spinors are sections of the spin bundle $\mathbb S$ associated with $\P$ via the spin representation $\rho$ of $H=S\!O(1,3)$ on $\CC^4$. We have then $\psi \in \Gamma(\mathbb S) \simeq \Omega^0_\text{eq}(\P, \CC^4)$. A point of the  field space $\Phi$ under consideration is $\phi =\{ \b A, \psi\}= \{A, e, \psi \}$ and the right action $\phi \rarrow R^\star_\gamma \phi$ of $\H$ on $\Phi$ is explicitly
 \begin{equation}
 \label{eq-phi-grav}
 \begin{aligned}
R^\star_\gamma \b A &= \b A^\gamma = \gamma\- \b A \gamma + \gamma\-d\gamma  \quad  \Rightarrow   \quad \begin{cases}\  R^\star_\gamma A=A^\gamma = \gamma\-  A \gamma + \gamma\-d\gamma, \\[-.5mm] \ R^\star_\gamma e= e^\gamma = \gamma\- e,  \end{cases} \\
R^\star_\gamma \psi &=\psi^\gamma =\rho(\gamma\-)\psi. 
 \end{aligned}
 \end{equation}
 \medskip
 It follows that, as special cases of  \eqref{Vert-dphi}-\eqref{Equiv-dphi}, the basis $\bs d \phi =\{\bs d\b A, \bs d \psi \}=\{ \bs d A, \bs d e, \bs d \psi \} \in \Omega^1_\text{eq}(\Phi)$ is s.t.
 \begin{equation}
 \begin{aligned}
 R^\star_\gamma \bs d \phi &= \uprho(\gamma)\- \bs d \phi \defeq \left( \gamma\- \bs d \b A \gamma\, ,\ \rho(\gamma)\- \bs d \vphi \right) = \left(  \gamma\- \bs d  A \gamma\, ,  \gamma\- \bs de\, ,\ \rho(\gamma)\- \bs d \psi  \right), \\
 \bs d \phi(\chi^v)&= \left( \bs d \b A(\chi^v)\,,\  \bs d \psi (\chi^v)\right)= \left( D^{\b A} \chi\,, \ -\rho_*(\chi)\psi \right)  =  \left( D^A \chi\, ,\ -\chi e \,, \ -\rho_*(\chi)\psi \right) \rdefeq \delta_\chi\phi.
 \end{aligned}
   \end{equation}
   From which follows, as a special case of \eqref{GT-dphi}, the field-dependent $\bs\H$-gauge transformation on $\Phi$
\begin{align}
\label{Field-dep-GT-dphi-grav}
\bs d \phi^{\bs \gamma} &= \uprho (\bs\gamma)\-    \left( \bs d \phi  + \delta_{ \bs{d\gamma\gamma}\-} \phi \right)
		 =   \begin{cases}  \ \bs d A^{\bs \gamma} = \bs{\gamma}\-   \left( \bs{d}A  + D^A\big\{ \bs{d}\bs{\gamma} {\bs{\gamma}\- }\big\} \right)    \bs{\gamma}, \\[1mm] 
		 			     \  \bs d e^{\bs \gamma} = \bs{\gamma}\-   \left( \bs{d}e  -\bs{d}\bs{\gamma} {\bs{\gamma}\- } e \right)    \bs{\gamma}, \\[1mm] 
		                               \  \bs d \psi^{\bs \gamma} =  \rho(\bs \gamma)\- \left( \bs d \psi -\rho_*(\bs{d\gamma\gamma}\-) \psi\right).
		               \end{cases} \
 \end{align}
  Similarly, as special  instances of \eqref{Vert-dF}-\eqref{GT-dF}, from $R^\star_\gamma \bs d \b F=\left(\gamma\-  \bs d F \gamma, \ \gamma\- T\right)$  and $\bs d \b F(\chi^v)=\big([F, \chi], \ -\chi T \big) $,  we have 
     \begin{align}
   \label{GT-dF-grav}
    \bs d \b F^{\bs \gamma}= \begin{cases} \  \bs{d}F^{\bs \gamma} =  \bs{\gamma}\-   \left( \bs{d}F  + \big[F, \bs{d}\bs{\gamma} {\bs{\gamma}\- }\big] \right)    \bs{\gamma} , \\[1mm]
					                          \   \bs{d}T^{\bs \gamma} =   \bs{\gamma}\-   \left( \bs{d}T   -  \bs{d}\bs{\gamma} {\bs{\gamma}\- T}\big] \right).   
	                                        \end{cases}
 \end{align}

To write the pure gravity sector of the theory in an index-free way, we consider the multilinear polynomial $P : \otimes^k M(2k, \RR) \rarrow \RR$ given by
\begin{align}
\label{Polyn-gravity}
P\big(A_1, \ldots, A_k \big)= A_1 \bullet\, \ldots\, \bullet A_k\defeq A^{i_1i_2}_1\, A^{i_3i_4}_2 \ldots \, A^{i_{2k-1}i_{2k}}_k \, \epsilon_{i_1 \ldots i_{2k}}, 
\end{align}
where the second equality defines the notation. Given $G \in GL(2k, \RR)$, it satisfies the identity
\begin{align}
\label{Prop1-P}
P\big(GA_1G^T, \ldots, GA_k G^T\big)&= GA_1G^T \bullet\, \ldots\, \bullet GA_kG^T, \notag\\
							   &= {G^{i_1}}_{j_1}A^{j_1j_2}_1 {G_{j_2}}^{i_2} \  {G^{i_3}}_{j_3}A^{j_3j_4}_2   {G_{j_4}}^{i_4} \ldots \    {G^{i_{2k-1}}}_{j_{2k-1}} A^{j_{2k-1}j_{2k}}_k   {G_{j_{2k}}}^{i_{2k}}\ \epsilon_{i_1 \ldots i_{2k}}, \notag \\
							   &= \det(G)\ A^{j_1j_2}_1\, A^{j_3j_4}_2 \ldots \  A^{j_{2k-1}j_{2k}}_k\, \epsilon_{j_1 \ldots j_{2k}}, \notag\\
							   &=  \det(G)\  A_1 \bullet\, \ldots\, \bullet A_k =  \det(G)\  P\big(A_1, \ldots, A_k \big).
\end{align}
Then, $P$ is $S\!O(2k)$-invariant, since for $S\in S\!O(2k)$, $S^T=S\-$, we have $P\big(S\-A_1S, \ldots, S\-A_k S\big)=P\big(A_1, \ldots, A_k \big)$. 
Also, given some matrix $M \in M(2k, \RR)$ decomposed as the sum of its symmetric and antisymmetric parts as $M=\sfrac{1}{2}(M+M^T) + \sfrac{1}{2}(M-M^T)\rdefeq {S}+  {\sf  A}$, we have
\begin{align}
\label{Prop2-P}
M\bullet A_2 \bullet \ldots \bullet A_k = \big(   \cancel{ S^{i_1i_2}}_{} \!+ {\sf A}^{i_1i_2} \big) \, A^{i_3i_4}_2 \ldots \, A^{i_{2k-1}i_{2k}}_k \, \epsilon_{i_1 i_2 \ldots i_{2k}}= {\sf A} \bullet A_2 \ldots \bullet A_k. 
\end{align}
We have then a $\Ad\big(S\!O(2k)\big)$-invariant map $P: \otimes^k \so(2k) \rarrow \RR$\footnote{Remark that the diagonal combination $P(A, \ldots, A)=\text{Pf}(A)$ is the Pfaffian of the $2k \times 2k$ antisymmetric matrix $A$, which is the square root of its determinant $\text{Pf}(A)^2=\det(A)$. Conversely, $P$ is the polarisation of the Pfaffian polynomial. }. 
For $X \in$ Lie$S\!O(r, s)$, $r+s=2k$, $X\eta\-$ is antisymmetric, and for $S \in S\!O(r, s)$ we have   $S\-XS\eta\-=S\- X\eta\-(S\-)^T$. Thus $P(X_1\eta\-, \cdots, X_k\eta\- )$ is a  $\Ad\big(S\!O(r, s)\big)$-invariant polynomial $P: \otimes^k \so(r,s ) \rarrow \RR$
 that one can use to write the Lagrangians of even dimensional gauge gravity theories. To lighten the notation, we will omit $\eta\-$ in front of Lie$S\!O(r,s)$-valued variables when writing expressions involving $P$, as it should be clear from the context that indices must be raised.
   
 \medskip
   
We consider the Lagrangian of General Relativity with  cosmological constant $\Lambda$, à la McDowell-Mansouri, coupled to Dirac spinors (massless to avoid unnecessary complications) to be, 
\begin{equation}
\label{GR-L}
\begin{aligned}
L_\text{\tiny GR}(\phi)=L_\text{\tiny MM}(\b A) + L_\text{\tiny Dirac}(\b A, \psi) &=  \tfrac{1}{2} F \bullet F + \langle \psi, \, \slashed D \psi \rangle,  \\[1mm]
				&= \tfrac{1}{2} R \bullet R -\tfrac{\epsilon}{\ell^2} \left(  R \bullet e \w e^T -  \tfrac{\epsilon}{2\ell^2} e\w e^T \bullet e\w e^T  \right) \ + \ \langle \psi, \, \upgamma \w * D \psi \rangle,  \\[1mm]
				&=  \tfrac{1}{2} R^{ab}R^{cd} \epsilon_{abcd} -\tfrac{\epsilon}{\ell^2}\left(  R^{ab}e^c e^d -  \tfrac{\Lambda}{6} e^a e^be^ce^d\right)\epsilon_{abcd}  \ + \ \langle \psi, \, \upgamma \w * D \psi \rangle.
\end{aligned}
\end{equation}
We have introduced convenient notations, and the gamma-matrices 1-form $\upgamma\defeq \upgamma_a e^a = \upgamma_a {e^a}_\mu dx^\mu \rdefeq \upgamma_\mu dx^\mu$ with $\{ \upgamma_a, \upgamma_b\}=\eta_{ab} \1_4$, from which follows  $\{ \upgamma_\mu, \upgamma_\nu\}=g_{\mu\nu} \1_4$, as $g_{\mu\nu}= {e_\mu}^a \eta_{ab}{e^b}_\nu$. It allows to define the top form $\slashed D \psi = \sqrt{|g|}\,d^4\!x\ \upgamma_\mu g^{\mu\nu} D_\nu \psi$ on $M$. Of course, $D\psi = d\psi + \rho_*(A)\psi$ is the minimal coupling of spinors to gravity via the spin connection. The bilinear map $\langle \ , \ \rangle : \Gamma(\mathbb S) \times \Gamma(\mathbb S) \rarrow \RR$ is $\rho(H)$-invariant. 

The first term in the Lagrangian is the Euler density, $L_\text{\tiny Euler}$,  a topological invariant of $M$: it doesn't change the field equation but contributes to the total presymplectic potential. Indeed it is easy to see that $\bs d L_\text{\tiny Euler}=\bs E_\text{\tiny Euler} +  d\bs \theta_\text{\tiny Euler} =\bs d A \bullet D^AR + d \left( \bs dA \bullet R\right)$, and the field equations vanish identically, being just the Bianchi identity $D^AR\equiv 0$. 
\medskip

It is clear that $R^\star_\gamma L_\text{\tiny GR}=L_\text{\tiny GR}$ for $\gamma \in \H=\SO(1,3)$, i.e.  $L_\text{\tiny GR} \in \Omega^0_\text{basic}(\Phi)$. The only computation needed is:
\begin{align}
\bs d L_\text{\tiny GR} &= \bs E_\text{\tiny GR} + d\bs\theta_\text{\tiny GR}   \quad  \in \Omega^1_\text{basic}(\Phi) \quad  \text{with}:  \notag \\[1.5mm]
\bs E_\text{\tiny GR}&= -\tfrac{2\epsilon}{\ell^2} \,       \bs d A \bullet T \w e^T + \Tr \big( | \rho_*(\bs d A)\psi \rangle\langle *\upgamma  \psi | \big)  \ \   -\tfrac{2\epsilon}{\ell^2} \,  \bs d e \w e^T \bullet \big( R -\tfrac{\epsilon}{\ell^2} e \w e^T  \big) +  \bs d e^a \sT_a      \label{E-GR} \\[1mm]
&\qquad   +\  \langle \bs d \psi, \upgamma \w *D\psi \rangle -  \langle D(*\upgamma \psi) , \bs d\psi\rangle         \quad \in \Omega^1_\text{inv}(\Phi) \notag\\[1mm]
\bs\theta_\text{\tiny GR}&  = \bs d A \bullet \big(  R -\tfrac{\epsilon}{\ell^2} e \w e^T \big) +  \Tr \big( | \bs d \psi  \rangle\langle *\upgamma  \psi | \big),  \quad \in \Omega^1_\text{inv}(\Phi).  \label{theta-GR}
\end{align}
The $\bs dA$-linear part of $\bs E_\text{\tiny GR}$ gives the coupling of the torsion to the spin density 3-form $\sS_{ab, c} *e^c= \langle \psi, \upgamma_c \sigma_{ab} \psi \rangle *e^c$, 
where $ \sigma_{ab} =\tfrac{1}{8}[\upgamma_a , \upgamma_b]$ is a representation ($\rho_*$) of the basis of Lie$S\!O(1,3)$. 
 The $\bs de$-linear part of $\bs E_\text{\tiny GR}$ gives of course  Einstein's equations, with the stress-energy tensor 3-form $\sT_a$ of the Dirac field, whose Hodge dual 1-form is $*\sT_a=\sT_{ab}e^b=  \left( \langle \psi, \upgamma_aD_b\psi \rangle - \eta_{ab} \eta^{ij} \langle \psi, \upgamma_iD_j\psi \rangle\right)e^b$ (and coincide with the traceless canonical stress-energy tensor). 
 The~\mbox{$\bs d\psi$-linear} part of $\bs E_\text{\tiny GR}$ gives Dirac's equation, $\slashed D\psi=0$. 

By the general formula \eqref{Noether-charge}, the Noether charge associated with $\chi \in$ Lie$\SO(1,3)$ is,
\begin{align}
Q^\text{\tiny GR}_\Sigma(\chi; \phi)&= \int_{\d\Sigma} \theta_\text{\tiny GR}(\chi; \phi) - \int_\Sigma E_\text{\tiny GR}(\chi; \phi), \notag \\
			   &= \int_{\d\Sigma}\chi \bullet \big( R -\tfrac{\epsilon}{\ell^2} e \w e^T  \big) +  \int_\Sigma  \tfrac{2\epsilon}{\ell^2}\, \chi \bullet T \w e^T - \Tr \big( | \rho_*( \chi)\psi \rangle\langle *\upgamma  \psi | \big),           \label{Noether-charge-GR} 
\end{align}
as can be checked by direct computation. As is now usual to remark, only the Lie$H$-linear piece of $\bs E_\text{\tiny GR}$ contributes, the Einstein and Dirac equations have no bearing on the result. On-shell, the charge is the same as in the pure gravity case $Q^\text{\tiny GR}_\Sigma(\chi; \phi)=Q^\text{\tiny GR}_\Sigma(\chi; \b A)_{\ |\S}$
 (see \cite{Francois2021} section 5.1.4) 
and vanishes on the  ground state of the theory, the homogeneous (anti-) de Sitter space $(A)dS$, which thus sets the zero mass-energy reference.

It would take some work to check directly what  \eqref{presympl-form-Noether-charge}  ensures, i.e. that the presymplectic 2-form  
\begin{align}
\label{Theta-GR}
\bs\Theta_\Sigma^\text{\tiny GR}=\int_\Sigma \bs{d\theta}\text{\tiny GR}= - \int_\Sigma \bs d A \bullet \bs d \big(  R -\tfrac{\epsilon}{\ell^2} e \w e^T \big) +  \Tr \big( | \bs d \psi  \rangle\langle \bs d( *\upgamma  \psi) | \big),  \quad \in \Omega^1_\text{inv}(\Phi),
\end{align}
relates to the charges via $\iota_{\chi^v} \bs \Theta^\text{\tiny GR}_\Sigma = - \bs d Q^\text{\tiny GR}_\Sigma(\chi; \phi)$. From which can also be verified that
\begin{align}
\big\{ Q^\text{\tiny GR}_\Sigma(\chi; \phi) ,  Q^\text{\tiny GR}_\Sigma(\eta; \phi)\big\} \defeq \bs\Theta_\Sigma^\text{\tiny GR}(\chi^v, \eta^v)=Q^\text{\tiny GR}_\Sigma([\chi, \eta]; \phi),
\end{align}
as the general formula \eqref{Poisson-bracket} allows to write immediately. By \eqref{Noether-charge-fied-dep-parameter},  \eqref{Non-integrable-charge} and  \eqref{Poisson-bracket-field-dep}, the above formulae holds for field-dependent Lorentz parameters, $\chi \rarrow \bs\chi$, so that both Lie$\H$ and Lie$\bs\H$ are  represented faithfully by the Poisson algebra of charges, even though in the field-dependent case the Lorentz charges \eqref{Noether-charge-GR}  are non-integrable.
\bigskip

Remark that, writing the charge in term of $F=R -\tfrac{\epsilon}{\ell^2} e \w e^T \in \Omega^2\left(M, \text{Lie}S\!O(1,3)\right)$,  it  is on-shell  
\begin{align}
\label{Charge-on-shell-grav}
Q^\text{\tiny GR}_\Sigma(\chi; \phi)  =Q^\text{\tiny GR}_\Sigma(\chi; \b A)  &= \int_{\d\Sigma}\chi \bullet F_{\ |\S} 
						         =\int_{\d\Sigma} \chi^{ab}F^{cd} \, \epsilon_{abcd}. 
\end{align} 
The striking similarity with the YM case  is no surprise as we  wrote the pure gravity sector  à la McDowell-Mansouri, $L_\text{\tiny MM}(\b A)= \tfrac{1}{2}\, F \bullet F =\tfrac{1}{2} F^{ab}F^{cd} \epsilon_{abcd}$.
Thus, the question of the physical interpretation of charges can be answered in essentially the same terms as in YM theory.  One first make the affine ansatz \eqref{Perturbation-ansatz} $\b A=\b A_0+ \b \alpha$ s.t. $D_0\chi \defeq D^{\b A_0} \chi \equiv 0$, i.e. $\chi$ is a Killing symmetry of the background gravitational field (Cartan connection)  $\b A_0= A_0 + e_0$, while $\b\alpha =\alpha+ \upepsilon$ is a perturbation. From it follows the analogue of expansion \eqref{Perturb-identities} for the Cartan curvature, $\b F = \b F_0 + \b f + \tfrac{1}{2}[\b \alpha, \b\alpha] $, whose Lorentz component gives
\begin{align}
 F =  F_0 + \left( D^{A_0}\alpha - \tfrac{\epsilon}{\ell^2} (e_0 \w \upepsilon^t + \upepsilon \w e_0^t )\right) +  \left( \tfrac{1}{2}[\alpha, \alpha] - \tfrac{\epsilon}{\ell^2} \upepsilon \w \upepsilon^t \right)
 	\rdefeq F_0+ f+  \tfrac{1}{2}[\alpha, \alpha]. 
\end{align}
 Plugging this into the charge \eqref{Noether-charge-GR}, it becomes on-shell
 \begin{align}
\label{Noether-charge-GR-perturb}
Q^\text{\tiny GR}_\Sigma(\chi; \phi)
			  &= \int_{\d\Sigma} \, \chi \bullet \big( F_0 + f + \tfrac{1}{2}[\alpha, \alpha] \big),  \notag \\ 
			  &= \int_{\d\Sigma} \, \chi \bullet F_0  + \chi \bullet  f  \rdefeq Q^\text{\tiny GR}_\Sigma(\chi; \b A_0) + Q^\text{\tiny GR}_\Sigma(\chi; \b \alpha),
\end{align}
where we used $\chi \bullet [\alpha, \alpha]= [\alpha, \chi] \bullet \alpha \equiv 0$ by the $\H$-invariance of $\alpha \bullet \alpha$. 
Here again,  $Q^\text{\tiny GR}_\Sigma(\chi; \b \alpha)$ is the charge measured against the background  $\b A_0$, which  in particular can chosen to be the (A)dS groundstate so that $Q^\text{\tiny GR}_\Sigma(\chi; \b A_0) \equiv 0$. 
\medskip

Finally, for  future expression of the basic presympletic structure, let us write the field-dependent transformations of the field equations and presymplecture structure of the theory. By \eqref{Field-depGT-FieldEq},  \eqref{Field-depGT-presymp-pot} and  \eqref{Field-depGT-presymp-form}, for $\bs \gamma \in \bs\H=\bs \SO(1,3)$:
\begin{align}
 \bs E_\text{\tiny GR}^{\bs \gamma}&= \bs E_\text{\tiny GR} + dE_\text{\tiny GR} \big(  \bs{d\gamma\gamma}\- ;  \phi \big) = \bs E_\text{\tiny GR} - d \left\{  \tfrac{2\epsilon}{\ell^2}\,  \big(  \bs{d\gamma\gamma}\-  \bullet T \w e^T  \big) - \Tr \left( | \rho_*( \bs{d\gamma\gamma}\-)\psi \rangle\langle *\upgamma  \psi | \right) \right\}, \label{SO-GT-E} \\[2mm]
(\bs\theta_\Sigma^\text{\tiny GR})^{\bs \gamma}&= \bs \theta_\Sigma^\text{\tiny GR} + \int_{\d\Sigma}  \theta_\text{\tiny GR} \big(  \bs{d\gamma\gamma}\- ;  \phi \big) - \int_{\Sigma} E_\text{\tiny GR} \big( \bs{d\gamma\gamma}\- ; \phi \big), \notag\\
				  &= \bs \theta_\Sigma^\text{\tiny GR} + \int_{\d\Sigma} \bs{d\gamma\gamma}\- \bullet F  +   \int_{\Sigma} \tfrac{2\epsilon}{\ell^2}\, \bs{d\gamma\gamma}\-\bullet T \w e^T   - \Tr \left( | \rho_*( \bs{d\gamma\gamma}\-)\psi \rangle\langle *\upgamma  \psi | \right),                           \label{SO-GT-presymp-pot} \\[2mm]
(\bs\Theta_\Sigma^\text{\tiny GR})^{\bs\gamma} &= \bs\Theta_\Sigma^\text{\tiny GR} +  \int_{\d\Sigma} \bs d  \theta_\text{\tiny GR}\big(\bs{d\gamma\gamma}\-; \phi \big) -  \int_{\d\Sigma} \bs d E_\text{\tiny GR}\big(\bs{d\gamma\gamma}\-; \phi \big), \notag\\
				    &= \bs\Theta_\Sigma^\text{\tiny GR} +  \int_{\d\Sigma}  \bs d  \left(\bs{d\gamma\gamma}\- \bullet  F \right) +       \int_{\Sigma}  \tfrac{2\epsilon}{\ell^2}\,  \bs d (\bs{d\gamma\gamma}\-\bullet T \w e^T) -  \bs d \Tr \left( | \rho_*( \bs{d\gamma\gamma}\-)\psi \rangle\langle *\upgamma  \psi | \right),         \label{SO-GT-presymp-form} 
\end{align}
which can be checked by explicit computation, with some work, via \eqref{Field-dep-GT-dphi-grav}-\eqref{GT-dF-grav} and \eqref{GT-dPhi} (which holds for $\vphi \rarrow \psi$). 
 Remark that for solutions of the theory that asymptotically decay to the $(A)dS$ ground state, $\b F=0$, both $\bs\theta_\Sigma^\text{\tiny GR}$ and $\bs\Theta_\Sigma^\text{\tiny GR}$ are $\bs\SO(1,3)$-invariant, and thus induce respectively a symplectic potential and 2-form on the physical phase space $\M_\S$.


  \section{Basic presymplectic structures}
\label{Basic presymplectic structures}

 As we have seen, the non-horizontality of $\bs\theta$ and $\bs\Theta$ is crucial to the very definition of  Noether currents and charges \eqref{Noether-current}-\eqref{Noether-charge}, and to the construction of the associated Poisson bracket \eqref{presympl-form-Noether-charge}-\eqref{Poisson-bracket}. But is also  results in their non-trivial  $\bs\H$-gauge transformations \eqref{Field-depGT-presymp-pot}-\eqref{Field-depGT-presymp-form}, which is a problem 
regarding the goal of associating a symplectic structure to a gauge theory over a bounded region. This lack of horizontality, thus of basicity, of $\bs\theta$ and $\bs\Theta$  is what we have called the \emph{boundary problem}. 

Yet, in the case at hand $L \in \Omega^0_\text{basic}(\Phi)$, thus $\bs d L \in \Omega^1_\text{basic}(\Phi)$. This means first that  $ \exists\ \b L \in \Omega^0(\M)$ s.t. $L=\pi^\star \b L$, by definition of a basic form. Then, by the same argument leading to \eqref{def-E-theta}, on $\M$ we have $\bs d_{\!_\M} \b L = \b{\bs E} + d \b{\bs\theta}$, with $\b{\bs E}, \b{\bs\theta} \in \Omega^1(\M)$.
Therefore, by naturality and linearity of the pullback we must have 
\begin{align}
\label{abstract-basic-presymp-struct}
\bs d L = \bs d \pi^\star \b L = \pi^\star \bs d_{\!_\M}  \b L = \pi^\star ( \b{\bs E} + d \b{\bs\theta})= \pi^\star  \b{\bs E} + d (\pi^\star \b{\bs\theta}) \rdefeq \bs E^b + d \bs \theta^b, 
\quad \text{with } \  \bs E^b, \bs \theta^b \in  \Omega^1_\text{basic}(\Phi). 
\end{align}
It would thus seem that we should be able to define a basic presymplectic potential $\bs \theta^b$, from which to  naturally derive a basic presymplectic 2-form $\bs \Theta^b \defeq \bs d\bs \theta^b \in \Omega^2_\text{basic}(\Phi)$ -- since $\bs d$ is the covariant derivative on basic forms, i.e. preserves the space. 
But how are we to find such a basic presymplectic structure starting  from the known $\bs\theta$ and $\bs\Theta$? 
In sections \ref{Variational connections on field space} and \ref{The dressing field method} we have already seen two methods to do so,  using respectively  variational connections or the DFM. In the following, we consider the results of each approach in turn.

 \subsection{Via variational connections}
\label{Via variational connections}

We use the general results of section   \ref{Variational Ehresmann connections} about the dual horizontalisation of forms relying on a variational Ehresmann connection $\bs\omega$. When a Lagrangian $L$  is invariant so are its associated field equations and presymplectic potential $\bs E, \bs\theta \in  \Omega^1_\text{inv}(\Phi)$, eq.\eqref{trivial-eq-E-theta}. So, applying the formula \eqref{horiz-1-form}, and using first $\iota_{\chi^v}\bs E= d E(\chi; \phi)$ -- stemming from $\bs L_{\chi^v} L=0$ and \eqref{Noether-current} -- we get the basic field equations:
\begin{align}
\label{omega-basic-E}
 \bs E^b_{\bs\omega} = \bs E - dE(\bs\omega; \phi)   \quad  \in  \Omega^1_\text{basic}(\Phi),
\end{align}
By \eqref{horiz-1-form} still, and given \eqref{Noether-current}, we get the $\bs\omega$-dependent basic presymplectic potential,
\begin{align}
\label{omega-basic-theta}
 \bs\theta^b_{\bs\omega} &= \bs\theta - J(\bs\omega; \phi),  \qquad  \in  \Omega^1_\text{basic}(\Phi),\\[1mm]
 				       &= \bs\theta - d \theta\big( \bs\omega; \phi \big) + E\big(\bs\omega; \phi \big).  \notag
\end{align}
The $\bs\H$-invariance of both $ \bs E^b_{\bs\omega}$ and $\bs\theta^b_{\bs\omega}$, although structurally garanteed,  is easily checked  explicitly knowing  \eqref{Field-depGT-FieldEq}-\eqref{Field-depGT-presymp-pot-current} and $\bs\omega^{\bs\gamma} =\bs\gamma\- \bs\omega \bs\gamma+ \bs\gamma\- \bs d \bs \gamma$. The basic presymplectic 2-form naturally associated with $\bs\theta^b_{\bs\omega}$ is then, 
\begin{align}
\label{omega-basic-Theta}
 \bs\Theta^b_{\bs\omega} \defeq&\, \bs d \bs\theta^b_{\bs\omega} \qquad  \in  \Omega^2_\text{basic}(\Phi),\\[1mm]
 						=&\, \bs\Theta - \bs d J(\bs\omega; \phi),  \notag\\
 				                 =&\, \bs\Theta - \bs d \left(  d \theta\big( \bs\omega; \phi \big) -   E\big(\bs\omega; \phi \big) \right),  \notag
\end{align}
whose $\bs\H$-invariance is again easily checked via \eqref{Field-depGT-Theta}. 
We may remark that contrary to what one could  be tempted to do, the correct approach is not to build the horizontal version of $\bs\Theta \in \Omega^2_\text{inv}(\Phi)$, as it is actually by definition the covariant derivative -- w.r.t. $\bs\omega$ - of $\bs\theta \in \Omega^1_\text{inv}(\Phi)$: 
\begin{align}
\bs\Theta^h\defeq \bs\Theta \circ |^h \defeq \bs d \bs\theta \circ |^h \rdefeq \bs D^{\bs\omega} \bs\theta, \quad \in \Omega^2_\text{basic}(\Phi). 
\end{align}
If it is basic indeed, it is not $\bs d$-closed, so cannot play the role of a presymplectic form. Actually, specialising eq.\eqref{Formula1} to this case and using again  \eqref{Noether-current}, we get
\begin{align}
\label{Diff-cov-der-theta-Theta-basic}
\bs D^{\bs\omega} \bs\theta &= \bs d \bs\theta^b_{\bs\omega} + \iota_{[\bs\Omega]^v} \bs\theta,  \notag \\[1mm]
                         \bs\Theta^h &= \bs\Theta^b_{\bs\omega} + J(\bs\Omega; \phi), \\
                         			  &= \bs\Theta^b_{\bs\omega} + d\theta (\bs\Omega; \phi) - E(\bs\Omega; \phi), \notag 
\end{align}
where of course $\bs\Omega \in \Omega^2_\text{tens}(\Phi)$ is the curvature of $\bs\omega$.
Thus, the actual basic presymplectic 2-form is the $\bs d$-exact part of the covariant derivative of the presymplectic potential $\bs\theta$. The above formula generalises the remark already made by Gomes \& Riello in the YM case -- \cite{Gomes-Riello2021}, corollary 3.2 and section 3.4, see also \cite{Riello2021} end of section 3.1. 
Manifestly, for a flat connections $\mathring{\bs\omega}$  the situation is degenerate, $\mathring{\bs\Theta}\ \!\! \phantom{|}^{\!h}\defeq \bs D^{\mathring{\bs\omega}} \bs\theta = \bs\Theta^b_{\mathring{\bs\omega}}$ (this is relevant to our  discussion of the approach via the DFM). 

The basic presymplectic structure is then given by
\begin{equation}
\begin{aligned}
\label{basic-presymp-struct}
 \bs\theta^b_{\bs\omega,\, \Sigma} &= \bs\theta_\Sigma  - \int_{\d\Sigma} \theta\big( \bs\omega; \phi \big) +  \int_{\Sigma} E\big(\bs\omega; \phi \big),  \quad \in \Omega^1_\text{basic}(\Phi),  \\[1mm]
  \bs\Theta^b_{\bs\omega,\, \Sigma} &= \bs\Theta_\Sigma  - \int_{\d\Sigma} \bs d \theta\big( \bs\omega; \phi \big) +  \int_{\Sigma}  \bs d E\big(\bs\omega; \phi \big),   \quad \in \Omega^2_\text{basic}(\Phi). 
 \end{aligned}
\end{equation}
which descend to $\M$, and on-shell turn $\M_\S$ into the desired reduced phase space associated with  the gauge theory $L$ over $\Sigma$. The presence of a boundary is no longer a problem. 

Notice how in the above construction of basic forms, only the Lie$\H$-linear part of $\bs E$ (and $\bs\theta$) contributes, so that the field equations for the matter fields are irrelevant to the scheme.\footnote{We could see this as another hint supporting the conceptual primacy/priority of the principal bundle $\P$ -- hence of gauge interactions -- over all associated bundles -- i.e. the matter fields. }

\paragraph{Ambiguity in the choice of connection} 

As the notation suggests though, the use of a connection $\bs\omega$ makes \eqref{omega-basic-E}, \eqref{omega-basic-theta} and \eqref{omega-basic-Theta}  ``coordinatisations" of the abstract basic objects $\bs E^b, \bs\theta^b$ and $\bs\Theta^b$. By the work done at the end of section \ref{Variational Ehresmann connections}, we easily find what happens under change of coordinatisation, i.e. under change of variational connection. 
As a special case of equation  \eqref{Ambiguity-basic-1-form}, and using again  \eqref{Noether-current}, we have that  basic presymplectic potentials built from connections $\bs \omega$ and $\bs\omega'$ s.t. $\bs\omega'=\bs\omega+\bs\beta$ with $\bs\beta \in \Omega^1_\text{tens}(\Phi, \text{Lie}\H)$, are related as
\begin{equation}
\begin{aligned}
\label{Ambig-basic-theta}
 \bs\theta^b_{\bs\omega'} &= \bs\theta^b_{\bs\omega}- J(\bs\beta; \phi), \\
  				       &= \bs\theta^b_{\bs\omega} - d \theta\big( \bs\beta; \phi \big) + E\big(\bs\beta; \phi \big).  \\[1mm]
\text{so that }\quad  \bs\theta^b_{\bs\omega'\!,\, \Sigma} &= \bs\theta^b_{\bs\omega, \, \Sigma}-   \int_{\d\Sigma} \theta\big( \bs\beta; \phi \big) +  \int_{\Sigma} E\big(\bs\beta; \phi \big).
\end{aligned}
\end{equation}
Similarly for basic field equation 1-forms, 
\begin{align}
\label{Ambig-basic-E}
 \bs E^b_{\bs\omega'} = \bs E^b_{\bs\omega} - dE(\bs\beta; \phi).
\end{align}
As a special case of  \eqref{Ambiguity-basic-2-form}, and following directly from \eqref{Ambig-basic-theta},  basic presymplectic 2-forms are related as, 
\begin{equation}
\begin{aligned}
\label{Ambig-basic-Theta}
 \bs\Theta^b_{\bs\omega'} &= \bs\Theta^b_{\bs\omega}- \bs d J(\bs\beta; \phi), \\
  				       &= \bs\Theta^b_{\bs\omega} - \bs d \left( d \theta\big( \bs\beta; \phi \big) - E\big(\bs\beta; \phi \big) \right).  \\[1mm]
\text{so that }\quad  \bs\Theta^b_{\bs\omega'\!,\, \Sigma} &= \bs\Theta^b_{\bs\omega, \, \Sigma}-   \int_{\d\Sigma}  \bs d \theta\big( \bs\beta; \phi \big) +  \int_{\Sigma} \bs d  E\big(\bs\beta; \phi \big).
\end{aligned}
\end{equation}
The ambiguity relations \eqref{Ambig-basic-theta}-\eqref{Ambig-basic-Theta} stems from the affine structure of the space of variational connections. It can be reasonably neglected if a preferred choice is available. Such would be the case in pure gauge theories since, as discussed in section \ref{Variational Ehresmann connections}, $\A$ has a distinguished connection $\bs\omega^{\mathring{\bs g}}$ associated with a natural bundle metric $\mathring{\bs g}$ (called the Singer-deWitt connection by Gomes-Riello). 

These ambiguity relations could also be interpreted as reflecting gluing properties: If one imagines that observers on regions $\Sigma'$ and $\Sigma$ separated by a boundary $\d\Sigma$  use different variational connections to build their respective basic presymplectic structures, then  \eqref{Ambig-basic-theta}-\eqref{Ambig-basic-Theta} -- with  $\Sigma$ on the left-hand side replaced by $\Sigma'$ -- are gluing relations between these structures. Thus understood, the above results generalise the discussion of section 6.7 in \cite{Gomes-Riello2021} on gluings of basic Yang-Mills presymplectic potentials built via  Singer-deWitt connections.


 \subsection{Via dressing fields}
\label{Via dressing fields}

We use the general results of section \ref{The dressing field method}  on the construction of basic forms relying on a field-dependent dressing fields $\bs u$. 
Applying the general formula \eqref{Dressed-variational-form}, and using our rule of thumb explained around  \eqref{GT-var-form} together with the results we obtained in section \ref{Presymplectic structure of invariant matter coupled gauge theories} for the $\bs\H$-gauge transformations of the field equations \eqref{Field-depGT-FieldEq}, the presymplectic potential \eqref{Field-depGT-presymp-pot-current}-\eqref{Field-depGT-presymp-pot} and the presymplectic 2-form \eqref{Field-depGT-Theta}-\eqref{Field-depGT-presymp-form}, we immediately get their associated dressed basic forms. First the dressed field equation,
\begin{align}
\label{dressed-E}
 \bs E^{\bs u} = \bs E + dE(\bs{duu}\-; \phi)   \quad  \in  \Omega^1_\text{basic}(\Phi).
\end{align}
Then the dressed presymplectic structure, 
\begin{equation}
\begin{aligned}
\label{dressed-presymp-struct}
 \bs\theta^{\bs u}_{\Sigma} &= \bs\theta_\Sigma  + \int_{\d\Sigma} \theta\big( \bs{duu}\-; \phi \big) -  \int_{\Sigma} E\big(\bs{duu}\-; \phi \big),  \quad \in \Omega^1_\text{basic}(\Phi),  \\[1mm]
  \bs\Theta^{\bs u}_{\Sigma} &= \bs\Theta_\Sigma  + \int_{\d\Sigma} \bs d \theta\big( \bs{duu}\-; \phi \big) -  \int_{\Sigma}  \bs d E\big(\bs{duu}\-; \phi \big),  \quad \in \Omega^2_\text{basic}(\Phi). 
 \end{aligned}
\end{equation}
As with \eqref{omega-basic-E}-\eqref{basic-presymp-struct}, these can be can be seen as realisations of the basic $\bs E^b, \bs\theta^b_\Sigma$ and $\bs\Theta^b_\Sigma$  associated with the invariant Lagrangian $L$ - as suggested in the  introduction to section \ref{Basic presymplectic structures}.
But a complementary viewpoint, central to the DFM philosophy, is available and worth stressing:  \eqref{dressed-E}-\eqref{dressed-presymp-struct}  are  the field equations and presymplectic structure associated with the \emph{dressed Lagrangian} 
\begin{align}
\label{dressed-inv-L}
L^{\bs u}\defeq \text{F}^\star_{\bs u} L = L \circ \text{F}_{\bs u} \quad  \in \Omega^0_\text{basic}(\Phi), \qquad \text{i.e.} \quad L^{\bs u}(\phi)=L(\phi^{\bs u}), 
\end{align}
and obtained in the standard way from $\bs d L^{\bs u}=\bs E^{\bs u}+ d\bs\theta^{\bs u}$, and $\bs\Theta^{\bs u}=\bs d \bs\theta^{\bs u}$. 
\begin{align}
\bs d L^{\bs u} = \bs E^{\bs u} + d \bs\theta^{\bs u} 
		      = E(\bs d\phi^{\bs u}; \phi ^{\bs u}) + d \theta(\bs d\phi^{\bs u}; \phi ^{\bs u})  \quad \in \Omega^1_\text{basic}(\Phi).  
\end{align}
The latter expression would allow to cross-check algebraically \eqref{dressed-E}-\eqref{dressed-presymp-struct} by inserting  \eqref{Dressed-dphi} and \eqref{Dressing-map}-\eqref{dressed-fields}  in~$E$~and~$\theta$. 

 This viewpoint is relevant to the question of ambiguity in the choice of dressing,  as we are about to discuss. But let us also remark that it clarifies the meaning of the \emph{edge mode} strategy as introduced by Donnelly \& Freidel \cite{DonnellyFreidel2016}, and applied in various contexts since  \cite{Geiller2017, Geiller2018, Speranza2018, Geiller2019,Teh-et-al2020, Freidel-et-al2020-1, Freidel-et-al2020-2, Freidel-et-al2020-3}, where dressing fields are known as `edge modes'.
As argued in \cite{Francois2021}, the DFM is the geometric foundation of this strategy. Taking indeed \eqref{dressed-presymp-struct}  on-shell, it may seem that  $\bs u$ needs only to live on $\d\Sigma$, hence the name `edge mode' it received elsewhere. But considering the boundary as a fictitious one,  being arbitrarily moved around, we see that actually $\bs u$ must in general be defined across $\Sigma$. This indeed makes all the more sense considering that $\bs u$ is built from $\phi$ which is  defined across $\Sigma$.
\medskip

The striking similarity between \eqref{dressed-E}-\eqref{dressed-presymp-struct} and \eqref{omega-basic-E}-\eqref{basic-presymp-struct} is of course no accident. As we observed in section \ref{Field-dependent dressing fields and variational connections}, the 1-form $\mathring{\bs\omega}\defeq -\bs{duu}\-$ is a flat variational Ehresmann connection. 
So, regarding the question of realising the basic presymplectic structure of invariant gauge theories, the dressed structure  \eqref{dressed-E}-\eqref{dressed-presymp-struct} can be seen as a special case of \eqref{omega-basic-E}-\eqref{basic-presymp-struct} involving a flat connection. This generalises the observation of Gomes-Riello \cite{Gomes-Riello2018} according to which the edge mode strategy of Donnelly-Freidel as applied to YM theory could be seen as a special case of their use of a connection. 

Yet, there is a noticeable difference in what can be done via the DFM that wouldn't be accessible through the use of a variational connection, and this relates to how ambiguities in the respective schemes arise.

\paragraph{Ambiguity in the choice of dressing field and residual $\bs\G$-transformations} 

Again, as the notation suggests, \eqref{dressed-E}-\eqref{dressed-presymp-struct}  are  ``coordinatisations" of the abstract basic objects $\bs E^b, \bs\theta^b$ and $\bs\Theta^b$. From sections \ref{The dressing field method} and  \ref{Field-dependent dressing fields and variational connections}, we know what happens under change of coordinatisation, i.e. under change of dressing field 
$\bs u', \bs u :\Phi \rarrow \D r[H, G]$. 

We have already seen that such a change is, in the most general case, of the form $\bs u '=\bs u \bs \xi$ for some $\phi$-dependent $\G$-valued map $\bs\xi$ s.t. $R^\star_\gamma \bs\xi=\bs\xi$. 
By application of \eqref{Ambiguity-basic-1-form-dressing} (or \eqref{Dressed-alpha-res-2nd-kind-inv2}),  we immediately get the relations 
\begin{equation}
\begin{aligned}
\label{Ambig-dressed-E-theta}
( \bs E^{\bs u})^{\bs \xi} &= \bs E^{\bs u} - dE(\mathring{\bs\beta}; \phi), \\
(\bs\theta^{\bs u}_\Sigma)^{\bs \xi} &= \bs\theta^{\bs u}_\Sigma -   \int_{\d\Sigma} \theta\big(\mathring{ \bs\beta}; \phi \big) +  \int_{\Sigma} E\big( \mathring{\bs\beta}; \phi \big),
\end{aligned}
\end{equation}
with $\mathring{ \bs\beta}=- \bs u \bs{d\xi\xi\-} \bs u\- \in \Omega^1_\text{tens}(\Phi)$. From which follows,
\begin{align}
\label{Ambig-dressed-Theta}
(\bs\Theta^{\bs u}_\Sigma)^{\bs \xi} = \bs\Theta^{\bs u}_\Sigma -   \int_{\d\Sigma} \bs d \theta\big(\mathring{ \bs\beta}; \phi \big) +  \int_{\Sigma} \bs d E\big( \mathring{\bs\beta}; \phi \big).
\end{align}
These can of course be interpreted as special cases of \eqref{Ambig-basic-theta}-\eqref{Ambig-basic-Theta} since change of dressing fields reflects a case of the affine character of connection space, $\mathring{\bs\omega}' = \mathring{\bs\omega} + \mathring{\bs\beta}$. But there is more to it. 
\medskip

In section \ref{The dressing field method}, we made the case that dressed variational forms $\bs\alpha^{\bs u}$, basic on $\Phi$, can be seen as forms on the $\G$-bundle of dressed fields $\Phi^{\bs u}$, arising from the ambiguity in the choice of dressing. 
From that point of view, the above relations are transformations under the gauge group  $\bs\G$ of $\Phi^{\bs u}$ -- remember the SES  \eqref{SESgroups-dressed}.
 Recalling indeed that the first version of  eq.\eqref{Dressed-alpha-res-2nd-kind-inv2}-\eqref{Ambiguity-basic-1-form-dressing} is the general equation \eqref{Dressed-alpha-res-2nd-kind-inv}, we apply the latter to rewrite \eqref{Ambig-dressed-E-theta}-\eqref{Ambig-dressed-Theta} as
 \begin{equation}
 \label{Ambig-dress-as-GT}
 \begin{aligned}
( \bs E^{\bs u})^{\bs \xi} &= \bs E^{\bs u} + dE( \bs{d\xi\xi\-} ; \phi^{\bs u}), \\
(\bs\theta^{\bs u}_\Sigma)^{\bs \xi} &= \bs\theta^{\bs u}_\Sigma +   \int_{\d\Sigma} \theta\big( \bs{d\xi\xi\-} ; \phi^{\bs u} \big) -  \int_{\Sigma} E\big(  \bs{d\xi\xi\-} ; \phi^{\bs u}\big), \\
(\bs\Theta^{\bs u}_\Sigma)^{\bs \xi} &= \bs\Theta^{\bs u}_\Sigma +   \int_{\d\Sigma} \bs d \theta\big( \bs{d\xi\xi\-} ; \phi^{\bs u} \big) -  \int_{\Sigma} \bs d E\big( \bs{d\xi\xi\-} ; \phi^{\bs u} \big).
\end{aligned}
\end{equation}
which, of course, look exactly like the $\bs\H$-transformations of $\bs E$ \eqref{Field-depGT-FieldEq}, $\bs\theta_\Sigma$  \eqref{Field-depGT-presymp-pot}  and $\bs\Theta_\Sigma$ \eqref{Field-depGT-presymp-form}. What is especially interesting though is that since the dressing ambiguity is encoded by a transformation group $\G$, which is the structure group of the bundle $\Phi^{\bs u}$, associated Noether charges and their Poisson bracket can be defined. 

Indeed,  as  by \eqref{Residual-2} we have  that $R_\xi \phi^{\bs u}$, $\xi \in \G$ is formally identical to $R_\gamma \phi$, $\gamma \in \H$, and since obviously $L$ and $L^{\bs u}$ have the same functional properties, among which invariance, it follows that $R^\star_\xi L^{\bs u}=L^{\bs u}$, i.e. $L^{\bs u} \in\Omega^0_\text{basic}(\Phi^{\bs u})$. Therefore, $\bs dL^{\bs u}  \in \Omega^1_\text{basic}(\Phi^{\bs u})$ and  on $\Phi^{\bs u}$ we get,
\begin{align}
\bs d L^{\bs u} = \bs E^{\bs u} + d \bs \theta^{\bs u} = E(\bs d \phi^{\bs u};  \phi^{\bs u}) + d \theta(\bs d \phi^{\bs u};  \phi^{\bs u}). 
\end{align}
From which is defined $\bs\Theta^{\bs u}\defeq \bs d \bs \theta^{\bs u}$. From there, the whole covariant phase space approach on $\Phi^u$ can be run through as in section \ref{Presymplectic structure of invariant matter coupled gauge theories}. 
We immediately get to write down the dressed Noether charges associate to $\vkappa \in$ Lie$\G$, generating the vertical vector fields $\vkappa^v \in \Gamma(V\Phi^{\bs u})$:
\begin{align}
\label{dressed-Noether-charge}
Q_\Sigma(\vkappa; \phi^{\bs u})= \int_{\d\Sigma} \theta(\vkappa; \phi^{\bs u}) - \int_\Sigma E(\vkappa; \phi^{\bs u}). 
\end{align}
These are related to the dressed presymplectic 2-form as $\iota_{\vkappa^v}\bs\Theta^{\bs u}_\Sigma = -\bs dQ_\Sigma(\vkappa; \phi^{\bs u})$ - $\vkappa$ is field-independent so the charges are integrable -
so that a Poisson bracket is defined the usual way:
\begin{align}
\label{dressed-Poisson-bracket}
\big\{ Q_\Sigma(\vkappa; \phi^{\bs u}) ,  Q_\Sigma(\vkappa'; \phi^{\bs u})\big\}\defeq\, \bs\Theta^{\bs u}_\Sigma\big(\vkappa^v, {\vkappa'}^v\big) =	 Q_\Sigma\big([\vkappa, \vkappa']; \phi^{\bs u}\big).
\end{align}
The Poisson algebra of dressed Noether charges is then isomorphic to Lie$\G$, and
 infinitesimal \mbox{$\G$-transformations} (of objects on $\Phi^{\bs u}$) can be  generated  via $\big\{ Q_\Sigma(\alpha; A^{\bs u}) , \  \ \big\}$.

 Equations \eqref{Ambig-dress-as-GT} are, as we said,  $\phi^{\bs u}$-dependent $\bs\G$-gauge transformations. They are used to extend the above to dressed charges associated with $\bs\vkappa \in$ Lie$\bs\G$,  for which we get 
 \begin{align}
\label{dressed-Noether-charge-field-dep}
Q_\Sigma(\bs\vkappa; \phi^{\bs u})= \int_{\d\Sigma} \theta(\bs\vkappa; \phi^{\bs u}) - \int_\Sigma E(\bs \vkappa; \phi^{\bs u}), 
\quad \text{which are s.t.} \quad \iota_{\bs\vkappa^v}\bs\Theta^{\bs u}_\Sigma& = -\bs dQ_\Sigma(\bs\vkappa; \phi^{\bs u}) + \bs dQ_\Sigma(\bs d \bs\vkappa; \phi^{\bs u}), \\[-2mm]
														      &=-\bs dQ_\Sigma( \munderline{red}{\bs\vkappa}; \phi^{\bs u}). \notag
\end{align}
These non-integrable dressed charges still satisfy a well-behaved Poisson bracket, 
\begin{align}
\label{dressed-Poisson-bracket-field-dep}
\big\{ Q_\Sigma(\bs\vkappa; \phi^{\bs u}) ,  Q_\Sigma(\bs\vkappa'; \phi^{\bs u})\big\}\defeq\, \bs\Theta^{\bs u}_\Sigma\big(\bs\vkappa^v, {\bs\vkappa'}^v\big) =	 Q_\Sigma\big([\bs\vkappa, \bs\vkappa']; \phi^{\bs u}\big).
\end{align}

As to the matter of the physical interpretation of dressed charges, and their observability, two cases occur. If~$\bs u=\bs u(\vphi)$, the ansatz \eqref{Perturbation-ansatz} implies its dressed version
\begin{align}
\label{Perturbation-ansatz-dressed}
A^{\bs u}=A_0^{\bs u} + \alpha^{\bs u} \ \left( \defeq (\bs u\- A_0 \bs u + \bs u\- d\bs u) + (\bs u\- \alpha \bs u) \right), 
\end{align}
which  can  simply  be plugged into \eqref{dressed-Noether-charge}, so that if we further declare the Killing equation $D^{A_0^{\bs u}} \vkappa \equiv 0$ valid, $Q_\Sigma(\vkappa; \phi^{\bs u})$ is interpretable as a  charge associated with the symmetries of the background field $A_0^{\bs u}$ conserved on-shell. It  splits as a constant contribution from $A_0^{\bs u}$, against which is measured/observed the contribution from $\alpha^{\bs u}$ (and $\vphi^{\bs u}$). 


In case  $\bs u=\bs u(A)$, special care must be taken as the dressing field is then itself affected by the affine ansatz \eqref{Perturbation-ansatz}. Splitting of the charge into  background and physical contributions may be trickier.  
\smallskip

None of this is  available when using connections to realise the basic presymplectic structure of $L$, for which the ambiguity -- or change of ``coordinatisation'' -- is not in general captured/parametrised by a group $\G$. 

\paragraph{Comments 1}

In the edge mode literature, the dressed presymplectic structure \eqref{dressed-presymp-struct} goes by the name of \emph{extended presymplectic structure}, as edge modes are seen as \emph{new} degrees of freedom living at the boundary. The interpretation being that $\d\Sigma$ breaks gauge invariance and that edge modes are kind of Goldstone bosons. Yet the DFM shows \cite{Francois2021} that if a $\phi$-independent dressing field $u$ is introduced by fiat in a theory -- as its interpretation as new d.o.f. implies -- it means that the underlying bundle one works with is trivial and that one actually has $\G\simeq \H$, i.e. the `ambiguity' symmetry $\G$ is just the original gauge symmetry $\H$ in another guise. This would be a challenge to the interpretation of $\G$ as a new symmetry stemming from the introduction of edge modes as new d.o.f.

Another notion found in the edge mode literature that we are here bound to challenge, is the interpretation of $\G$ -- and $\bs\G$ -- as a \emph{physical} transformation group, usually referred to as \emph{surface} or \emph{boundary symmetry},\footnote{Again, because $\bs u$ is usually seen as living on $\d\Sigma$ when introduced to restore the invariance (horizontality) of on-shell quantities, so that $\xi\in \G$ is seen as a map $\xi:\d\Sigma \rarrow G$.}
 insofar as it is seen as a permutation group of physical degrees of freedom. 
Seing that $\phi^{\bs u}$ are $\H$-invariant fields, thus potentially physical d.o.f., eq.\eqref{Residual-2}  would  indeed seem to suggest that $\G$ transforms physical field configurations, making it literally a physical symmetry. Yet as we have argued in section \ref{The dressing field method}  and reminded above, $\G$ is the structure group of $\Phi^{\bs u}$, whose base $\M^{\bs u}$ is isomorphic to the physical configuration space $\M$ (the base of the $\H$-bundle $\Phi$). And as the SES  \eqref{SESgroups-dressed} shows, $\G$ acts trivially on $\M^{\bs u}\simeq\!\M$ and therefore cannot be a physical symmetry as understood in the edge mode literature. Physical symmetries, understood as permutation of physical d.o.f., rather belong to (subgroups of) $\bs \Diff(\M^{\bs u}) \simeq \bs \Diff(\M)$. 
This is not in contradiction with the above discussion on the physical interpretability of charges associated with $\G$. It simply means that the problem of the physical relevance of $\G$ as a symmetry group is of the same nature as that of the original gauge group $\H$, and does not enjoy a more immediate physical interpretation. 

\paragraph{Comments 2}

Regarding the aim of building a basic symplectic structure associated with a theory $L$,  the existence of $\G$ as a symmetry of the dressed theory $L^{\bs u}$ spoils everything: the boundary problem posed by the $\H$-symmetry of $L$ as been solved, but as is clear from  \eqref{Ambig-dress-as-GT}, a boundary problem reemerges w.r.t. $\G$. 
This reflects the two options we had advertised (below\eqref{SESLieAlg-dressed}):   Either  1) the constructive procedure of building a dressing field $\bs u$ from the field content $\phi$ is free (enough) of ambiguity so that $\G$ is `small', perhaps reduced to a rigid or discret group, in which case the boundary problem can be considered solved.  Or 2) $\G$ is indeed a new meaningful gauge symmetry giving rise to its own boundary problem, but whose physical relevance may manifests through charges.

Yet even in the first case an important remark should be raised. It the constructive procedure ends-up producing a dressing field $\bs u$ which is \emph{local} (in the sense of field theory), then the theory can be  rewritten so that each individual field variable $\phi^{\bs u}$ is a gauge singlet yet still remains local. If the gauge symmetry of a theory can be thus eliminated without losing locality, it is said to be an \emph{artificial gauge symmetry} \cite{Pitts2009} (the terminology ``fake symmetry" of Jackiw \& Pi \cite{Jackiw-Pi2015} covers the same notion). On the contrary, if it happens that no dressing field can be found, or that only \emph{non-local} ones can be produced, then the dressed variables $\phi^{\bs u}$ are $\H$-invariant but non-local. In such theories, the gauge symmetry is only eliminated at the cost of locality, and this is usually recognised as hallmark and physical signature of \emph{substantial gauge symmetries}.  This distinction between two classes of gauge symmetries, one physically relevant the other not, generalises the well-know distinction between artificial and substantive general covariance familiar in the foundation of general relativistic physics \cite{Pitts2009}. The DFM methods can then be used as  a tool to assess the nature of the gauge symmetry of a theory  \cite{Francois2018, Berghofer-et-al2021}.  

Arguably there is no genuine boundary problem in a theory with an artificial gauge symmetry, or if there is, the dressed presymplectic structure \eqref{dressed-presymp-struct} solves it in case 1). 
A  genuine boundary problem arises only for substantial gauge symmetries -- be it the original $\H$ or the new $\G$, case 2) -- and the fact that it is only solved by sacrificing the locality of the theory may be seen as yet another signal of the non-local -- or non-separable -- character of the physics described by (true) gauge field theories  \cite{Lyre2004, Guay2008, Healey2009, Dougherty2017, Nguyen-et-al2017, Teh2020}. 
\bigskip

Let us know consider immediate applications of the last two sections to Yang-Mills theory and 4D gauge gravity, which will illustrate in particular the above discussion.


  \subsection{Applications}
\label{Applications}

\subsubsection{Yang-Mills theory}
\label{Yang-Mills theory}

We here start from, and rely on, the results of section \ref{Coupled Yang-Mills theory}. We have  $H=S\!U(n)$, $\H=\SU(n)=\G$. 

\paragraph{Basic with connections}
As the Lagrangian of the theory is basic, $L_\text{\tiny YM}\in \Omega^0_\text{basic}(\Phi)$, the field equations and the presymplectic structure are invariant, 
$\bs E_\text{\tiny YM}, \bs\theta^\text{\tiny YM}_\Sigma,  \bs\Theta^\text{\tiny YM}_\Sigma \in \Omega^\bullet_\text{inv}(\Phi)$. 
Therefore, given a connection $\bs\omega \in \Omega^1_\text{eq}(\Phi, \text{Lie}\H)$, one can proceed and write down the corresponding basic versions. 

By application of \eqref{omega-basic-E} and given \eqref{YM-E}, we get the basic field equations
\begin{align}
\label{Basic-YM-E}
\bs E^\text{\tiny YM}_{\bs\omega}&= \bs E_\text{\tiny YM} - d E_\text{\tiny YM}(\bs\omega; \phi), \notag\\
						  &=\bs E^\text{\tiny YM} - d \Tr\big(  \bs\omega \,\{D*\!F-J \}  \big), \quad  \in \Omega^1_\text{basic}(\Phi).
\end{align}
By \eqref{basic-presymp-struct} and given  \eqref{YM-E}-\eqref{YM-theta}, we obtain immediately the basic presymplectic structure, 
\begin{align}
 \bs\theta^\text{\tiny YM}_{\bs\omega,\, \Sigma} &= \bs\theta_\Sigma  - \int_{\d\Sigma} \theta_\text{\tiny YM}\big( \bs\omega; \phi \big) +  \int_{\Sigma} E_\text{\tiny YM}\big(\bs\omega; \phi \big)  \quad \in \Omega^1_\text{basic}(\Phi), \notag \\
 							&= \bs \theta^\text{\tiny YM}_\Sigma - \int_{\d \Sigma }\Tr\big(\bs\omega *\!F \big) + \int_\Sigma\Tr\big( \bs\omega\, \{D\!*\!F-J\}\big),  \label{YM-basic-theta}  \\[1.5mm]
  \bs\Theta^\text{\tiny YM}_{\bs\omega,\, \Sigma} &= \bs\Theta_\Sigma  - \int_{\d\Sigma} \bs d \theta_\text{\tiny YM}\big( \bs\omega; \phi \big) +  \int_{\Sigma}  \bs d E_\text{\tiny YM}\big(\bs\omega; \phi \big)   \quad \in \Omega^2_\text{basic}(\Phi), \notag \\
  							&= \bs \Theta^\text{\tiny YM}_\Sigma - \int_{\d \Sigma } \bs d \Tr\big(\bs\omega *\!F \big) + \int_\Sigma \bs d \Tr\big( \bs\omega\, \{D\!*\!F-J\}\big),  \label{YM-basic-Theta} 
\end{align}
This reproduces  eq.(6.28)-(6.33) of \cite{Gomes-et-al2018} (who consider coupling to spinors rather than to scalar fields, which changes nothing of substance). 
Eq.\eqref{YM-basic-theta}-\eqref{YM-basic-Theta} would be one coordinatisation of the  phase space of YM theory over $\Sigma$,  $(\M_\S, \bs\Theta^b_\Sigma)_\text{\tiny YM}$. Others are obtained under change of connection $\bs\omega'=\bs\omega + \bs \beta$, as by \eqref{Ambig-basic-E}, \eqref{Ambig-basic-theta} and \eqref{Ambig-basic-Theta} we have: 
\begin{equation}
\begin{aligned}
\label{Ambig-basic-presymp-struc-YM}
\bs E^\text{\tiny YM}_{\bs\omega'}  &=\bs E^\text{\tiny YM}_{\bs\omega} - d \Tr\big(  \bs\beta \,\{D*\!F-J \}  \big),  \\[1mm]
 \bs\theta^\text{\tiny YM}_{\bs\omega',\, \Sigma} &=  \bs\theta^\text{\tiny YM}_{\bs\omega,\, \Sigma} - \int_{\d \Sigma }\Tr\big(\bs\beta\, *\!F \big) + \int_\Sigma\Tr\big( \bs\beta\, \{D\!*\!F-J\}\big),   \\[1.5mm]
  \bs\Theta^\text{\tiny YM}_{\bs\omega',\, \Sigma} &=  \bs\Theta^\text{\tiny YM}_{\bs\omega,\, \Sigma} - \int_{\d \Sigma } \bs d \Tr\big(\bs\beta\, *\!F \big) + \int_\Sigma \bs d \Tr\big( \bs\beta\, \{D\!*\!F-J\}\big). 
\end{aligned}
\end{equation}
As we commented at the end of section \ref{Via variational connections}, these can also be seen a gluing relations of a sort between the basic objects constructed by two observers on each side of a region partitioned in two subregions by a boundary $\d\Sigma$. Thus interpreted, \eqref{Ambig-basic-presymp-struc-YM} -- the second line in particular -- reproduces (on-shell) the results of section 6.7 of \cite{Gomes-Riello2021} and 
section 5.3 of \cite{Riello2021bis} (see in particular eq.(83), in the free abelian case). 

Let us remark that the relation between $\bs{D^\omega}\bs\theta^\text{\tiny YM}_\Sigma$ and $ \bs\Theta^\text{\tiny YM}_{\bs\omega, \Sigma}$, both basic 2-forms, is immediately read-off \eqref{Diff-cov-der-theta-Theta-basic} -- itself a special case of \eqref{Formula1} -- to be:
\begin{align}
\label{YM-bidule}
\bs{D^\omega}\bs\theta^\text{\tiny YM}_\Sigma &=  \bs\Theta^\text{\tiny YM}_{\bs\omega, \Sigma} + \int_{\d \Sigma } \theta_\text{\tiny YM} (\bs\Omega; \phi) - \int_{\Sigma}E_\text{\tiny YM}(\bs\Omega; \phi),  \notag\\
									&=  \bs\Theta^\text{\tiny YM}_{\bs\omega, \Sigma} +  \int_{\d \Sigma }\Tr\big(\bs\Omega\, *\!F \big) - \int_\Sigma\Tr\big( \bs\Omega\, \{D\!*\!F-J\}\big),
\end{align}
with $\bs\Omega \in \Omega^2_\text{tens}(\Phi, \text{Lie}\H)$ the curvature of $\bs\omega$. This reproduces (on-shell) the corollary 3.2 and the equation in section 3.4 of  \cite{Gomes-Riello2021} (also found in footnote 27 of \cite{Riello2021}), see also eq.(6.31)-(6.32) of \cite{Gomes-et-al2018}.

\paragraph{Basic with dressing fields}
As we know by now, the existence of a $\phi$-dependent dressing field $\bs u : \Phi \rarrow \D r[H,H]$ induces the existence of a flat connection $\mathring{\bs\omega}\defeq -\bs{duu}\-$, and another choice of dressing field $\bs u'=\bs u \bs\xi$ induces $\mathring{\bs\omega}'=\mathring{\bs\omega} + \mathring{\bs \beta}$. So,~we could simply say that all of the above formulae specialise to $\bs\omega, \bs\beta \rarrow \mathring{\bs\omega},  \mathring{\bs\beta}$,\footnote{With in particular the remark that \eqref{YM-bidule} specialises to $\bs D^{\mathring{\bs\omega}} \bs\theta^\text{\tiny YM}_\Sigma =  \bs\Theta^\text{\tiny YM}_{\mathring{\bs\omega}, \Sigma}$, i.e. the dressed presymplectic 2-form (right) coincides with the $\mathring{\bs\omega}$-covariant derivative of the original potential $\bs\theta^\text{\tiny YM}_\Sigma$.}
and leave it there. 

But there is of course more to say. By \eqref{dressed-E}-\eqref{dressed-presymp-struct} and given  \eqref{YM-E}-\eqref{YM-theta},\footnote{Or using \eqref{GT-E-YM} and \eqref{GT-thetaYM}-\eqref{GT-ThetaYM} together with the rule of thumb $\bs \gamma \rarrow \bs u$.} we have
\begin{align}
 \bs E_\text{\tiny YM}^{\bs u}&= \bs E_\text{\tiny YM} + dE_\text{\tiny YM} \big(  \bs{duu}\- ;  \phi \big) \quad \in \Omega^1_\text{basic}(\Phi),  \notag \\
 					   &= \bs E_\text{\tiny YM}+ d\Tr\big(  \bs{duu}\- \{D*\!F-J \}  \big), \label{Dressed-E-YM} \\
(\bs\theta^\text{\tiny YM}_\Sigma)^{\bs u}&= \bs \theta^\text{\tiny YM}_\Sigma + \int_{\d\Sigma}\theta_\text{\tiny YM} \big(  \bs{duu}\- ;  \phi \big) - \int_\Sigma E_\text{\tiny YM} \big( \bs{duu}\- ; \phi \big)  \quad \in \Omega^1_\text{basic}(\Phi),\notag \\
            						     &= \bs \theta^\text{\tiny YM}_\Sigma + \int_{\d \Sigma }\Tr\big(\bs{duu}\- *\!F \big) - \int_\Sigma\Tr\big( \bs{duu}\-\, \{D\!*\!F-J\}\big),   \label{Dressed-theta-YM}  \\[1mm]
(\bs\Theta^\text{\tiny YM}_\Sigma)^{\bs u} &= \bs\Theta^\text{\tiny YM}_\Sigma +  \int_{\d \Sigma}\bs d \theta_\text{\tiny YM}\big(\bs{duu}\-; \phi\big) - \int_\Sigma \bs d E_\text{\tiny YM}\big(\bs{duu}\-; \phi\big)  \quad \in \Omega^2_\text{basic}(\Phi), \notag\\
				                    &= \bs\Theta^\text{\tiny YM}_\Sigma + \int_{\d \Sigma} \bs d \Tr\big( \bs{duu}\- *\!F \big)  - \int_\Sigma \bs d  \Tr\big(  \bs{duu}\-\, \{D\!*\!F-J\} \big).  \label{Dressed-Theta-YM} 
\end{align}
Observe again how only the Lie$\H$-linear part of $\bs E$ (and $\bs\theta$) contributes here, so that the field equations for the matter fields are irrelevant. 
Eq.\eqref{Dressed-theta-YM}-\eqref{Dressed-Theta-YM} generalise, on-shell, eq.(2.19) and eq.(2.22)-(2.23) in \cite{DonnellyFreidel2016}. 
As observed in section \ref{Via dressing fields},  these are the field equations and presymplectic structure of the dressed Lagrangian,
\begin{align}
\label{Dressed-L-YM}
L^{\bs u}_\text{\tiny YM} = \text{F}_{\bs u}^\star L_\text{\tiny YM} \quad \text{i.e.}\quad L^{\bs u}_\text{\tiny YM}(\phi)= L_\text{\tiny YM}(\phi^{\bs u})=\tfrac{1}{2}\Tr(F^{\bs u} *\!F^{\bs u})+ \tfrac{1}{2}\langle D^{\bs u}\vphi^{\bs u}, *D^{\bs u}\vphi^{\bs u} \rangle, 
\end{align}
seen as an invariant 0-form on $\Phi^{\bs u}$, and obtained via 
\begin{align}
\label{dressed-dL-YM}
\bs d L^{\bs u}_\text{\tiny YM} = \bs E^{\bs u}_\text{\tiny YM} + d \bs\theta^{\bs u}_\text{\tiny YM}
						= E_\text{\tiny YM}(\bs d\phi^{\bs u}; \phi ^{\bs u}) + d \theta_\text{\tiny YM}(\bs d\phi^{\bs u}; \phi ^{\bs u}).  
\end{align}


Now, the a priori ambiguity in the choice/construction of a dressing is parametrised by a group $\G=\SU(n)$ (still) which is the  structure group of $\Phi^{\bs u}$ and a a symmetry of $L^{\bs u}_\text{\tiny YM}$. We can thus associate to it Noether charges which, for $\vkappa \in$ Lie$\G$ inducing $\vkappa^v \in \Gamma(V\Phi^{\bs u})$, are immediately given by \eqref{dressed-Noether-charge}
\begin{equation}
\label{YM-dressed-charge}
\begin{aligned}
Q^\text{\tiny YM}_\Sigma(\vkappa; \phi^{\bs u})&= \int_{\d\Sigma} \theta_\text{\tiny YM}(\vkappa; \phi^{\bs u}) - \int_\Sigma E_\text{\tiny YM}(\vkappa; \phi^{\bs u}), \qquad \text{s.t.} \quad  \iota_{\vkappa^v}  (\bs\Theta^\text{\tiny YM}_\Sigma)^{\bs u} = - \bs d Q^\text{\tiny YM}_\Sigma(\vkappa; \phi^{\bs u}),  \\
						&= \int_{\d\Sigma} \Tr\big(\vkappa *\!F^{\bs u} \big) - \int_\Sigma \Tr\big( \vkappa\,  \{D^{\bs u}\!*\!F^{\bs u} -  J^{\bs u} \}  \big).
\end{aligned}
\end{equation}
where $ J^{\bs u}=  |\!*\!D^{\bs u} \vphi^{\bs u}\rangle \langle \vphi^{\bs u}| =  | \rho(\bs u)\-\!*\!D\vphi\rangle \langle \rho(\bs u)\-\vphi|=u\- | \!*\!D\vphi\rangle \langle \vphi| u =\bs u\- J \bs u$.  
Quite naturally, the dressed 2-form $(\bs\Theta^\text{\tiny YM}_\Sigma)^{\bs u}$ induce a Poisson bracket for these \emph{dressed} charges which is by \eqref{dressed-Poisson-bracket}  
\begin{align}
\label{dressed-Poisson-bracket-YM}
\big\{ Q^\text{\tiny YM}_\Sigma(\vkappa; \phi^{\bs u}) ,  Q^\text{\tiny YM}_\Sigma(\vkappa'; \phi^{\bs u})\big\}\defeq\, (\bs\Theta^\text{\tiny YM}_\Sigma)^{\bs u}\big(\vkappa^v, {\vkappa'}^v\big) = Q^\text{\tiny YM}_\Sigma\big([\vkappa, \vkappa']; \phi^{\bs u}\big).
\end{align}
This generalises eq.(2.35)-(2.36)-(2.38) of \cite{DonnellyFreidel2016}. Comparable formulae holds for field-dependent gauge parameter $\bs\vkappa \in$ Lie$\bs\G$, despite the fact that this time the dressed charges are non-integrable so that,
\begin{align}
 \iota_{\bs\vkappa^v}  (\bs\Theta^\text{\tiny YM}_\Sigma)^{\bs u} 
 		&= - \bs d Q^\text{\tiny YM}_\Sigma(\bs\vkappa; \phi^{\bs u}) +\bs d Q^\text{\tiny YM}_\Sigma( \bs d \bs\vkappa; \phi^{\bs u})= -\bs dQ^\text{\tiny YM}_\Sigma( \munderline{red}{\bs\vkappa}; \phi^{\bs u}), \\
		 &=-\int_{\d\Sigma} \Tr\big(\bs\vkappa *\!\bs dF^{\bs u} \big) + \int_\Sigma \Tr\big( \bs\vkappa\, \bs d  \{D^{\bs u}\!*\!F^{\bs u} -  J^{\bs u} \}  \big). \notag
\end{align}
The dressed charge \eqref{YM-dressed-charge} could gain a clearer physical status by plugging the affine ansatz $A^{\bs u}=A^{\bs u}_0+\alpha^{\bs u}$ and declaring $\vkappa$ a symmetry of the background $A^{\bs u}_0$, so that similarly to \eqref{Noether-charge-YM-perturb} we get on-shell
 \begin{align}
\label{Noether-charge-YM-perturb-dress}
 Q^\text{\tiny YM}_\Sigma(\vkappa; \phi^{\bs u})= \int_{\d\Sigma} \theta_\text{\tiny YM}(\vkappa; \phi^{\bs u})_{\ |\S} 
			  &= \int_{\d\Sigma} \Tr\big(\vkappa *\! \{ F_0^{\bs u} + f^{\bs u} + \tfrac{1}{2}[\alpha^{\bs u}, \alpha^{\bs u}] \} \big),  \notag \\ 
			  &= \int_{\d\Sigma} \Tr\big(\vkappa *\! F_0^{\bs u}\big)  + \Tr\big(\vkappa *\! f^{\bs u}\big) \rdefeq Q^\text{\tiny YM}_\Sigma(\vkappa; A_0^{\bs u}) + Q^\text{\tiny YM}_\Sigma(\vkappa; \alpha^{\bs u}),
\end{align} 
with the second term interpreted as the contribution of the perturbation measured against the background. 
Of course, if the dressing field is built from the connection, it is  itself affected by the affine ansatz \eqref{Perturbation-ansatz} and special care must be applied as the charge may not be as simply written as in \eqref{Noether-charge-YM-perturb-dress}.

Seen as forms on the $\G$-bundle  $\Phi^{\bs u}$, \eqref{Dressed-E-YM}-\eqref{Dressed-Theta-YM}  of course transform under its gauge group $\bs\G$ by  \eqref{Ambig-dress-as-GT} as, 
\begin{equation}
\label{Ambig-dressing-YM}
\begin{aligned}
 [\bs E_\text{\tiny YM}^{\bs u}]^{\bs\xi} &= \bs E_\text{\tiny YM}^{\bs u} + d\Tr\big(  \bs{d\xi \xi}\- \{D^{\bs u}\!*\!F^{\bs u}-J^{\bs u} \}  \big),  \\
							 &= \bs E_\text{\tiny YM}^{\bs u} - d\Tr\big(  \mathring{\bs\beta}\, \{D\!*\!F-J \}  \big), \\[2mm]
[(\bs\theta^\text{\tiny YM}_\Sigma)^{\bs u}]^{\bs \xi} &= (\bs \theta^\text{\tiny YM}_\Sigma)^{\bs u} + \int_{\d \Sigma }\Tr\big(\bs{d\xi\xi}\- *\!F^{\bs u} \big) - \int_\Sigma\Tr\big( \bs{d\xi\xi}\-\, \{D^{\bs u}\!*\!F^{\bs u}-J^{\bs u}\}\big),    \\[1mm]
										 &= (\bs \theta^\text{\tiny YM}_\Sigma)^{\bs u} - \int_{\d \Sigma }\Tr\big(\mathring{\bs\beta}\, *\!F \big) + \int_\Sigma\Tr\big(\mathring{\bs\beta}\, \{D\!*\!F-J \}\big),    \\[2mm]
[(\bs\Theta^\text{\tiny YM}_\Sigma)^{\bs u}]^{\bs \xi} &= (\bs\Theta^\text{\tiny YM}_\Sigma)^{\bs u} + \int_{\d \Sigma} \bs d \Tr\big( \bs{d\xi\xi}\- *\!F^{\bs u} \big)  - \int_\Sigma \bs d  \Tr\big(  \bs{d\xi\xi}\-\, \{D^{\bs u}\!*\!F^{\bs u}-J^{\bs u}\} \big), \\[1mm]
										&= (\bs\Theta^\text{\tiny YM}_\Sigma)^{\bs u} - \int_{\d \Sigma} \bs d \Tr\big(\mathring{\bs\beta}\, *\!F \big)  + \int_\Sigma \bs d  \Tr\big( \mathring{\bs\beta}\, \{D\!*\!F-J\} \big). 
\end{aligned}
\end{equation}
The display of each second line, where $\mathring{\bs\beta}= - \bs u \bs{d\xi\xi\-} \bs u\-$,  invites comparison with \eqref{Ambig-basic-presymp-struc-YM} and illustrates \eqref{Ambig-dressed-E-theta}-\eqref{Ambig-dressed-Theta}. 
These show that a boundary problem may reappear w.r.t. $\G$. Comments are thus in order. 

\paragraph{Comments}

We here have an occasion to illustrate the discussion held at the end of section \ref{Via dressing fields} on the local \emph{vs} non-local dressing fields in a model and the nature of its gauge symmetry.

In pure YM theory, there are no local dressing fields that can be built from the gauge potential (as far as is known). Only non-local dressings $\bs u=\bs u(A)$ seem possible, and basically related to holonomies of the connection. This is also true in YM theory coupled to spinors, as no local  $\SU(n)$-dressing field can be constructed from a spinor field. In the special case of case abelian YM theory coupled to spinors, i.e. spinorial electromagnetism (EM), the well-known Dirac  phase \cite{Dirac55} (see also \cite{Dirac58} section 80) is an instance of non-local dressing field $\bs u=\bs u(A)$.\footnote{Dirac's eq.(16) and (21) in  \cite{Dirac55} -- and eq.(110) in \cite{Dirac58} -- are abelian instances of the invariant dressed fields \eqref{dressed-fields} above, and his non-local dressing field built from the gauge potential is given in eq.(18)-(19). Eq.(111) in \cite{Dirac58} is the spinorial version of our $J^{\bs u}$ after \eqref{YM-dressed-charge}. } In these cases, one would then conclude that the $\SU(n)$-gauge symmetry is \emph{substantial}, as it is killed only at the cost of the locality of the theory. 
\medskip

In YM theory coupled to scalar fields, things are different. Considering for example the simple case of abelian YM theory coupled to a $\CC$-scalar field, i.e. scalar electromagnetism, one can extract a local dressing field from the complex field by polar decomposition of the latter: $\vphi=\rho \bs u(\vphi) \in \CC$, where $\rho=|\vphi| \in \RR^+$ and $\bs u(\vphi)=e^{i \theta} \in U(1)$. Obviously the phase carries the gauge transformation, $\bs u(\vphi)^{\gamma}= \bs u(\vphi^\gamma)=\gamma\- \bs u(\vphi)$. That is, the dressing field is simply the phase of the scalar field, which is local. Thus the invariant dressed field  \eqref{dressed-fields} $A^{\bs u} = A + \bs u\-d\bs u$ and $\vphi^{\bs u}=\rho$ are both local and $\U(1)$-invariant. Any $\U(1)$-invariant Lagrangian $L(A, \vphi)$ can thus be dressed as in   \eqref{Dressed-L-YM}, giving a local theory with no gauge symmetry.  The $\U(1)$-symmetry of $L$ is then \emph{artificial}. Furthermore, one may consider that the polar decomposition if rather unambiguous, so that no $\G$-transformations arise. 
This has noteworthy interpretive consequences for models couched in the framework of $\CC$-scalar EM. Let us consider two: a semi-classical description of the Aharonov-Bohm (AB) effect and the abelian Higgs model. 

The AB effect is usually seen as the prototypical phenomenon illustrating the non-locality (or non-separability) inherent to gauge theories, 
as the phase shift in the interference pattern cannot be explained by the local interaction of electrons described by $\vphi$ with the gauge potential $A$, which are \emph{both} non-invariant fields.\footnote{Curiously,  emphasis is often on the non-invariance of the gauge potential only, as if the wave function of the electron was  unproblematic regarding gauge invariance.    }  Yet as we just saw, the theory can be rewritten so that electrons and the EM potential can be described by $\rho$ and $A^{\bs u}$ respectively, which are invariant fields. Formulated within scalar EM, the AB effect is thus entirely non-problematic, as the phase shift can be explained via local interaction of invariant fields (as noted by philosopher D. Wallace \cite{Wallace2014}). 
Of course, in the more realistic framework of spinorial EM, since as stated above there is no extracting local dressings from spinors, the $\U(1)$-gauge symmetry is truly substantial, so the AB effect does actually highlights the non-local/non-separable character of EM phenomena.

The abelian Higgs model is often given as the simplest illustration of the notion of Spontaneous Symmetry Breaking (SSB): the gauge potential $A$ is minimally coupled to a $\CC$-field $\vphi$  embedded in a potential $V(\vphi)=\mu^2 \vphi^*\vphi + \lambda  (\vphi^*\vphi)^2$ whose minima are $\vphi_0=0$ and $\{\vphi_0\}=\{ |\vphi_0|=\sqrt{\sfrac{-\mu^2}{2\lambda}} \}$. The first is unique and $\U(1)$-invariant, while the others form a $\U(1)$-orbit. The theory has two phases given by the sign of $-\mu^2$: One in which the only vacuum solution is $\vphi_0=0$, $\U(1)$ preserved, and in which $A$ is massless. Another where the vacuum is degenerate, so that upon spontaneous selection of one point in $\{\vphi_0\}$,  $\U(1)$ is broken and  $A$ gains a mass proportional to  $\sqrt{\sfrac{-\mu^2}{2\lambda}}$. 

But, upon dressing, the Lagrangian is rewritten as in \eqref{Dressed-L-YM} with the field $A^{\bs u}$ minimally coupled to the $\RR^+$-field $\rho$ embedded in the potential $V(\rho)=\mu^2 \rho^2 + \lambda  \rho^4$ with only two minima $\rho_0=0$ and $\rho_0=\sqrt{\sfrac{-\mu^2}{2\lambda}}$. This theory has no $\U(1)$-symmetry, yet still two phases according to the sign of $-\mu^2$: one in which  $A^{\bs u}$ is massless, one in which it has a mass proportional to  $\sqrt{\sfrac{-\mu^2}{2\lambda}}$. 
One can thus appreciate that the notion of SSB is not what does the heavy lifting in the mass acquisition mechanism.
 Rather, the true operative notion is that of a phase transition between two vacuum structures (which are non-degenerate),  and this doesn't necessarily coincide with a symmetry breaking -- as is clearly the case here since  the artificial $\U(1)$-symmetry is eliminated, and the physical d.o.f. exhibited, in both phases of the theory. 
 \medskip
 
 Coming back to general YM theories coupled to scalar fields, the same considerations applies to the electroweak (EW) model, where a local $\SU(2)$-dressing field $\bs u=\bs u(\vphi)$ is extracted from a polar decomposition of the $\CC^2$-scalar field $\vphi$, coupled minimally to the $U(1) \times S\!U(2)$ gauge potential $A=a + B$. One may conclude that $\SU(2)$ is artificial and eliminate it, thereby exhibiting physical d.o.f,  in both the massless and massive phases of the theory, leaving $\U(1)$ as the only substantial gauge symmetry of the model.  SSB is thus bypassed and the EW vacuum phase transition shown to be the operative phenomenon. For technical details on the DFM treatment of the EW model, and discussions on attending philosophical issues, see \cite{Attard_et_al2017, Francois2018}, and \cite{Berghofer-et-al2021} (to appear) for the inclusion of chiral fermions. The polar decomposition of $\vphi \in \CC^2$ may be seen as suffering from some ambiguity, giving rise to residual $\G$-transformations, but \cite{Masson-Wallet} gives arguments as to why this ambiguity might be reduced to a discret choice. 
 
 Thus challenging the SSB interpretation of the EW model may seem an heretical thing to do. But it turns out that gauge-invariant treatments of Yang-Mill-Higgs models have a long history, starting very early on with Higgs and Kibble themselves in 1966 and 1967 (before the papers by Weinberg and Salam): one can easily see that eq.(23)  in \cite{Higgs66} and eq.(66)  \cite{Kibble67}  are  instances of \eqref{Dressed-L-YM}. Before the conclusion of his paper, Kibble explicitly says ``\emph{It is perfectly possible to describe [the theory] without ever introducing the notion of symmetry breaking}".  After the advent of the EW model, invariant treatments independently emerged several times, e.g. in \cite{Banks-Rabinovici1979} (compare eq.(75) and (77) to \eqref{dressed-fields} and \eqref{Dressed-L-YM})  or in \cite{Frohlich-Morchio-Strocchi81} (compare eq.(6.1) to \eqref{dressed-fields}), and more recently \cite{Buchmuller-1994, McMullan-Lavelle95, Faddeev2009, Ilderton-Lavelle-McMullan2010, Masson-Wallet}. 
 The review \cite{Maas2019} provides an extensive  list of references on recent developments in particle physics looking at such gauge-invariant accounts of electroweak physics. 
Philosopher of physics have also seize the subject in the past fifteen years \cite{Earman2004a,Smeenk2006,Lyre2008,Struyve2011,Friederich2013, Friederich2014}, pointing to the fact that a better understanding of the electroweak model, and gauge symmetries, might be necessary to make genuine progress beyond the current best established theories  \cite{Berghofer-et-al2021}. 

In any event, \eqref{Dressed-E-YM}-\eqref{Dressed-Theta-YM} give the dressed presymplectic structure of $\CC$-EM and $\CC^2$-YM theory, stemming from \eqref{Dressed-L-YM}-\eqref{dressed-dL-YM}. From these are read-off the on-shell dressed presymplectic structure of both the abelian Higgs and  EW models since, as we observed  at the end of \ref{Coupled Yang-Mills theory}, the potential term does not contribute.

\subsubsection{4D gauge gravity}
\label{4D gauge gravity}

We now turn to a last illustration of the general framework: gravity.
We will rely on the results of section \ref{Coupled 4D gauge gravity}, where $H=S\!O(1,3)$ and $\H=\SO(1,3)$. 
Espousing the same template as the previous section, we first give a very swift description of the basic structure that would be obtained via variational connections, then engage in a more intricate discussion on what can -- or could -- be done via the DFM.

\paragraph{Basic with connections}
The Lagrangian of the theory is basic, $L_\text{\tiny GR}\in \Omega^0_\text{basic}(\Phi)$, so the field equations and the presymplectic structure are invariant, 
$\bs E_\text{\tiny GR}, \bs\theta^\text{\tiny GR}_\Sigma,  \bs\Theta^\text{\tiny GR}_\Sigma \in \Omega^\bullet_\text{inv}(\Phi)$. 
Then, given a connection $\bs\omega \in \Omega^1_\text{eq}(\Phi, \text{Lie}\H)$,  one can write their basic counterparts.
By application of \eqref{omega-basic-E} and given \eqref{E-GR}, we get the basic field equations
\begin{align}
\label{Basic-GR-E}
\bs E^\text{\tiny GR}_{\bs\omega}&= \bs E_\text{\tiny GR} - d E_\text{\tiny GR}(\bs\omega; \phi), \notag\\
						  &=\bs E_\text{\tiny GR} + d \left\{  \tfrac{2\epsilon}{\ell^2} \, \bs\omega \bullet T \w e^T - \Tr \big( | \rho_*(\bs\omega)\psi \rangle\langle *\upgamma  \psi | \big) \right\}, \quad  \in \Omega^1_\text{basic}(\Phi).
\end{align}
By \eqref{basic-presymp-struct} and given  \eqref{E-GR}-\eqref{theta-GR},  the basic presymplectic structure is immediately found to be
\begin{align}
 \bs\theta^\text{\tiny GR}_{\bs\omega,\, \Sigma} &= \bs\theta_\Sigma  - \int_{\d\Sigma} \theta_\text{\tiny GR}\big( \bs\omega; \phi \big) +  \int_{\Sigma} E_\text{\tiny GR}\big(\bs\omega; \phi \big)  \quad \in \Omega^1_\text{basic}(\Phi), \notag \\
 							&= \bs \theta^\text{\tiny GR}_\Sigma - \int_{\d \Sigma } \bs\omega \bullet F  - \int_\Sigma \tfrac{2\epsilon}{\ell^2}\, \bs\omega \bullet T \w e^T   - \Tr \big( | \rho_*( \bs\omega)\psi \rangle\langle *\upgamma  \psi | \big),  \label{GR-basic-theta}  \\[1.5mm]
  \bs\Theta^\text{\tiny GR}_{\bs\omega,\, \Sigma} &= \bs\Theta_\Sigma  - \int_{\d\Sigma} \bs d \theta_\text{\tiny GR}\big( \bs\omega; \phi \big) +  \int_{\Sigma}  \bs d E_\text{\tiny GR}\big(\bs\omega; \phi \big)   \quad \in \Omega^2_\text{basic}(\Phi), \notag \\
  							&= \bs \Theta^\text{\tiny GR}_\Sigma - \int_{\d \Sigma } \bs d \big(\bs\omega \bullet F \big) - \int_\Sigma  \tfrac{2\epsilon}{\ell^2}\,  \bs d ( \bs\omega \bullet T \w e^T) -  \bs d \Tr \left( | \rho_*( \bs\omega)\psi \rangle\langle *\upgamma  \psi | \right),    \label{GR-basic-Theta} 
\end{align}
Eq.\eqref{GR-basic-theta}-\eqref{GR-basic-Theta} would be one coordinatisation of the  phase space of GR over $\Sigma$,  $(\M_\S, \bs\Theta^b_\Sigma)_\text{\tiny GR}$. Others would of course be obtained under change of connection $\bs\omega'=\bs\omega + \bs \beta$, as by \eqref{Ambig-basic-E}, \eqref{Ambig-basic-theta} and \eqref{Ambig-basic-Theta} we have: 
\begin{equation}
\begin{aligned}
\label{Ambig-basic-presymp-struc-GR}
\bs E^\text{\tiny GR}_{\bs\omega'}  &=\bs E^\text{\tiny GR}_{\bs\omega} +  d \left\{ \tfrac{2\epsilon}{\ell^2} \, \bs\beta \bullet T \w e^T - \Tr \big( | \rho_*(\bs\beta)\psi \rangle\langle *\upgamma  \psi | \big) \right\},  \\[1mm]
 \bs\theta^\text{\tiny GR}_{\bs\omega',\, \Sigma} &=  \bs\theta^\text{\tiny GR}_{\bs\omega,\, \Sigma} - \int_{\d \Sigma } \bs\beta \bullet F  - \int_\Sigma \tfrac{2\epsilon}{\ell^2}\, \bs\beta \bullet T \w e^T   - \Tr \big( | \rho_*( \bs\beta)\psi \rangle\langle *\upgamma  \psi | \big),   \\[1.5mm]
  \bs\Theta^\text{\tiny GR}_{\bs\omega',\, \Sigma} &=  \bs\Theta^\text{\tiny GR}_{\bs\omega,\, \Sigma} - \int_{\d \Sigma } \bs d \big(\bs\beta \bullet F \big) - \int_\Sigma \tfrac{2\epsilon}{\ell^2}\,  \bs d ( \bs\beta \bullet T \w e^T) -  \bs d \Tr \left( | \rho_*( \bs\beta)\psi \rangle\langle *\upgamma  \psi | \right). 
\end{aligned}
\end{equation}
As per section \ref{Via variational connections}, these can be seen a gluing relations the basic structures constructed by two observers on each side of a region partitioned in two subregions by a boundary $\d\Sigma$. 

Finally, let us notice the relation between the basic 2-forms $\bs{D^\omega}\bs\theta^\text{\tiny GR}_\Sigma$ and $ \bs\Theta^\text{\tiny GR}_{\bs\omega, \Sigma}$, which by \eqref{Diff-cov-der-theta-Theta-basic} -- or \eqref{Formula1} -- is:
\begin{align}
\label{YM-bidule}
\bs{D^\omega}\bs\theta^\text{\tiny GR}_\Sigma &=  \bs\Theta^\text{\tiny GR}_{\bs\omega, \Sigma} + \int_{\d \Sigma } \theta_\text{\tiny GR} (\bs\Omega; \phi) - \int_{\Sigma}E_\text{\tiny GR}(\bs\Omega; \phi),  \notag\\
									&=  \bs\Theta^\text{\tiny GR}_{\bs\omega, \Sigma} +  \int_{\d \Sigma }\bs\Omega\bullet F  + \int_\Sigma \tfrac{2\epsilon}{\ell^2}\, \bs \Omega \bullet T \w e^T   - \Tr \left( | \rho_*( \bs\Omega)\psi \rangle\langle *\upgamma  \psi | \right),
\end{align} 
with $\bs\Omega \in \Omega^2_\text{tens}(\Phi, \text{Lie}\H)$ the curvature of $\bs\omega$. Both coincide when $\bs\omega=\mathring{\bs\omega}$ is flat, as would be the case if using the DFM was an option.

\paragraph{Basic with dressing fields}

The use of the DFM in the gravitational case is slightly more subtle than in the YM context. It is instructive to address first the case of the pure gravitational theory, before coming back to the coupling to spinors. 
\medskip

\noindent $\bullet$ {\bf Pure gauge gravity:} 
 Let us quickly set the stage.
The pure gravity sector is given by the McDowell-Mansouri term in \eqref{GR-L}, a basic 0-form on the $\H$-bundle  space of Cartan connections $\Phi=\b \A$: 
\begin{align}
&L_\text{\tiny MM}(\b A) =  \tfrac{1}{2} F \bullet F, \quad \in \Omega^0_\text{basic}(\b\A), 
\quad \text{so} \quad
 \bs d L_\text{\tiny MM} = \bs E_\text{\tiny MM} + d \bs\theta_\text{\tiny MM} \quad \in \Omega^1_\text{basic}(\b\A)  \quad \text{with}, \notag \\[1mm]
&\bs E_\text{\tiny MM} =  -\tfrac{2\epsilon}{\ell^2}  \left\{     \bs d A \bullet T \w e^T + \bs d e \w e^T \bullet F)   \right\}  \quad \in \Omega^1_\text{inv}(\b\A) \quad \text{so} \quad
\bs\theta_\text{\tiny MM} =  \bs d A \bullet F  \quad \in \Omega^1_\text{inv}(\b\A), \label{MM-E-theta}
\end{align}
which is also read-off  \eqref{E-GR}-\eqref{theta-GR}. The presymplectic 2-form  is $\bs\Theta^\text{\tiny MM}_\Sigma=- \int_\Sigma \bs dA \bullet  \bs d F$. From \eqref{Noether-charge},  the Lorentz charges are 
\begin{align}
\label{charge-MM}
Q^\text{\tiny MM}_\Sigma(\chi; \b A)= \int_{\d\Sigma} \theta_\text{\tiny MM}(\chi; \b A) - \int_\Sigma E_\text{\tiny MM}(\chi; \b A)
			   = \int_{\d\Sigma}\chi \bullet F +  \int_\Sigma  \tfrac{2\epsilon}{\ell^2}\, \chi \bullet T \w e^T ,
\end{align}
as is also seen from \eqref{Noether-charge-GR}. In  pure GR, the restriction on-shell is too strong, and imposing normality of $\b A$, that is $T=0$, is enough to have the charge written as a pure boundary term. In full generality, $\bs\Theta^\text{\tiny MM}_\Sigma$ induces a Poisson bracket for the above charges $\big\{ Q^\text{\tiny MM}_\Sigma(\chi; \b A) ,  Q^\text{\tiny MM}_\Sigma(\eta; \b A)\big\} \defeq \bs\Theta_\Sigma^\text{\tiny MM}(\chi^v, \eta^v)=Q^\text{\tiny MM}_\Sigma([\chi, \eta]; \b A)$. The $\bs\H=\bs\SO(1,3)$ field-dependent gauge transformations of $ E_\text{\tiny MM}$, $\bs\theta^\text{\tiny MM}_\Sigma$ and $\bs\Theta^\text{\tiny MM}_\Sigma$ are of course found to be special cases of \eqref{SO-GT-E}-\eqref{SO-GT-presymp-form} (without $\psi$-terms).
\medskip


We may notice that  a local $S\!O(1,3)$-dressing field is readily available in gravity.
 Indeed we have that the soldering part of the Cartan connection $\b A$ transforms as $R_\gamma e\defeq e^\gamma= \gamma\- e$. 
 Given a coordinate system $\{ x^{\, \mu}\}$ on $U \subset M$, the soldering is $e^a={ e^a}_\mu\, dx^{\,\mu}$, or $e=\bs e \cdot dx$, so the map $\bs e \defeq {e^a}_\mu: U \rarrow GL(4)$  is s.t. $\bs e^\gamma  = \gamma\- \bs e$. The tetrad is  thus  a field-dependent local Lorentz dressing field
$ \bs u : \b \A \rarrow D r\big[S\!O, GL\big]$, 
 $           \b A  \mapsto	\bs u(\b A)=\bs e$,
s.t. $R^\star _\gamma \bs u(\b A)= \bs u(R_\gamma \b A)=\bs u(\b A^\gamma)= \bs e^\gamma=\gamma\- \bs e=\gamma\- \bs u(\b A)$.

Using the Lorentz dressing $\bs u(\b A)=\bs e$, written as the $5\times 5$ matrix $\b{\bs u} =  \begin{psmallmatrix}  \bs u & 0 \\  0 & 1 \end{psmallmatrix}$,
the~$\SO$-invariant dressed Cartan connection is 
\begin{align}
\label{dressed Cartan connection}
\b A^{\b{\bs u}}= \b{\bs u}\- \b A \b{\bs u} + \b{\bs u}\- d\b{\bs u} = \begin{pmatrix}A^{\bs u} &  \tfrac{1}{\ell}\bs e^{\bs u} \\  \tfrac{-\epsilon}{\ell}(e^t)^{\bs u}  &0  \end{pmatrix} 
									= \begin{pmatrix}   \bs e\- A \bs e + \bs e \- d\bs e &  \tfrac{1}{\ell} dx \\ \tfrac{-\epsilon}{\ell}dx^T\cdot \bs g & 0  \end{pmatrix} 
 								\rdefeq   \begin{pmatrix}  \Gamma &  \tfrac{1}{\ell} dx \\ \tfrac{-\epsilon}{\ell}dx^T\cdot \bs g& 0  \end{pmatrix} = \b\Gamma.
\end{align}
where  $\Gamma={\Gamma^\mu}_\nu ={\Gamma^\mu}_{\nu, \,\rho}\,dx^{\,\rho}$  is the familiar linear connection with values in $M(4, \RR)=$ Lie$GL(4)$,  $dx= {\delta^{\,\mu}}_\rho\, dx^{\, \rho}$ and
 and $dx^T\cdot \bs g = dx^{\,\mu} g_{\mu\nu}$. 
The metricity condition is automatic, as we have  $\nabla \bs g\defeq d\bs g - \Gamma^T\bs g - \bs g \Gamma = -\bs e^T\big( A^T\eta + \eta\, A \big)\bs e =0$.
A~similar matrix computation  for the dressed Cartan curvature gives,
\begin{align}
\label{dressed Cartan curvature}
\b F^{\bs u}= \b{\bs u}\- \b F \b{\bs u} \quad \Rightarrow \quad \left\{ \begin{array}{l} F^{\bs u} =  \bs e\- F \bs e   \rdefeq  {\sf F} = {\sf R} - \tfrac{\epsilon}{\ell^2}\, dx\w dx^T\cdot \bs g,  \\[1mm] T^{\bs u}=\bs e \- T \rdefeq {\sf T} = \Gamma \w dx, \end{array} \right.
\end{align}
where  ${\sf R}= d\,\Gamma +\tfrac{1}{2}[\Gamma, \Gamma]=\tfrac{1}{2} {{\sf R}^{\,\mu}}_{\nu, \, \rho\sigma} \, dx^{\,\rho} \w dx^{\,\sigma}$ is  $M(4, \RR)$-valued (with components the usual Riemann tensor), and ${\sf T} = {\sf T}^{\,\mu}=\tfrac{1}{2}{{\sf T}^{\,\mu}}_{\rho\sigma} \, dx^{\,\rho} \w dx^{\,\sigma} = {\Gamma^\mu}_{\rho\sigma}\, dx^{\,\rho}\w dx^{\,\sigma}$ is the known expression for the torsion.

Yet another simple matrix computation shows that, by \eqref{Dressed-dphi},  
\begin{align*}
\bs d \b A^{\b{\bs u}}= \b{\bs u}\- \left(  \bs d \b A  + D^{\b A} \left\{ \bs d \b{\bs u}\b{\bs u}\- \right\}  \right)  \b{\bs u}  
				= \begin{pmatrix} \bs d A^{\bs u} &  \tfrac{1}{\ell}\bs d e^{\bs u} \\[1mm]  \tfrac{-\epsilon}{\ell} \bs d(e^t)^{\bs u}  &0  \end{pmatrix} 
				= \begin{pmatrix} \bs e\-  \left( \bs{d}A  + D^A\left\{ \bs{dee}\- \right\} \right)  \bs e & 0 \\[1mm]  \tfrac{-\epsilon}{\ell} \left(\bs d e^T\eta \bs e + e^T \eta \bs{de} \right)   &0  \end{pmatrix} 
				\rdefeq\begin{pmatrix} \bs d \Gamma & 0 \\[1mm] \tfrac{-\epsilon}{\ell}  dx^t \cdot \bs{dg}  &0  \end{pmatrix}  = \bs d \b \Gamma.  
\end{align*}
where  
 one uses 
that $\bs{dee}\- e =\bs d { e^a}_\mu { (e\-)^{\,\mu}}_b\ e = \bs d { e^a}_\mu \, dx^{\, \mu} = \bs d e$ to have the top right component vanish (as also heuristically expected from \eqref{dressed Cartan connection} and  $\bs d\, dx =0$).

As the dressing field takes values in a group larger than the structure group $S\!O$(1,3), to apply our general results we need  to clarify the following technical point:
The polynomial \eqref{Polyn-gravity} we used to write the pure gravity sector of the Lagrangian for 4D gravity in sections \ref{Coupled 4D gauge gravity} is $S\!O$-invariant by \eqref{Prop1-P}. It~is the restriction of the $GL$-invariant polynomial $\b P : \otimes^k M(2k, \RR) \rarrow \RR$ given by
\begin{align}
\label{Polyn-gravity2}
\b P\big(M_1, \ldots, M_k \big)= \sqrt{|\det(\bs g)|}\ M_1 \bullet\, \ldots\, \bullet M_k\defeq \sqrt{|\det(\bs g)|}\ M^{\mu_1 \mu_2}_1\, M^{\mu_3\mu_4}_2 \ldots \, M^{\mu_{2k-1}\mu_{2k}}_k \, \epsilon_{\mu_1 \ldots \mu_{2k}},
\end{align}
The $GL$-invariance under the substitution 
$\bs g \rarrow G^{T} \bs g\, G$ and $M\rarrow G\- M\,G^{-1T}$, with $G={G^{\alpha}}_{\beta} \in GL(4)$, is easily checked (by a computation analogue to  \eqref{Prop1-P}). 
One obtains the $S\!O$-invariant polynomial $P$ by the substitution  $\bs g \rarrow \eta$. 
Conversely, if in $P$ one plugs variables $\bs e\, M\,\bs e\- \eta\-=\bs e\, M\, \bs g\- \bs e^T $ (restoring on the left $\eta\-$ that was kept tacit) then by \eqref{Prop1-P} again we get
\begin{align}
\label{Polyn-gravity4}
P\big(\bs e\, M_1 \bs e\- \eta\-, \ldots, \bs e \,M_k\,\bs e\- \eta\-  \big) &= \bs e\, M_1 \bs g\- \bs e^T \bullet  \ldots \bullet \bs e \,M_k\bs g\-  \bs e^T,  \notag\\
													   & = \det(\bs e)\  M_1 \bs g\- \bs \bullet  \ldots \bullet M_k\bs g\-  =\b P\big(M_1 \bs g\-, \ldots, M_k \bs g\- \big).
\end{align}
To lighten the notation, we will omit $\bs g\-$ in front of variables in expressions involving $\b P$, as it should be clear from the context that indices must be raised. 

Now, the dressed field equations and  presymplectic structure associated with $L_\text{\tiny MM}$ are, by \eqref{dressed-E}-\eqref{dressed-presymp-struct}, 
\begin{align}
\bs E_\text{\tiny MM}^{\bs u}&= E_\text{\tiny MM} + dE(\bs{duu}\-; \b A) =   E_\text{\tiny MM} - \tfrac{2\epsilon}{\ell^2}d\left(  \bs{dee}\-   \bullet T\w e^T \right), \label{Dressed-E-MM} \\[1.5mm]
(\bs\theta_\Sigma^\text{\tiny MM})^{\bs u}  
					&=  \bs \theta^\text{\tiny MM}_\Sigma + \int_{\d\Sigma} \theta( \bs{duu}\- ; \b A) - \int_\Sigma E(\bs{duu}\-; \b A),   \notag\\
					&=  \bs \theta^\text{\tiny MM}_\Sigma + \int_{\d\Sigma} \bs{dee}\- \bullet F + \tfrac{2\epsilon}{\ell^2} \int_\Sigma  \bs{dee}\-  \bullet T \w e^T,   \label{dressed-theta-MM} \\
(\bs\Theta_\Sigma^\text{\tiny MM})^{\bs u} &=\bs\Theta^\text{\tiny MM}_\Sigma + \int_{\d\Sigma}   \bs d \theta\big(\bs{duu}\-; \b A\big) - \int_\Sigma \bs d E\big(\bs{duu}\-; \b A\big).    \notag  \\   
					&=  \bs \Theta_\Sigma^\text{\tiny MM} + \int_{\d\Sigma} \bs d\left(  \bs{dee}\- \bullet F  \right) + \tfrac{2\epsilon}{\ell^2} \int_\Sigma  \bs d\left( \bs{dee}\-  \bullet T \w e^T \right).   \label{dressed-Theta-MM} 
\end{align}

As per the DFM philosophy, section \ref{Via dressing fields}, these are none other than the field equation and presymplectic structure of the dressed Lagrangian  
\begin{align}
\label{dressed-L-GR-e}
L^{\bs u}_\text{\tiny MM} (\b A)=\b L_\text{\tiny MM}(\b\Gamma)  &= \sqrt{|\det(\bs g)|} \  \ \tfrac{1}{2}\,  {\sf F}  \bullet  {\sf F} , \\
				&= \sqrt{|\det(\bs g)|} \  \ \tfrac{1}{2}\,  {\sf R}  \bullet  {\sf R}   -  \tfrac{\epsilon}{\ell^2}  \left( {\sf R}  \bullet dx \w dx^T  - \tfrac{\epsilon}{2\ell^2} dx \w dx^T \bullet dx \w dx^T \right),  \notag
\end{align}
where $\b L_\text{\tiny MM}$ is  based on the polynomial \eqref{Polyn-gravity2}. This is manifestly just the Lagrangian of GR in the `metric' formulation. Which means in particular that $(\bs\theta_\Sigma^\text{\tiny MM})^{\bs u}$ is simply the presymplectic potential of the metric formulation, 
\begin{align}
(\bs\theta_\Sigma^\text{\tiny MM})^{\bs u}= \b \theta_\text{\tiny MM}( \bs d \b \Gamma; \b \Gamma)= \int_\Sigma \sqrt{|\det(\bs g)|} \ \bs d \Gamma \bullet {\sf F}= \int_\Sigma \sqrt{|\det(\bs g)|} \ \bs d \Gamma \bullet \left( {\sf R} - \tfrac{\epsilon}{\ell^2}\, dx\w dx^T\right).
\end{align}
The associated 2-form is easily deduced. Equation \eqref{dressed-theta-MM} then gives  the relation between the metric and tetrad potentials of pure GR, and its boundary term generalises in particular the (aptly named from the DFM viewpoint) ``dressing 2-form'' of DePaoli-Speziale \cite{De-Paoli-Speziale2018} --  see also \cite{Oliveri-Speziale2020, Oliveri-Speziale2020-II}.
\medskip

In the situation at hand, we have an occasion to see that the ambiguity in the choice of dressing field encodes a relevant gauge symmetry of the dressed theory: coordinate changes. 
Indeed, we identified the components of the soldering form  $\bs e ={e^a}_{\,\mu}$ in a given coordinate system $\{ x^{\,\mu}\}$ as a good $S\!O$-dressing field. In another coordinate system we have of course  $\bs e ' = \bs e\, \xi$, i.e. ${{e'}^a}_{\,\nu} = {e^a}_{\,\mu}\, {\xi^{\,\mu}}_\nu$, where  $\xi={\xi^{\,\mu}}_\nu \in GL(4)$ is the Jacobian of the coordinate change. 
The ambiguity in the choice of Lorentz dressing is thus  parametrised by the group of local coordinate transformations $\G =\GL \defeq \left\{ \xi :U \rarrow GL(4)\, |\, \xi^\gamma=\xi \right\}$ acting as $\bs u^\xi = \bs u \xi$ and of course trivially on the Cartan connection $\b A^\xi=\b A$. 
 The space of dressed Cartan connections $\b \A^{\bs u}=\b \Gamma$ is then a $\G$-principal bundle, with a right action of  $\G$ given by 
 \begin{align}
 R_\xi {\b A}^{\bs u} = (\b A^{\bs u})^\xi \defeq \xi \- \b A^{\bs u} \xi + \xi\- d\xi \quad \Rightarrow \quad \left\{ \begin{array}{l} (A^{\bs u})^\xi = \Gamma^\xi =  \xi \-   \Gamma \xi + \xi\- d\xi ,  \\[1mm] (e^{\bs u})^\xi= (dx)^\xi = \xi\- dx. \end{array} \right.
 \end{align}
 By \eqref{Polyn-gravity2}, the dressed Lagrangian  \eqref{dressed-L-GR-e} has trivial $\G$-equivariance $R^\star_\xi \b L_\text{\tiny MM}=\b L_\text{\tiny MM}$. So, for $\vkappa={\vkappa^{\, \mu}}_{\!\nu} \in $ Lie$\G$ generating $\vkappa^v \in V(\b \A^{\bs u})$ and by  \eqref{dressed-Noether-charge}, the dressed Noether charge is:
\begin{align}
\b Q^\text{\tiny MM}_\Sigma(\vkappa; \b A^{\bs u})  &=  \int_{\d\Sigma} \b\theta_\text{\tiny MM}(\vkappa; {\b A}^{\bs u}) - \int_\Sigma \b E_\text{\tiny MM}(\vkappa; {\b A}^{\bs u}) ,  \notag\\
					            &=  \int_{\d\Sigma} \sqrt{|\det(\bs g)|} \ \ \vkappa   \bullet   {\sf F}   + \tfrac{2\epsilon}{\ell^2} \int_\Sigma  \sqrt{|\det(\bs g)|} \ \ \vkappa   \bullet {\sf T} \w dx^T.   \label{dressed-Noether-charge-MM}
\end{align}
It satisfies  $\iota_{\vkappa^v}(\bs\Theta^\text{\tiny MM}_\Sigma)^{\bs u} = -\bs d\b Q^\text{\tiny MM}_\Sigma(\vkappa; \b A^{\bs u})$ and via \eqref{dressed-Poisson-bracket} the dressed presymplectic 2-form induces the Poisson bracket 
\begin{align}
\label{dressed-Poisson-bracket-MM}
\big\{ \b Q^\text{\tiny MM}_\Sigma(\vkappa; \b A^{\bs u}) ,  \b Q^\text{\tiny MM}_\Sigma(\vkappa'; \b A^{\bs u})\big\}\defeq\, (\bs\Theta^\text{\tiny MM}_\Sigma)^{\bs u}\big(\vkappa^v, {\vkappa'}^v\big) =	 \b Q^\text{\tiny MM}_\Sigma\big([\vkappa, \vkappa']; \b A^{\bs u}\big),
\end{align}
so that the Poisson algebra of dressed (metric) charges is  isomorphic to the Lie algebra of coordinate changes Lie$\G$. 
The same relations hold for field-dependent parameters $\bs\vkappa \in$ Lie$\bs\G$, where $\bs\G=\bs\GL$ is the gauge group of the $\G$-bundle $\b\A^{\bs u}$, even though in this case the charge is non-integrable $\iota_{\bs\vkappa^v}(\bs\Theta^\text{\tiny MM}_\Sigma)^{\bs u} = -\bs d\b Q^\text{\tiny MM}_\Sigma(\munderline{red}{\bs\vkappa}; \b A^{\bs u})$.

Charges \eqref{charge-MM} of $L_\text{\tiny MM}$, like those \eqref{Charge-on-shell-grav} of  $L_\text{\tiny GR}$ in section \ref{Coupled 4D gauge gravity},  could be interpreted via the affine ansatz \eqref{Perturbation-ansatz} $\b A=\b A_0+ \b \alpha$. But when attempting to do the same with the charges \eqref{dressed-Noether-charge-MM} above, echoing  the caveats at the end of section \ref{Via dressing fields}, we must exercise care as the ansatz affects the dressing field $\bs u(\b A )=\bs e$ as well. Splitting in particular the tetrad as background and fluctuation, $\bs e =\bs e_0 + \upepsilon$, induces a corresponding splitting of the metric $\bs g =\bs g_0 + \bs h$ which must then be plugged in the on-shell expression for $\b Q^\text{\tiny MM}_\Sigma(\vkappa; \b A^{\bs u})$. 
Although doable, it is cumbersome. One may rely on a simpler heuristics to interpret the dressed charge: on-shell (or simply in the normal case) these are
\begin{align}
\b Q^\text{\tiny MM}_\Sigma(\vkappa; \b A^{\bs u}) =  \int_{\d\Sigma} \sqrt{|\det(\bs g)|} \ \ \vkappa   \bullet   {\sf F}_{\ | \S}
					            &=  \int_{\d\Sigma} \sqrt{|\det(\bs g)|} \  \ \vkappa   \bullet \left(  {\sf R} -  \tfrac{\epsilon}{\ell^2} dx \w dx^T \right),  \label{Gen-Komar-mass} \\
					            &=  \int_{\d\Sigma} \sqrt{|\det(\bs g)|} \ \ \left( \epsilon_{\, \mu\nu\sigma\rho} \, \alpha^{\,\mu\nu} \, \tfrac{1}{2} {{\sf R}^{\sigma\rho}}_{\alpha\beta}-  \tfrac{\epsilon}{\ell^2}\, \vkappa^{\,\mu\nu} \, {\epsilon_{\mu\nu\alpha\beta}} \right) dx^{\,\alpha}\!\! \w\!dx^{\,\beta}. \notag
\end{align}
Now if one considers $\vkappa = \d \zeta = \d_{\,\nu}\zeta^{\mu}$ with $\zeta$ the components of a Killing vector field of $\bs g$, the above expression is a generalised Komar integral. 
It reproduces the result for 4D Zumino-Lovelock theory gravity  (also known as Gauss-Bonnet gravity) obtained in  \cite{Kastor2008} eq.(17)-(20), and
generalises the usual Komar mass as given  in \cite{Choquet-Bruhat2009} (definition 4.6, eq.(4.8) p. 460) -- known to coincide with the Newtonian mass and ADM mass for (stationnary) asymptotically flat spacetimes $M$ and  to vanish if and only if $M$ is  flat (Lemma 4.10, Theorem 4.13 and Theorem 4.11 in \cite{Choquet-Bruhat2009}). 
 The charge  \eqref{Gen-Komar-mass} gives a good notion of mass and angular momentum (according to the nature of $\zeta$) in gravity with $\Lambda \neq0$ as it vanishes on the (A)dS groundstate of the theory, i.e. the homogeneous space of the underlying Cartan geometry.
 
\medskip

The general formulae  \eqref{Ambig-dress-as-GT} applied here give the field-dependent coordinate transformations of the field equation and presymplectic structure of $\b L_\text{\tiny MM}$
\begin{equation}
\label{Ambig-dressingMM}
\begin{aligned}
 [\bs E_\text{\tiny MM}^{\bs u}]^{\bs\xi} &= \bs E_\text{\tiny MM}^{\bs u} - d\left(  \tfrac{2\epsilon}{\ell^2}\, \sqrt{|\det(\bs g)|}\  \bs{d\xi\xi}\-  \bullet {\sf T}\w dx^T \right),  \\
							 &= \bs E_\text{\tiny YM}^{\bs u} +  d\left(  \tfrac{2\epsilon}{\ell^2}\,  \mathring{\bs\beta} \bullet T\w e^T  \right), \\[2mm]
[(\bs\theta^\text{\tiny MM}_\Sigma)^{\bs u}]^{\bs \xi} &= (\bs \theta^\text{\tiny MM}_\Sigma)^{\bs u} + \int_{\d \Sigma }  \sqrt{|\det(\bs g)|}\  \bs{d\xi\xi}\- \bullet {\sf F}  +  \tfrac{2\epsilon}{\ell^2}  \int_\Sigma  \sqrt{|\det(\bs g)|}\ \bs{d\xi\xi}\-\bullet {\sf T}\w dx^T,    \\[1mm]
										 &= (\bs \theta^\text{\tiny MM}_\Sigma)^{\bs u} - \int_{\d \Sigma } \mathring{\bs\beta} \bullet F  -  \tfrac{2\epsilon}{\ell^2} \int_\Sigma  \mathring{\bs\beta} \bullet T\w e^T ,    \\[2mm]
[(\bs\Theta^\text{\tiny MM}_\Sigma)^{\bs u}]^{\bs \xi} &= (\bs \Theta^\text{\tiny MM}_\Sigma)^{\bs u} + \int_{\d \Sigma }  \bs  d \left\{  \sqrt{|\det(\bs g)|}\  \bs{d\xi\xi}\- \bullet {\sf F}  \right\} +  \tfrac{2\epsilon}{\ell^2}  \int_\Sigma  \bs d \left\{  \sqrt{|\det(\bs g)|}\ \bs{d\xi\xi}\-\bullet {\sf T}\w dx^T \right\},    \\[1mm]
										 &= (\bs \Theta^\text{\tiny MM}_\Sigma)^{\bs u} - \int_{\d \Sigma } \bs d ( \mathring{\bs\beta} \bullet F )  -  \tfrac{2\epsilon}{\ell^2} \int_\Sigma  \bs d ( \mathring{\bs\beta} \bullet T\w e^T ).
\end{aligned}
\end{equation}
We provide the second line, where $\mathring{\bs\beta}= - \bs e \bs{d\xi\xi\-} \bs e\-$, for comparison with \eqref{Ambig-basic-presymp-struc-GR}. The similarity is of course no accident as using the dressing $\bs u(\b A)=\bs e$ is equivalent to using the flat connection $\mathring{\bs\omega} \defeq -\bs{duu}\-$ on the $\H$-bundle $\b\A$. One checks easily the two defining properties  \eqref{var-connection1}-\eqref{var-connection2} of an Ehresmann connection:  
\begin{align*}
R^\star_\gamma\mathring{\bs\omega} &= -\bs d R^\star_\gamma \bs e (R^\star_\gamma e)\-= - \gamma\- \bs{dee}\- \gamma= \gamma\-  \mathring{\bs\omega} \gamma, \quad \text{ for }  \gamma \in \H=\SO(1,3),\\[1mm]
\mathring{\bs\omega}(\chi^v)&= -\bs{dee}\-(\chi^v) = - \chi^v(\bs e) \bs e \- = -(-\chi \bs e) \bs e\- = \chi \ \in \text{Lie}\SO(1,3).
\end{align*}
And the flatness condition is trivial $\bs d \mathring{\bs\omega} +\tfrac{1}{2}[\mathring{\bs\omega},\mathring{\bs\omega}]= \bs{dede}\- + \bs{dee}\-\bs{dee}\- \equiv 0$. A field-dependent coordinate change $\bs u' = \bs u \bs \xi$ induces an affine shift  of flat connection $ \mathring{\bs\omega}' = \mathring{\bs\omega} +  \mathring{\bs\beta}$.

In good illustration of the general {\bf Comment 1} of section \ref{Via dressing fields}, we have no trouble appreciating that $\G=\GL$ is not a physical transformation group, and doesn't permute points in the physical phase space $\S/\SO \simeq \S^{\bs u}/\G$. The question of the physical relevance (or signature) of $\G$ is of the same nature as that of the original $\SO$ gauge symmetry, as illustrated above in the discussion of the interpretation of dressed charges. Finally, \eqref{Ambig-dressingMM} shows that as a boundary problem for $\SO$ is solved by dressing, another emerges for $\G=\GL$ which is very likely a substantial symmetry of the dressed/metric theory, suggesting that the boundary problem signals the non-locality, or non-separability, of gravitational physics. 
\bigskip

\noindent $\bullet$ {\bf Theory coupled to spinors:} 
According to the DFM philosophy, the presence of the tetrad as a local dressing field that can be used to rewrite the theory in terms of Lorentz-invariant variables makes $\SO(1,3)$ an \emph{artificial} symmetry of \emph{both} pure GR \emph{and} of GR coupled with bosonic and scalar fields (EM field, fluids, dust...). 
This is an interesting contrast with, say, the EM case: In the pure gauge theory no local dressing built from the connection exists so $\U(1)$ is substantial, while in $\CC$-scalar EM a local dressing is extracted from the matter sector so that $\U(1)$ is artificial. 

Yet, in the same way that the coupling of  the EM field to spinors makes  $\U(1)$ is substantial again, the coupling of gravity to spinors a priori changes the verdict on $\SO(1,3)$. 
The tetrad being $GL$-valued, and for lack of finite dimensional spin representations of $GL$, it cannot be used to produce a Lorentz-invariant spinor: we cannot write $\psi^{\bs u}=\rho(\bs u)\-\psi= \rho(\bs e)\-\psi$. This lack of local dressing for the whole theory strongly suggest that the Lorentz gauge symmetry $\SO(1,3)$ is \emph{substantial} in GR $+$ spinors, and the associated boundary problem a reflection of the non-locality/non-separability of gravitational physics.
\medskip

We may nuance this conclusion in  light of the fact that,  
 so long as one works on $U \subset \M$, one can decompose the tetrad as $\bs e=\bs{ut}$, where $\bs u\in H=S\!O(1,3)$ and $\bs t={t^b}_\mu$ has the same d.o.f as the metric field and is s.t.  $\bs g=\bs t^T\eta \bs t$. For details and references on such a decomposition, which relies on the Schweinler-Wigner orthogonalization procedure, we refer to section 4.3 of \cite{GaugeInvCompFields} -- see also the end of section 2 and footnote 12 of \cite{Francois2018}. It is akin to the polar decomposition of the $\CC$-scalar field in EM:  $\bs u$ carries the gauge representation, $\bs u^\gamma=\gamma^{-1}\bs u$ with $\gamma \in \SO(1,3)$, so that  $\bs u =\bs u(e) : \Phi \rarrow \D r[H, H]$   is a (minimal) local Lorentz dressing field. 

As per section \ref{Via dressing fields}, it can be used to obtain the dressed Lagrangian,
\begin{align}
\label{Dressed-L-GR}
L^{\bs u}_\text{\tiny GR} = \text{F}_{\bs u}^\star L_\text{\tiny GR} \quad \text{i.e.}\quad L^{\bs u}_\text{\tiny GR}(\phi)=L_\text{\tiny MM}(\b A^{\bs u}) + L_\text{\tiny Dirac}(\b A^{\bs u}, \psi^{\bs u}) &=  \tfrac{1}{2} F^{\bs u} \bullet F^{\bs u} + \langle \psi^{\bs u}, \, \slashed D^{\bs u} \psi^{\bs u}\rangle \quad \in \Omega^0_\text{basic}(\Phi).
\end{align}
 where  invariant spinor fields $\psi^{\bs u}:=\rho(\bs u)^{-1} \psi$  couples  to gravity via the invariant gauge field $A^{\bs u}:=\bs u\- A \bs u+\bs u\-d\bs u$ (the Lorentz part of the dressed Cartan connection $\b A^{\bs u}$ whose soldering part is $e^{\bs u}=t=\bs t \cdot dx$). 
 From the  variation of \eqref{Dressed-L-GR}, 
$\bs d L^{\bs u}_\text{\tiny GR} = \bs E^{\bs u}_\text{\tiny GR} + d \bs\theta^{\bs u}_\text{\tiny GR} = E_\text{\tiny GR}(\bs d\phi^{\bs u}; \phi ^{\bs u}) + d \theta_\text{\tiny GR}(\bs d\phi^{\bs u}; \phi ^{\bs u})$,
one gets the associated field equations and basic presymplectic structure, which 
by \eqref{dressed-E}-\eqref{dressed-presymp-struct} and given  \eqref{E-GR}-\eqref{theta-GR} -- or by \eqref{SO-GT-E}-\eqref{SO-GT-presymp-form}  together with the rule of thumb $\bs \gamma \rarrow \bs u$ -- are:
\begin{align}
 \bs E_\text{\tiny GR}^{\bs u}&= \bs E_\text{\tiny GR} + dE_\text{\tiny GR} \big(  \bs{duu}\- ;  \phi \big),  \quad \in \Omega^1_\text{basic}(\Phi),  \notag \\
 						&= \bs E_\text{\tiny GR} - d \left\{\tfrac{2\epsilon}{\ell^2}\, \big(  \bs{duu}\-  \bullet T \w e^T  \big) + \Tr \left( | \rho_*( \bs{duu}\-)\psi \rangle\langle *\upgamma  \psi | \right) \right\}, \label{dressed-E-GR} \\[2mm]
(\bs\theta_\Sigma^\text{\tiny GR})^{\bs u}&= \bs \theta_\Sigma^\text{\tiny GR} + \int_{\d\Sigma}  \theta_\text{\tiny GR} \big(  \bs{duu}\- ;  \phi \big) - \int_{\Sigma} E_\text{\tiny GR} \big( \bs{duu}\- ; \phi \big),  \quad \in \Omega^1_\text{basic}(\Phi), \notag\\
				  &= \bs \theta_\Sigma^\text{\tiny GR} + \int_{\d\Sigma} \bs{duu}\- \bullet F  +   \int_{\Sigma} \tfrac{2\epsilon}{\ell^2}\, \bs{duu}\-\bullet T \w e^T   - \Tr \left( | \rho_*( \bs{duu}\-)\psi \rangle\langle *\upgamma  \psi | \right),                           \label{dressed-theta-GR} \\[2mm]
(\bs\Theta_\Sigma^\text{\tiny GR})^{\bs u} &= \bs\Theta_\Sigma^\text{\tiny GR} +  \int_{\d\Sigma} \bs d  \theta_\text{\tiny GR}\big(\bs{duu}\-; \phi \big) -  \int_{\d\Sigma} \bs d E_\text{\tiny GR}\big(\bs{duu}\-; \phi \big),   \quad \in \Omega^2_\text{basic}(\Phi), \notag\\
				    &= \bs\Theta_\Sigma^\text{\tiny GR} +  \int_{\d\Sigma}  \bs d  \left(\bs{duu}\- \bullet  F \right) +       \int_{\Sigma}  \tfrac{2\epsilon}{\ell^2}\,  \bs d (\bs{duu}\-\bullet T \w e^T) -  \bs d \Tr \left( | \rho_*( \bs{duu}\-)\psi \rangle\langle *\upgamma  \psi | \right),         \label{dressed-Theta-GR} 
\end{align}

Unfortunately, \eqref{dressed-theta-GR}-\eqref{dressed-Theta-GR} do not really solve the boundary problem, as these are at best valid only locally, i.e. on a single coordinate patch. 
Indeed, $\bs u$ and by extension the composite fields and other quantities built from it, depend on the coordinate chart $\{x^\mu\}$ on $U\subset M$ in such a way that they have no determined well-behaved transformation law under coordinate changes (as explained in \cite{GaugeInvCompFields}). Which makes them ill-defined as global geometrical objects on $M$. This also means that our general discussion about the ambiguity in the choice of dressing doesn't apply in this case, and in particular no charges associated with $\GL(4)$ can be assigned via $(\bs\theta_\Sigma^\text{\tiny GR})^{\bs u}$. 

This construction of $\bs u$ from a decomposition of the tetrad seems to bear some relation to the attempts -- pioneered in the `50s and `60s by DeWitt, Ogievetsky and Polubarinov -- to build spinors without introducing Lorentz gauge symmetry. See \cite{Pitts2012} for a  review with an extensive bibliography.  Upon mild restrictions on the admissible coordinate systems, the coordinate transformation law for such spinors is formally attainable, but only in the weak field regime around a flat background, i.e when $\bs g$ is a small perturbation around $\eta$. Even then, the transformation law is metric dependent and highly non-linear. It is thus not  obvious that such a framework could be satisfactory in the strong field regime of GR, or that it can be accommodated to QFT in curved spacetime. A prudent commitment to the initial assessment that $\SO(1, 3)$ is substantial in spinorial gravity therefore seems reasonable.

\medskip

The absence of dressing for the theory `gravity $+$ spinors' is less of a problem if one is mainly interested in charges and their Poisson algebra: on-shell, charges in the coupled theory \eqref{Noether-charge-GR} are the same as charges of pure gravity \eqref{charge-MM} and reduce to boundary terms: $Q^\text{\tiny GR}_\Sigma(\chi; \phi) = \int_{\d\Sigma}\chi \bullet F_{\,|\S}$. Which makes sense as the matter sources presumably have compact support while their gravitational field propagates and reaches infinity. But then the on-shell charges of the coupled theory can be dressed as in  \eqref{dressed-Noether-charge-MM} giving $Q^\text{\tiny GR}_\Sigma(\vkappa; \phi) = \int_{\d\Sigma}  \sqrt{|\det (\bs g)|}\, \vkappa \bullet {\sf F}_{\,|\S}$, and interpreted as in the free case, while their Poisson bracket is on-shell \eqref{dressed-Poisson-bracket-MM}.


%
%
%

\section{Conclusion}
\label{Conclusion}

In this note we made decisive use of the bundle geometry of field space which, articulated with covariant phase space methods, allows to give a series of general results on the presymplectic structure of invariant matter coupled gauge theories. That is, given the Lagrangian $L$ of any such theory, we gave the off-shell expression of the Noether charges for field-independent \eqref{Noether-charge} and field-dependent \eqref{Noether-charge-fied-dep-parameter} parameters, as well as their Poisson bracket \eqref{Poisson-bracket}-\eqref{Poisson-bracket-field-dep} induced by the presymplectic 2-form. We also gave the general field-dependent gauge transformations of the presymplectic potential \eqref{Field-depGT-presymp-pot} and 2-form \eqref{Field-depGT-presymp-form}, which exhibit the boundary problem in full generality. We stress that  the only computation needed to apply these results in any given example, is to derive the field equation $\bs E$ and the presymplectic potential $\bs\theta$ from $L$. Which we did in the case of YM theory (section \ref{Coupled Yang-Mills theory}) and GR in its Cartan geometric formulation (section \ref{Coupled 4D gauge gravity}), thereby recovering standard results.  
 In passing, we noticed that using the affine structure of the space of connections $\A$, it is possible to split the Noether charge as a background contribution and a measurable contribution associated with a Killing symmetry of the background gauge field. This  generalises the approach of Abbott \& Deser \cite{Abbott-Deser1982, Abbott-Deser1982bis}.

Emphasis on the bundle geometry of field space $\Phi$ allows to appreciate that solving the boundary problem boils down to one thing only: finding ways to build the basic counterpart of a given variational form on $\Phi$. We reminded that variational connections and the DFM (a.k.a edge modes) are means  to do just that, and conducted a systematic comparative analysis of both in sections \ref{Variational connections on field space} and \ref{The dressing field method}. 
We could apply this to produce the general basic presymplectic structure of an invariant theory as obtained via connections \eqref{basic-presymp-struct} and via dressings \eqref{dressed-presymp-struct}. 
Special applications to YM theory and GR reproduce several results of the literature, e.g. \cite{DonnellyFreidel2016, Gomes-et-al2018, Gomes-Riello2021}, and  in particular  the DFM gives from first principle the unambiguous relations \eqref{dressed-theta-MM}-\eqref{dressed-Theta-MM} between the presymplectic structures of GR in the tetrad and metric formulations, generalising the dressing 2-form of \cite{De-Paoli-Speziale2018} -- see also \cite{Oliveri-Speziale2020, Oliveri-Speziale2020-II}. 

The most relevant conceptual difference between the two approaches -- beside the fact that the existence of dressing fields impose stronger topological constraints on field space than (non-flat) connections  --  resides in how their respective ambiguities arise. Any two choices of connections are related via an affine shift by a $\Ad$-tensorial 1-form on $\Phi$, leading to relations like \eqref{Ambiguity-basic-1-form}/\eqref{Ambig-basic-theta}, which is not associated with any relevant new symmetries. On~the contrary, the ambiguity in the choice of dressings, while indeed manifesting itself as a special case of affine shift of the associated connections \eqref{Ambiguity-basic-1-form-dressing},  is more generally encoded by a  group $\G$ that may be a relevant  symmetry  of the dressed theory~$L^{\bs u}$. This, the DFM shows  via \eqref{SESgroups-dressed}, is a \emph{gauge} symmetry that doesn't enjoy a more direct physical interpretation than the original gauge symmetry $\H$ of $L$, but whose relevance may show through the associated dressed charges \eqref{dressed-Noether-charge}. As a matter of fact, in  GR  these dressed charges give (generalised) Komar integrals  \eqref{Gen-Komar-mass}. 
 This, we argue, also clarifies the status of the so-called ``surface symmetries" of the edge mode literature, which are exactly instances of $\G$-symmetries. The corollary is of course that $\G$ may manifest its own boundary problem, unless the construction of a dressing field in the theory under consideration is free enough of any ambiguities -- as it arguably is the case for example  in $\CC$-scalar electrodynamics, as discussed at the end of section \ref{Yang-Mills theory}. 
 \medskip

We remark that following the template of sections \ref{Geometry of field space} and \ref{Presymplectic structures of matter coupled gauge theories over bounded regions} above,  one may analyse the presymplectic structure of \emph{twisted} gauge theories. 
These  are gauge theories  whose configuration space is $\t \Phi\defeq \t\A \times \Gamma(\t E)$ with $\t \A$ the space of twisted connections on a standard $H$-principal bundle $\P$ -- generalising Ehresmann and Cartan connections -- and $\Gamma(\t E)$ the space of sections of twisted bundles associated with $\P$ built via 1-cocycles of the structure group~$H$ -- extensions of standard associated bundles built via representations. 
We briefly evoked these notions in sections \ref{Field space as a principal bundle} and  \ref{Variational twisted connections} in the context of  the bundle $\Phi$, but these have finite dimensional precursors whose  elementary theory is described in \cite{Francois2019_II}. 
As shown there, conformal gravity is an unexpected example of twisted gauge theory, the (dressed) conformal Cartan connection -- otherwise known as the tractor or twistor connection -- being a twisted connection transforming via a cocycle of the group of Weyl rescalings (a subgroup of the structure group of the conformal Cartan bundle \cite{Sharpe}). 
The presymplectic structure of conformal gravity -- possibly coupled to twistors (or conformal tractors \cite{curry_gover_2018}) -- would then follow as an immediate application of the suggested extension of the present work to twisted gauge theories

A more immediate direction we wish to explore is to consider the case of non-invariant theories, i.e. to include classical gauge anomalies. In \cite{Francois2021}, this was done for pure gauge theories, and only charges for field-independent gauge parameters were considered.  We endeavor to maximally generalise the latter work to encompass non-invariant coupled theories, include the case where the symmetry under consideration is $\Diff(M)$ rather than an internal gauge group, and extend the discussion to charges for field-dependent gauge parameters and their Poisson bracket. 
The~latter should be a generalisation of the Barnich-Troessaert bracket proposed in \cite{Barnich_Troessaert2011}. This, by the way, should provide a cross-check of the recent \cite{Freidel-et-al2021} who suggest they have found such a generalised bracket that is  not centrally extended, while generically we expect ours to be.
 As a matter of direct applications, we expect our Poisson algebras of charges in 3D and 4D gravity with $\Lambda \neq 0$ to extend the $\mathfrak{bms}3$ and  $\mathfrak{bms}4$ algebras,  likely making  contact with the $\Lambda$-$\mathfrak{bms}4$  algebra of \cite{Compere-et-al2020}. Working out the Poisson algebra of charge of conformal gravity, in the framework alluded to above, would certainly reproduce or a least make contact with the so-called Weyl-$\mathfrak{bms}$ algebra of \cite{freidel-et-al2021bis}.

\section*{Acknowledgment}  

J.F. and N.B. are supported by the Fonds de la
Recherche Scientifique -- FNRS under grant PDR No. T.0022.19 (``Fundamental issues in extended
gravitational theories''), and J.F. is also supported by the FNRS grant MIS No.\ F.4503.20
(``HighSpinSymm'').


\appendix

\section{Appendix}
\label{Appendix}



\subsection{Proof of formula \eqref{Formula1}}
\label{Proof of formula 1}

Given $\bs\alpha \in \Omega^1_\text{inv}(\Phi)$, the infinitesimal version of its trivial $\H$-equivariance reads  $\bs L_{\chi^v} \bs \alpha = \iota_{\chi^v} \bs d \bs \alpha + \bs d \iota_{\chi^v} \bs \alpha \equiv 0$ for any $\chi \in \text{Lie}\H$. So, using $[\bs L_{\bs X}, \iota_{\bs Y}]=\iota_{[\bs X, \bs Y]}$, we have the identity
\begin{align}
\label{A1}
\bs d \bs \alpha(\chi^v, \eta^v)= \iota_{\eta^v} \iota_{\chi^v} \bs{d\alpha}= - _{\eta^v} \bs d \iota_{\chi^v} \bs \alpha = - \bs L_{\eta^v} \iota_{\chi^v} \bs \alpha = -\iota_{\chi^v} \bs L_{\eta^v} \bs\alpha - \iota_{[\eta^v, \,\chi^v]} \bs \alpha = \iota_{[\chi, \eta]^v} \bs \alpha.
\end{align}
Its covariant derivative is $\bs{D^\omega \alpha} \defeq (\bs d \bs \alpha)\circ |^h \in \Omega^2_\text{basic}(\Phi)$.
Its horizontal counterpart is $\bs\alpha^h\defeq \bs\alpha\circ|^h \in \Omega^1_\text{basic}(\Phi)$. And since on basic forms the covariant derivative reduces to the exterior derivative, $\bs{D^\omega \alpha}^h =\bs d \bs\alpha^h \in \Omega^2_\text{basic}(\Phi)$. We want to prove that 
 \begin{align}
 \label{Formula1bis}
 \bs{D^\omega\alpha}= \bs d \bs\alpha^h + \iota_{[\bs\Omega]^v}\bs\alpha.
 \end{align}
 
Now, we have on the one hand, by the Kozsul formula and \eqref{horiz-proj-map}:
\begin{align}
\label{A2}
\bs d\bs \alpha^h_{|\phi} (\bs X_\phi, \bs Y_\phi) &= \bs X \cdot \bs\alpha_{|\phi}^h (\bs Y_\phi) - \bs Y \cdot \bs\alpha_{|\phi}^h (\bs X_\phi) - \bs\alpha_{|\phi}^h \big([\bs X, \bs Y]_\phi\big), \notag\\
							 &= \munderline{ForestGreen}{ \bs X \cdot \bs\alpha_{|\phi} } \left( \munderline{ForestGreen}{\bs Y_\phi} - [\bs\omega_{|\phi}(\bs Y_\phi)]^v_\phi \right) 
							    -  \munderline{ForestGreen}{ \bs Y \cdot \bs\alpha_{|\phi} } \left( \munderline{ForestGreen}{\bs X_\phi} - [\bs\omega_{|\phi}(\bs X_\phi)]^v_\phi \right) 
							     - \munderline{ForestGreen}{ \bs\alpha_{|\phi}} \left( \munderline{ForestGreen}{ [\bs X, \bs Y]_\phi} - [\bs\omega_{|\phi}([\bs X, \bs Y_\phi)]^v_\phi \right), \notag\\
							 &= \munderline{ForestGreen}{ \bs d \bs\alpha_{|\phi}( \bs X_\phi, \bs Y_\phi)} \ 
							 			-  \bs X \cdot \bs\alpha_{|\munderline{red}{\phi}}\left(  [\bs\omega_{|\munderline{red}{\phi}}(\bs Y_{\munderline{red}{\phi}})]^v_{\munderline{red}{\phi}}   \right) 
							 			+  \bs Y \cdot \bs\alpha_{|\munderline{red}{\phi}}\left(  [\bs\omega_{|\munderline{red}{\phi}}(\bs X_{\munderline{red}{\phi}})]^v_{\munderline{red}{\phi}}   \right) 
										+ \bs\alpha_{|\phi}\left(  [\bs\omega_{|\phi}([\bs X, \bs Y]_\phi)]^v_\phi   \right). 
\end{align}
The underlined $\munderline{red}{\phi}$ in the two central terms means that the variational vector fields acting as a differential operator `see' all the $\phi$'s. On the other hand, we have
\begin{align*}
\bs{D^\omega \alpha}_{|\phi} \big(\bs X_\phi, \bs Y_\phi \big) &\defeq \bs d\bs \alpha_{|\phi} (\bs X^h_\phi, \bs Y^h_\phi) 
									  =   \bs d\bs \alpha_{|\phi}\left( \bs X_\phi -  [\bs\omega_{|\phi}(\bs X_\phi)]^v_\phi,  \bs Y_\phi -  [\bs\omega_{|\phi}(\bs Y_\phi)]^v_\phi \right), \\
									&=  \bs d \bs\alpha_{|\phi}( \bs X_\phi, \bs Y_\phi) -  \bs d \bs\alpha_{|\phi}(  [\bs\omega_{|\phi}(\bs X_\phi)]^v_\phi , \bs Y_\phi)
									  									    -  \bs d \bs\alpha_{|\phi}(  \bs X_\phi,  [\bs\omega_{|\phi}(\bs Y_\phi)]^v_\phi)
																		  + \bs d \bs\alpha_{|\phi}\left(  [\bs\omega_{|\phi}(\bs X_\phi)]^v_\phi ,   [\bs\omega_{|\phi}(\bs Y_\phi)]^v_\phi\right), \\
									&=  \bs d \bs\alpha_{|\phi}( \bs X_\phi, \bs Y_\phi) + \iota_{\bs Y} \bs d \left( \bs\alpha_{|\phi}\big( [\omega_{|\phi}(\bs X_\phi) ]^v_\phi  \big) \right)
																		      - \iota_{\bs X} \bs d \left( \bs\alpha_{|\phi}\big( [\omega_{|\phi}(\bs Y_\phi) ]^v_\phi  \big) \right)
																		 +\bs d \bs\alpha_{|\phi}\left(  [\bs\omega_{|\phi}(\bs X_\phi)]^v_\phi ,   [\bs\omega_{|\phi}(\bs Y_\phi)]^v_\phi \right), \\
						&=  \bs d \bs\alpha_{|\phi}( \bs X_\phi, \bs Y_\phi) + \bs Y \cdot \left( \bs\alpha_{|\munderline{red}{\phi}}\big( [\omega_{|\phi}(\bs X_\phi) ]^v_{\munderline{red}{\phi}}  \big) \right)
															     - \bs X \cdot \left( \bs\alpha_{|\munderline{red}{\phi}}\big( [\omega_{|\phi}(\bs Y_\phi) ]^v_{\munderline{red}{\phi}}  \big) \right)
															     +  \bs\alpha_{|\phi} \left(  [\bs\omega_{|\phi}(\bs X_\phi), \bs\omega_{|\phi}(\bs Y_\phi)]^v_\phi \right).
\end{align*}
In the  equality before last above, we have used $\bs L_{\chi^v} \bs\alpha=0$ to rewrite the two central terms -- and the last by \eqref{A1} --  thus considering $[\omega_{|\phi}(\bs Y_\phi)]$ as $\phi$-independent. 
Hence the fact that in the last equality, the  variational vector fields only see the underlined $\munderline{red}{\phi}$'s. 
To further rewrite the last term we use,
\begin{align*}
\bs\Omega_{|\phi}\big( \bs X_\phi, \bs Y_\phi \big)&= \bs d\bs\omega_{|\phi} \big( \bs X_\phi, \bs Y_\phi \big) + \tfrac{1}{2}[\bs\omega_{|\phi}, \bs\omega_{|\phi} ]\big( \bs X_\phi, \bs Y_\phi \big), \\
						  &= \bs X \cdot \bs\omega_{|\phi}(\bs Y_\phi) - \bs Y \cdot \bs\omega_{|\phi}(\bs X_\phi) -\bs\omega_{|\phi}\big( [\bs X, \bs Y]_\phi \big) + [ \bs\omega_{|\phi}(\bs X_\phi),  \bs\omega_{|\phi}(\bs Y_\phi)]. 
\end{align*}
Then we get, 
\begin{align*}
\bs{D^\omega \alpha}_{|\phi} \big(\bs X_\phi, \bs Y_\phi \big) =  \bs d \bs\alpha_{|\phi}( \bs X_\phi, \bs Y_\phi)
													&+ \bs Y \cdot \left( \bs\alpha_{|\munderline{red}{\phi}}\big( [\omega_{|\phi}(\bs X_\phi) ]^v_{\munderline{red}{\phi}}  \big) \right)
												            - \bs X \cdot \left( \bs\alpha_{|\munderline{red}{\phi}}\big( [\omega_{|\phi}(\bs Y_\phi) ]^v_{\munderline{red}{\phi}}  \big) \right) \\
										                 	& + \bs\alpha_{|\phi} \left( [\bs Y \cdot \bs\omega_{|\munderline{red}{\phi}}(\bs X_{\munderline{red}{\phi}}]^v_\phi ) \right) 
										                 	    -  \bs\alpha_{|\phi} \left( [\bs X \cdot \bs\omega_{|\munderline{red}{\phi}}(\bs Y_{\munderline{red}{\phi}}]^v_\phi ) \right) 
											                                         +  \bs\alpha_{|\phi} \left( [\bs\omega_{|\phi}\big( [\bs X, \bs Y]_\phi]^v_\phi  \big)  \right) \\
											                 & \hspace{7cm}                        +  \bs\alpha_{|\phi} \left( [\bs\Omega_{|\phi}\big( \bs X_\phi, \bs Y_\phi \big)]^v_\phi \right), \\
											   =  \bs d \bs\alpha_{|\phi}( \bs X_\phi, \bs Y_\phi)  
										& +  \bs Y \cdot \bs\alpha_{|\munderline{red}{\phi}}\left(  [\bs\omega_{|\munderline{red}{\phi}}(\bs X_{\munderline{red}{\phi}})]^v_{\munderline{red}{\phi}}   \right)	   
							 			    -  \bs X \cdot \bs\alpha_{|\munderline{red}{\phi}}\left(  [\bs\omega_{|\munderline{red}{\phi}}(\bs Y_{\munderline{red}{\phi}})]^v_{\munderline{red}{\phi}}   \right) 
										   + \bs\alpha_{|\phi}\left(  [\bs\omega_{|\phi}([\bs X, \bs Y]_\phi)]^v_\phi   \right) \\
										 & \hspace{7cm}                        +  \bs\alpha_{|\phi} \left( [\bs\Omega_{|\phi}\big( \bs X_\phi, \bs Y_\phi \big)]^v_\phi \right), \\
										 =\bs d\bs \alpha^h_{|\phi} (\bs X_\phi, \bs Y_\phi) &+  \bs\alpha_{|\phi} \left( [\bs\Omega_{|\phi}\big( \bs X_\phi, \bs Y_\phi \big)]^v_\phi \right).
\end{align*}
By \eqref{A2} in the last equality, which proves \eqref{Formula1bis}. 


\subsection{The Abbott-Deser derivation of charges in YM theory}
\label{The Abbott-Deser derivation of charges in YM theory}

One starts with the Yang-Mills equation for a gauge potential $A$ sourced by an external current $(n-1)$-form. 
\begin{align}
\label{YM-eq}
D*\!F = J.
\end{align}
Then one introduces the \emph{first ansatz}, i.e. that $A$ is written a (potentially large) perturbation around a background configuration,
\begin{align}
\label{Ans1}
A=A_0 + \alpha.
\end{align}
From this one as the expansion of the field-strength of $A$,
\begin{align}
F=F_0 + f + \tfrac{1}{2}[\alpha, \alpha], 
\end{align}
with $f\defeq D^{A_0}\alpha=D_0\alpha$. It is further supposed that the background satisfies the source-free YM equation $D_0*\!F_0 = 0$, with $F_0$ the field-strength of $A_0$, so that \eqref{YM-eq} is rewritten,
\begin{align}
\label{YM-eq-perturb}
&\cancel{D_0*\!F_0} + \underbrace{[\alpha, *F_0] + D_0*\!f}_{\text{linear in $\alpha$}}+ \underbrace{ [\alpha, *\! f] + D^A *\! \tfrac{1}{2}[\alpha, \alpha]}_{\text{$\defeq \{D*F\}^R$, order $\geq 2$ in $\alpha$}}=J, \notag \\
&D_0*\!f +[\alpha, *F_0] = J - \{D*F\}^R \defeq j.
\end{align}
The point is  to prove the fact that, on-shell, the newly defined current satisfies a covariant conservation law w.r.t to the background. It is easily found that, on the one hand $D_0D_0 *\!f=[F_0, *\!f]$, and on the other hand 
\begin{align}
D_0[\alpha, *F_0] = [D_0\alpha, *F_0] - [\alpha, \cancel{D_0*\!F_0}]=[f, *F_0]. 
\end{align}
Thus, applying $D_0$ on \eqref{YM-eq-perturb}, one indeed get the on-shell relation $D_0j \approx 0$. 

Now the \emph{second ansatz} is introduced, i.e. that one consider the Killing equation 
\begin{align}
\label{Killing-eq}
D_0\chi = 0,
\end{align}
so that $\chi$ is a symmetry of the background field $A_0$, and from which follows of course $D_0D_0\chi =[F_0, \chi]=0$. Then, from the $D_0$-conservation of $j$ and \eqref{Killing-eq} one obtains
\begin{align}
d\Tr(\chi j)= \Tr (d\chi\, j) + \Tr(\chi\, dj) + \underbrace{\Tr ([A_0, \chi], j) + \Tr(\chi\, [A_0, j])}_{\text{$=0$ by $\H$-invariance of  $\Tr$}} = \Tr (D_0\chi\, j) + \Tr(\chi D_0j) \approx 0,
\end{align}
We have then the conserved Noether current $J_\chi\defeq \Tr(\chi j)$, and integrated over the codimension 1 surface $\Sigma$, it gives the Noether charge $Q_\chi\defeq \int_\Sigma \Tr(\chi j)$ associated with the background Killing symmetry $\chi$. 

But this is not over yet: using again the field equation \eqref{YM-eq-perturb}, we get 
\begin{align}
\label{charge-AB}
Q_\chi = \int_\Sigma \Tr(\chi j)\approx   \int_\Sigma  \Tr(\chi\, D_0*\!f) +  \underbrace{\Tr(\chi\, [\alpha, *F_0])}_{-\Tr([\alpha, \chi] *F_0 )=0} =  \int_\Sigma d  \Tr(\chi *\!f) -  \Tr(\cancel{D_0 \chi} *\!f)
	    =\int_{\d \Sigma} \Tr(\chi *\!f). 
\end{align}
Which may be compared to eq.\eqref{Noether-charge-YM-perturb} in section \ref{Coupled Yang-Mills theory}. 
A final step can be taken to get, 
\begin{align}
\label{charge-AB-bis}
Q_\chi =\int_{\d \Sigma} \Tr(\chi *\!f) = \int_{\d \Sigma} *\Tr(\chi D_0\alpha ) = \int_{\d \Sigma} *\, d\Tr(\chi \alpha), 
\end{align}
using again the Killing equation. Abbott and Deser \cite{Abbott-Deser1982} remark that this, in component, generalises the electric charge in electrodynamics. It reduces to it indeed in the abelian case, as then the Killing equation is $d\chi=0$, so the constant gauge parameter exits the integral -- and we recover the notion that the conservation of the electric charge results from a global (instead of gauge) abelian symmetry. 


\subsection{Commutation relations with the extended bracket \eqref{extended-bracket}}
\label{Commutation relations with the extended bracket}

The bracket under consideration is defined by  $\{\bs\chi, \bs\eta\}\defeq [\bs\chi, \bs\eta] + {\bs\chi}^v(\bs\eta) - {\bs\eta}^v(\bs\chi)$ for $\phi$-dependent gauge parameters. In this paper we have $\bs\chi, \bs \eta \in$ Lie$\bs\H$, which actually reduces the bracket to $-[\bs \chi, \bs \eta]$. But we do not make this simplification so as to keep the calculations valid for field-dependent diffeomorphisms.  
Let us compute, using \eqref{Vert-dphi},
\begin{align}
[\iota_{\bs \chi^v}, \iota_{\bs{d\eta}^v}] \bs d \phi= \iota_{\bs \chi^v} \delta_{\bs{d\eta}} \phi - \cancel{\iota_{\bs{d\eta}^v} \delta_{\bs \chi}\phi} = \delta_{\bs{d\eta}(\bs \chi^v)} \phi= \iota_{[\bs\chi^v(\bs\eta)]^v} \bs d \phi.  
\end{align}
As $ \iota_{\bs{d\eta}^v}$ is manifestly a derivation of degree $0$, $[\iota_{\bs \chi^v}, \iota_{\bs{d\eta}^v}] $ is a derivation of degree $-1$. The above result thus extends to arbitrary variational form: $[\iota_{\bs \chi^v}, \iota_{\bs{d\eta}^v}] =\iota_{[\bs\chi^v(\bs\eta)]^v}$. 

Now let us consider, 
\begin{align}
[\bs L_{\bs \chi^v}, \iota_{\bs\eta^v}] \bs d \phi &= \bs L_{\bs \chi^v} \delta_{\bs \eta} \phi - \iota_{\bs\eta^v}\bs d \left( \iota_{\bs\chi^v}\bs d \phi \right) 
								       = \iota_{\bs \chi^v}  \bs d \left( \delta_{\bs \eta} \phi \right) -  \iota_{\bs\eta^v}\bs d \left( \delta_{\bs \chi} \phi \right), \notag \\
								      &= \iota_{\bs \chi^v}   \left(\delta_{ \bs d\bs \eta} \phi + \delta_{\bs \eta} \bs d \phi \right) - \iota_{\bs \eta^v}   \left(\delta_{ \bs d\bs \chi} \phi  + \delta_{\bs \chi} \bs d \phi \right), \notag\\
								      &= \delta_{\bs\chi^v(\bs\eta)} \phi + \delta_{\bs\eta} \delta_{\bs\chi} \phi -  \delta_{\bs\eta^v(\bs\chi)} \phi - \delta_{\bs\chi} \delta_{\bs\eta} \phi, \notag \\
								      &=\delta_{[\bs\chi, \,\bs \eta]} \phi +  \delta_{\bs\chi^v(\bs\eta)} \phi-  \delta_{\bs\eta^v(\bs\chi)} \phi, \notag \\
								      &= \iota_{ \{ \bs\chi, \bs\eta \}^v} \bs d\phi. 
\end{align}
The bracket of $\iota_{\bs\chi^v}$ and  $\iota_{\bs{d\eta}^v}$  is a derivation of degree $-1$ that extends to any form, so $[\bs L_{\bs\chi^v}, \iota_{\bs\eta^v}] =\iota_{\{\bs\chi, \bs \eta\}^v}$. 
From this  and $[\bs L_{\bs\chi^v}, \bs d]=0$ follows that, 
\begin{align}
[\bs L_{\bs\chi^v}, \bs L_{\bs\eta^v}] &= [\bs L_{\bs\chi^v}, \iota_{\bs\eta^v}\bs d + \bs d \iota_{\bs\eta^v}], \notag\\
						       &=\bs L_{\bs\chi^v} \iota_{\bs\eta^v}\bs d - \iota_{\bs\eta^v} \bs d \bs L_{\bs\chi^v}  \ + \ \bs L_{\bs\chi^v} \bs d \iota_{\bs\eta^v} - \bs d \iota_{\bs\eta^v} \bs L_{\bs\chi^v}, \notag\\
						       &= \bs L_{\bs\chi^v} \iota_{\bs\eta^v}\bs d  - \iota_{\bs\eta^v}  \bs L_{\bs\chi^v}  \bs d  \ + \  \bs d  \bs L_{\bs\chi^v}\iota_{\bs\eta^v}- \bs d \iota_{\bs\eta^v} \bs L_{\bs\chi^v}, \notag\\
						       &= [\bs L_{\bs\chi^v}, \iota_{\bs\eta^v} ] \bs d +  \bs d [\bs L_{\bs\chi^v} \iota_{\bs\eta^v}]= \iota_{ \{\bs\chi, \bs \eta \}^v  } \bs d + \bs d  \iota_{ \{\bs\chi, \bs \eta \}^v  }, \notag \\
						       &=\bs L_{ \{\bs\chi, \bs \eta \}^v  }. 
\end{align}

{
\normalsize 
 \bibliography{Biblio2.5}
}

\end{document}